\documentclass[aps,pra,reprint,superscriptaddress]{revtex4-2}
\usepackage{amsmath}
\usepackage{amssymb}
\usepackage{graphicx}
\usepackage{placeins}
\usepackage{epsfig}

\usepackage{color}

\DeclareMathAlphabet\mathbfcal{OMS}{cmsy}{b}{n} %

\begin{document}
\title{Exceptional points in perturbed dielectric spheres: A resonant-state expansion study}
\author{K. S. Netherwood}
\email{netherwoodks@cardiff.ac.uk}
\affiliation{School of Physics and Astronomy, Cardiff University, Cardiff CF24 3AA, United Kingdom}
\author{H. K. Riley}
\email{hannah.riley-6@postgrad.manchester.ac.uk}
\affiliation{School of Physics and Astronomy, Cardiff University, Cardiff CF24 3AA, United Kingdom}
\affiliation{Department of Physics and Astronomy,  University of Manchester, Manchester M13 9PL, United Kingdom}
\author{E. A. Muljarov}
\email{egor.muljarov@astro.cf.ac.uk}
\affiliation{School of Physics and Astronomy, Cardiff University, Cardiff CF24 3AA, United Kingdom}

\begin{abstract}
Exceptional points (EPs) in open optical systems are rigorously studied using the resonant-state expansion (RSE). A spherical resonator, specifically a homogeneous dielectric sphere in a vacuum,  perturbed by two point-like defects which break the spherical symmetry and bring the optical modes to EPs, is used as a worked example.
The RSE is a non-perturbative approach encoding the information about an open optical system in matrix form in a rigorous way, and thus offering a suitable tool for studying its EPs. These are simultaneous degeneracies of the eigenvalues and corresponding eigenfunctions of the system, which are rigorously described by the RSE and illustrated for perturbed whispering-gallery modes (WGMs). An exceptional arc, which is a line of adjacent EPs, is obtained analytically for perturbed dipolar WGMs. Perturbation of high-quality WGMs with large angular momentum and their EPs are found by reducing the RSE equation to a two-state problem by means of an orthogonal transformation of a large RSE matrix. WGM pairs have opposite chirality in spherically symmetric systems and equal chirality at EPs. This chirality at EPs can be observed in circular dichroism measurements, as it manifested itself in a squared-Lorentzian part of the optical spectra, which we demonstrate here analytically and numerically in the Purcell enhancement factor for the perturbed dipolar WGMs.

\end{abstract}

\maketitle

\section{Introduction}
An exceptional point (EP), originally named by Kato (1966) \cite{kato1966}, is a simultaneous degeneracy of the eigenvalues and the corresponding eigenfunctions of a system. An EP of $N$th-order has $N$ degenerate eigenvalues and eigenfunctions. EPs are a typical feature of open systems, which are characterized by the presence of gain and/or loss of energy and information, and can be described by non-Hermitian matrices which have generally complex eigenvalues \cite{heiss2012physics}.

\begin{figure}
\centering
\includegraphics[width = \linewidth]{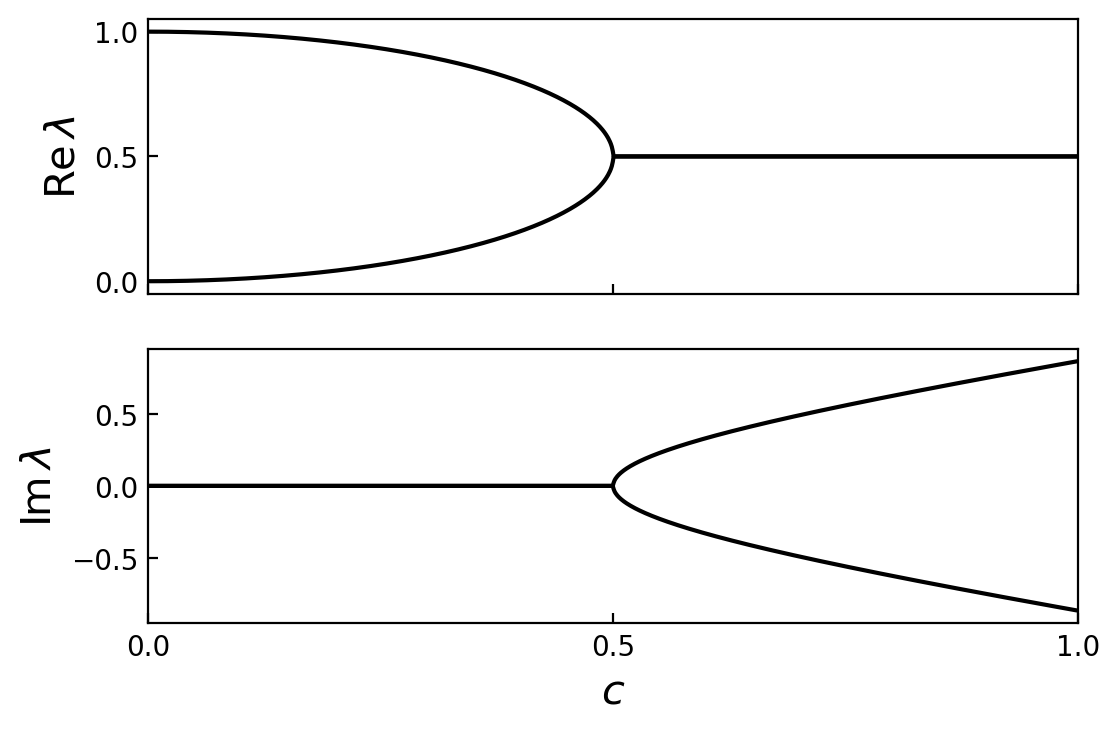}
\caption{Eigenvalues of Eq.(\ref{EP2 matrix}), where $a=0$, $b=ic$, and $d=1$, varied against parameter $c$, taking a value of $c=1/2$ at an EP.}
\label{ex EP}
\end{figure}

Matrices allow a mathematically rigorous and simultaneously the most straightforward investigation of EPs as a special case of their eigenvalues and eigenvectors. To give a mathematical example of an EP, we introduce the $2\times2$ symmetric matrix
\begin{equation}\label{EP2 matrix}
M =
\begin{pmatrix}
a & b\\
b & d
\end{pmatrix}
\end{equation}
where $a$, $b$, and $d$ are complex numbers. The matrix $M$ has the eigenvalues
\begin{equation}\label{EP2 eigenvalue}
\lambda = \frac{a+d}{2} \pm \frac{1}{2}\sqrt{(a-d)^2 + 4b^2}\, .
\end{equation}
To find a point where the eigenvalues are degenerate, we let the square-root term in Eq.(\ref{EP2 eigenvalue}) vanish. This gives the degeneracy condition
\begin{equation}\label{EP2 condition}
b = \pm \frac{i (a-d)}{2}\, .
\end{equation}
If $b \neq 0$ and Eq.(\ref{EP2 condition}) is satisfied, $a$, $b$, and $d$ are the matrix elements of $M$ at an EP. If Eq.(\ref{EP2 condition}) is satisfied but $b=0$, the degeneracy is called a diabolic point (DP) which is a degeneracy of eigenvalues but not eigenvectors. DPs are equivalent to any degeneracies in a Hermitian system, but in a non-Hermitian system they are only the degeneracies that arise due to symmetry, and they generally do not have the characteristic shape of an EP. This characteristic shape along with other features of EPs can be demonstrated, for example, by setting the matrix elements of Eq.(\ref{EP2 matrix}) to $a=0$, $b=ic$, and $d=1$ where $c$ is a real variable. Using Eq.(\ref{EP2 eigenvalue}), the eigenvalues of this example matrix around a second-order EP at $c=1/2$ are plotted in Fig.\ref{ex EP}.

Figure \ref{ex EP} shows the characteristic shape of the eigenvalues in the proximity of an EP. This shape is due to the fact that eigenvalues vary non-linearly with respect to the parameters, and are instead proportional to the square root of the varied parameter in a second-order EP's proximity. More generally, eigenvalues are proportional to the $N$th root of the parameter in the proximity of an $N$th-order EP \cite{seyranian2005coupling}. Furthermore, the eigenvalues have an infinite derivative with respect to the varied parameter when precisely at the EP \cite{heiss2012physics}.

It has been suggested that the extra sensitivity near an EP could be exploited in sensing applications \cite{wiersig2014enhancing,wiersig2016sensors,parto2021non,yu2021whispering}, however it has been shown that while the eigenvalues are non-linear near an EP, the quantum-limited signal to noise ratio is linear, and there is no enhanced sensor precision for an EP sensor compared with a DP sensor \cite{langbein2018no,geng2021discrepancy,duggan2022limitations}. On the other hand, response functions in the proximity of an EP are a topic of recent research \cite{wiersig2022response,hashemi2022linear}, and their potential features are yet to be fully understood, so enhanced precision at an EP is still an open question \cite{zhang2019quantum,wiersig2020robustness,wiersig2020review,peters2022exceptional,he2010ultrasensitive,kuhl2008resonance}. In spite of this, EPs have already been involved in sensing \cite{CHEN2022128534,wiersig2020review}, and new applications for EP sensors have been proposed \cite{jin2018high,goryachev2019probing,jian2020parity,liu2019enhanced,liu2020gravitational}, due to the sensitive splitting of degenerate eigenfrequencies \cite{wiersig2014enhancing}. There are also other features of EPs that can be exploited such as an enhanced spontaneous emission rate which is often attributed to the self-orthogonality of eigenfunctions at an EP \cite{lin2016enhanced,pick2017general}. Spontaneous emission at an EP has a squared-Lorentzian lineshape \cite{yoo2011quantum,takata2021observing} which is useful for increasing the linewidth of a laser \cite{zhang2018phonon}. Contrary to suggestions by \cite{pick2017general,yoo2011quantum,takata2021observing}, different eigenfunctions at an EP are still orthogonal \cite{heiss2012physics} but they are considered parallel at the same time \cite{langbein2018no}.

EPs can occur in any system that can be described by a non-Hermitian matrix and therefore can be met in different fields of physics. For example, the critical damping of a classical harmonic oscillator occurs at an EP \cite{fernandez2018exceptional}. In quantum mechanics, a transformation of a pair of conjugate resonant states (RSs) into a bound-antibound state pair occurs at an EP \cite{moiseyev2011non,TanimuJPC18}. Hydrogen atoms in crossed electric and magnetic fields \cite{cartarius2007exceptional}, graphene metamaterials \cite{liu2017exceptional}, and elastodynamic metamaterials \cite{gupta2022requisites} have been shown to feature EPs. EPs can also be used in topological metasurfaces \cite{song2021plasmonic}. Furthermore, EPs have been studied in non-Hermitian extensions of quantum field theories \cite{alexandre2020discrete}, the standard model of particle physics \cite{millington2021non}, and in evolutionary game theory where a non-Hermitian matrix featuring an EP has been used to describe changes in strategy in rock-paper-scissors games \cite{yoshida2022non}.

A common way to look at EPs is by using parity-time (PT) symmetric Hamiltonians which describe open systems with equal gain and loss. These effective Hamiltonians have real eigenvalues \cite{bender1998real} which is known as pseudo-Hermiticity \cite{mostafazadeh2002pseudo}. An EP occurs at the point of broken PT symmetry where the eigenvalues cease to be real \cite{el2018non,ozdemir2019parity}. In contrast to these studies, the system considered in the present paper has (radiative) loss but no gain so it is not PT symmetric. Passive systems, such as this, can be an advantage in sensing applications \cite{gupta2022requisites}, since loss is inherent to the system, but gain is created actively. Systems with gain thus often require large, complicated, and expensive external components \cite{tsoy2017coupled,wu2019asymmetric,schindler2012symmetric,bender2013observation}.

This work focuses on the EPs of whispering-gallery modes (WGMs) in a dielectric microsphere perturbed by point-like defects (perturbers). WGMs are high-quality RSs of a microsphere formed due the total internal reflection. However, they are only a small part of the whole set of the RSs forming the electromagnetic spectrum of such an open optical system. In a spherically symmetric case, the RSs have either a transverse electric (TE) or transverse magnetic (TM) polarization \cite{doost2014resonant}, and the modes of a given angular momentum number $l$ are $2l+1$ degenerate at a DP. Such degenerate modes with equal wavenumber are usually counted by the magnetic quantum number $m$ \cite{sztranyovszky2022optical}. A perturbation that breaks the spherical symmetry results in some of the states no longer being degenerate at a DP, and the variation of the eigen-wavenumbers by the perturbation is generally different for different eigenstates. However, at specific parameters of the perturbation, the wavenumber for some modes will become degenerate again at an EP \cite{wiersig2011structure,ramezanpour2021generalization} rather than a DP. This kind of EP is the focus of this work.

The aim of this work is to rigorously investigate EPs for perturbed WGMs and to develop a stronger mathematical framework for their study. This is made possible by using the resonant-state expansion (RSE), a method which maps Maxwell's equations onto a linear matrix eigenvalue problem \cite{muljarov2011brillouin} and thus allows direct access to the study of EPs. WGMs were chosen because they are the fundamental modes of the system and have generally high quality factor, however they are only used to illustrate the underlying properties of open systems at EPs and to demonstrate that the RSE is a powerful tool for their study. This paper begins with finding the states of the unperturbed system and then uses the RSE to treat perturbations in a form of point-like defects. The parameter space of this perturbation is explored for locating EPs, illustrating their features, such as the chirality of the RSs and the non-Lorentzian optical spectrum at an EP, and demonstrating at the same time that the majority of the RS wavenumbers are unaffected by these perturbations.

An article on a similar system by Wiersig \cite{wiersig2011structure}, that has also looked at EPs in an optical system without PT symmetry, treats WGMs in an optical microdisk (rather than a microsphere) perturbed by nanoparticles. This study shows that a perturbation that breaks the rotational symmetry of a microdisk can also break the standing-wave nature of degenerate pairs of WGMs. At an EP, the WGMs' eigenvectors become parallel and the corresponding waves propagate in the same direction. One result of that study is a two-mode approximation leading to an effective Hamiltonian that phenomenologically treats the chiral states and EPs. We note, however, that the matrix formulation introduced in the present paper and used for the study of EPs is rigorously derived from the RSE. Some conclusions of Wiersig's study, namely the chirality of EPs and the $2\times2$ matrix approximation of a second-order EP, are reproduced in Sec.\ref{l=20}.

Optical microdisks and microspheres are promising candidates for applications in modern sensing due to the presence of WGMs \cite{gorodetsky1996ultimate}.
An extra sensitivity of WGMs comes from their high quality factors and low mode volume leading also to a strong Purcell enhancement \cite{muljarov2016exact}. As a result, the light-matter coupling of a WGM is strong, and the sensor can be made highly responsive \cite{yu2021whispering}. WGM sensing is advantageous because glass microspheres are easy to fabricate, although liquid water microdroplets are also a promising candidate due to their near perfect spherical symmetry \cite{maayani2016water,giorgini2017fundamental,canales2023self}. WGM sensors have been used for protein \cite{vollmer2002protein}, DNA \cite{vollmer2003multiplexed}, single molecule \cite{armani2007label,vollmer2008whispering}, single virus \cite{vollmer2008single,shao2013detection}, and single nanoparticle \cite{shao2013detection,zhu2010chip} detections. All of these can be treated as point-like defects perturbing the WGMs.

The paper is organized as follows. The electric fields and wavenumbers of the optical modes of a homogeneous sphere are calculated in Sec.\ref{2.2}. Section \ref{3} introduces point-like defects to the system with Sec.\ref{3.1} summarizing the RSE, Sec.\ref{truncation} discussing the truncation of the RSE matrix, and Sec.\ref{2 perturb} looking at the case of only two point-like defects. EPs of perturbed WGMs, with $l=1$ (corresponding to a $3\times 3$ and reducible to a $2\times 2$ matrix problem),  along with the contribution of nearby modes are discussed in Sec.\ref{l=1}. The spectral properties of the system in a two-mode approximation are studied in Sec.\ref{Purcell}, using the Purcell factor of $l=1$ WGMs for illustration.  Sec.\ref{20.1} discusses a WGM EP, with $l=20$ (corresponding to a $41\times 41$ matrix problem), investigating its eigenfunctions and chirality. Section \ref{Orthogonal} uses an orthogonal transformation to represent an EP of the same system by a smaller matrix, equivalent to the original formulation.

\section{Electromagnetic modes of a homogeneous sphere}
\label{2.1}\label{2.2}\label{degenerate}
In order to calculate the wavenumbers and electric fields of the RSs of a microsphere, we first analytically solve Maxwell's equations following the approach in \cite{doost2014resonant}, using Gaussian units and the speed of light in a vacuum equal to 1. The basis system, which is further used in the RSE, is an uncharged non-magnetic homogeneous dielectric sphere of radius $R$, surrounded by vacuum, with magnetic permeability $1$ and the electric permittivity given by
\begin{equation}\label{basis}
\varepsilon(\mathbf{r}) =
\begin{cases}
	\epsilon & \text{for} \; r \leqslant R\,,\\
	1 & \text{for} \;  r > R\,,
\end{cases}
\end{equation}
where $\mathbf{r}$ is the position vector and $r= \lvert \mathbf{r} \rvert$ is the radial position.
The refractive index of the microsphere is therefore $n_r=\sqrt{\epsilon}$.

Assuming the time dependent factor of the fields is $e^{-ikt}$ \cite{stratton2007electromagnetic}, Faraday's and Ampere's laws take the form
\begin{equation}\label{Max}
\nabla \times \mathbf{E} = ik\mathbf{H}\,,\;\;\;\;\;\; \nabla \times \mathbf{H} = -ik\varepsilon(\mathbf{r})\mathbf{E}\,,
\end{equation}
respectively, where $\mathbf{E}$ is the electric field, $\mathbf{H}$ is the magnetic field, and $k$ is the RS wavenumber in vacuum which is {\it complex} in finite optical systems.
By combining Maxwell's equations (\ref{Max}), we obtain Maxwell's wave equation for the electric field
\begin{equation}\label{Max wave}
\nabla \times \nabla \times \mathbf{E}(\mathbf{r}) = k^2 \varepsilon(\mathbf{r}) \mathbf{E}(\mathbf{r})\,.
\end{equation}
Using the spherical symmetry, Eq.(\ref{Max wave}) is solved by separating variables in spherical coordinates with $\mathbf{r}=(r,\theta,\varphi)$ and introducing a scalar function
\begin{equation}
f(\mathbf{r}) = {\cal R}_l(r) Y_{lm}(\theta,\varphi)\,,
\end{equation}
where $\theta$ is the polar angle, $\varphi$ is the azimuthal angle, and $Y_{lm}(\theta,\varphi)$ are the real-valued  spherical harmonics
\begin{equation}\label{Y}
Y_{lm}(\theta,\varphi) = \sqrt{\frac{2l+1}{2}\frac{(l-\lvert m \rvert)!}{(l+\lvert m \rvert)!}}P_l^{\lvert m\rvert}(\cos\theta) \chi_m(\varphi)
\end{equation}
with $P_l^{\lvert m\rvert}(\cos\theta)$ being the associated Legendre polynomials and
\begin{equation}\label{chi}
\chi_m(\varphi) =
	\begin{cases}
		\pi^{-\frac{1}{2}}\sin(m\varphi) & \text{for $m<0$}\,,\\
		(2\pi)^{-\frac{1}{2}} & \text{for $m=0$}\,,\\
		\pi^{-\frac{1}{2}}\cos(m\varphi) & \text{for $m>0$}\,.
	\end{cases}
\end{equation}
The radial functions, for the RSs satisfying the outgoing boundary conditions, take the form
\begin{equation}\label{Rl}
{\cal R}_l(r) =
	\begin{cases}
	\dfrac{j_l(n_r kr)}{j_l(n_r kR)}   & \text{for} \; r \leq R\,,\\
	\dfrac{h_l^{(1)}(kr)}{h_l^{(1)}(kR)}   & \text{for} \; r > R\,,
	\end{cases}
\end{equation}
where $j_l(z)$ and $h_l^{(1)}(z)$ are the spherical Bessel function and the Hankel function of the first kind, respectively.

As found in Appendix A, the solutions of Eq.(\ref{Max wave}) in spherical coordinates, for the electric fields of the RSs are given by
\begin{equation}\label{ETE}
\mathbf{E}^{\text{TE}}(\mathbf{r}) = A_l^{\text{TE}}{\cal R}_l(r)
\begin{pmatrix}
0\\
\dfrac{1}{\sin\theta}\dfrac{\partial}{\partial\varphi}Y_{lm}(\theta,\varphi)\\
-\dfrac{\partial}{\partial\theta}Y_{lm}(\theta,\varphi)
\end{pmatrix}
\end{equation}
in TE polarization and
\begin{equation}\label{ETM}
\mathbf{E}^{\text{TM}}(\mathbf{r}) = \frac{A_l^{\text{TM}}}{\varepsilon(\mathbf{r}) k r}
\begin{pmatrix}
l(l+1) {\cal R}_l(r) Y_{lm}(\theta,\varphi)\\
\dfrac{\partial}{\partial r} r {\cal R}_l(r) \dfrac{\partial}{\partial \theta} Y_{lm}(\theta,\varphi)\\
\dfrac{\partial}{\partial r} \dfrac{r {\cal R}_l(r)}{\sin\theta} \dfrac{\partial}{\partial\varphi} Y_{lm}(\theta,\varphi)
\end{pmatrix}
\end{equation}
in TM polarization, with the normalization factors \cite{doost2014resonant}
\begin{equation}\label{Anorm}
\begin{split}
A_l^{\text{TE}} &= \sqrt{\frac{1}{l(l+1)R^3(n_r^2-1)}},\\
A_l^{\text{TM}}  = n_r A_l^{\text{TE}} &\left(\left[\frac{j_{l-1}(n_r kR)}{j_l(n_r kR)} - \frac{l}{n_r kR}\right]^2 + \frac{l(l+1)}{k^2 R^2} \right)^{-\frac{1}{2}}\, .
\end{split}
\end{equation}

By imposing Maxwell's boundary conditions, we obtain the secular equation \cite{muljarov2020full}
\begin{equation}\label{secular}
\beta\frac{J_l^\prime(n_r k R)}{J_l(n_r k R)} =\frac{H_l^{\prime}(k R)}{H_l(k R)}\,,
\end{equation}
where $\beta=n_r$ for TE polarization and $\beta=n_r^{-1}$ for TM polarization, $J_l(z)=zj_l(z)$, $H_l(z)=zh_l^{(1)}(z)$, and the prime denotes the derivative with respect to the argument. Solving Eq.(\ref{secular}) numerically gives the wavenumbers $k$ of the RSs of the dielectric microsphere.\\

\begin{figure}
\includegraphics[width=1\linewidth]{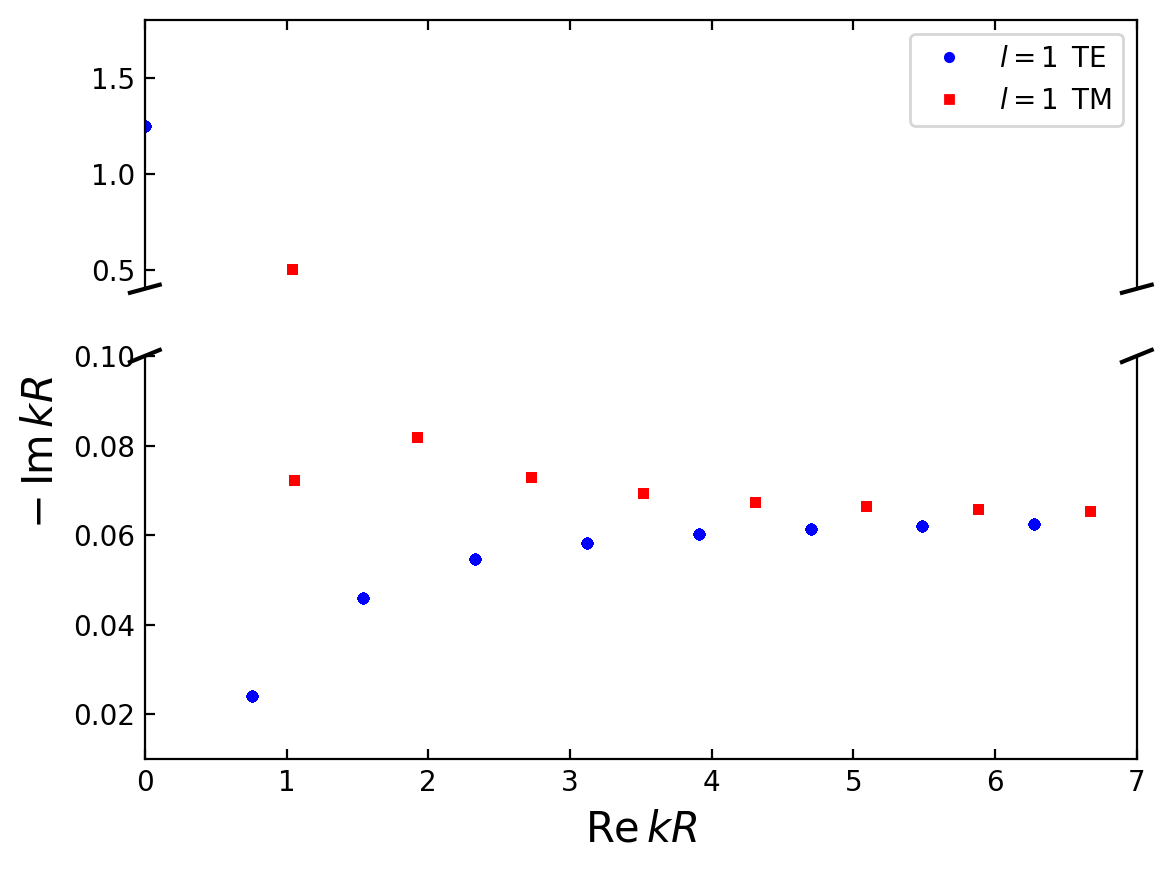}
\caption{(Color online) Wavenumbers of the TE (blue circles) and TM (red squares) $2l+1$ degenerate modes of a homogeneous dielectric sphere in a vacuum with refractive index $n_r=4$ { and angular momentum number $l=1$}.  }
\label{l=1 modes}
\end{figure}

\begin{figure*}
\includegraphics[width=0.49\linewidth]{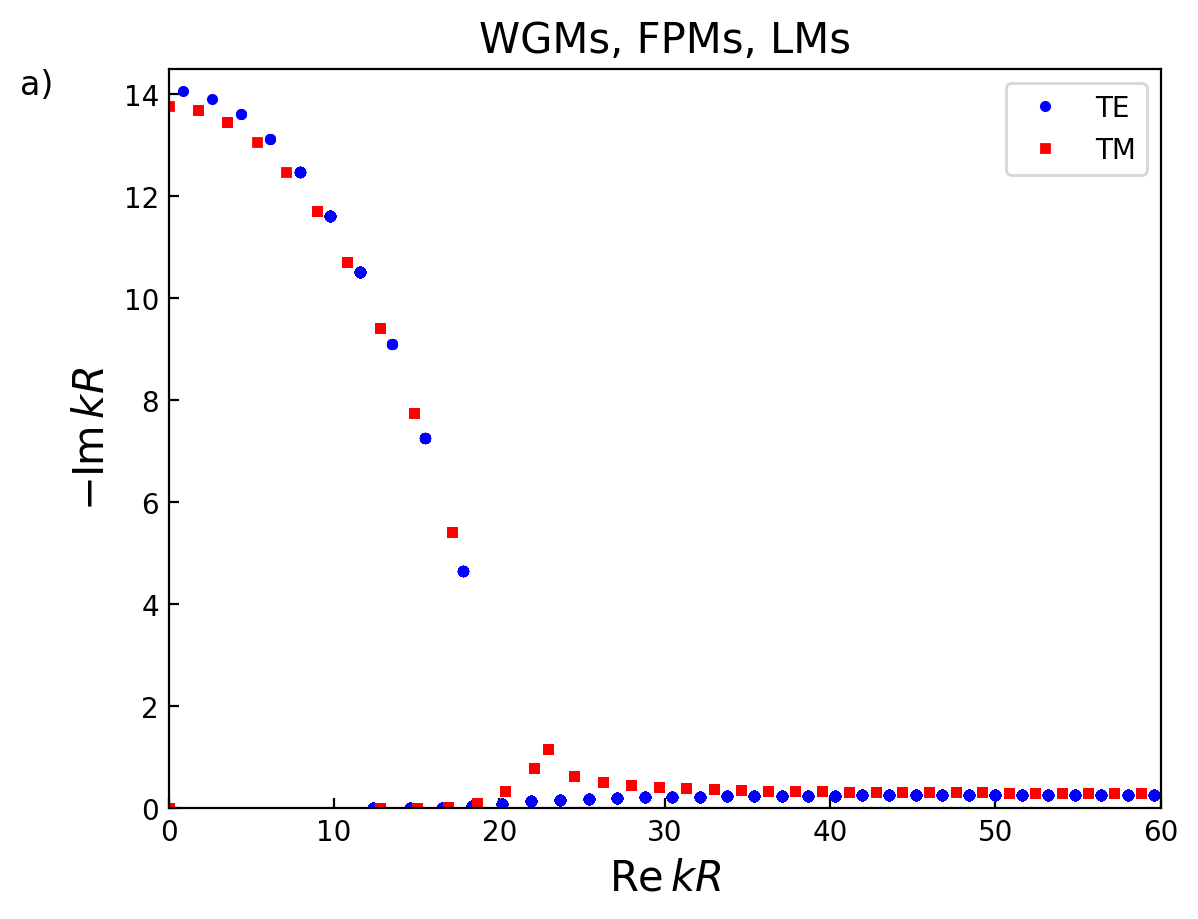}
\includegraphics[width=0.49\linewidth]{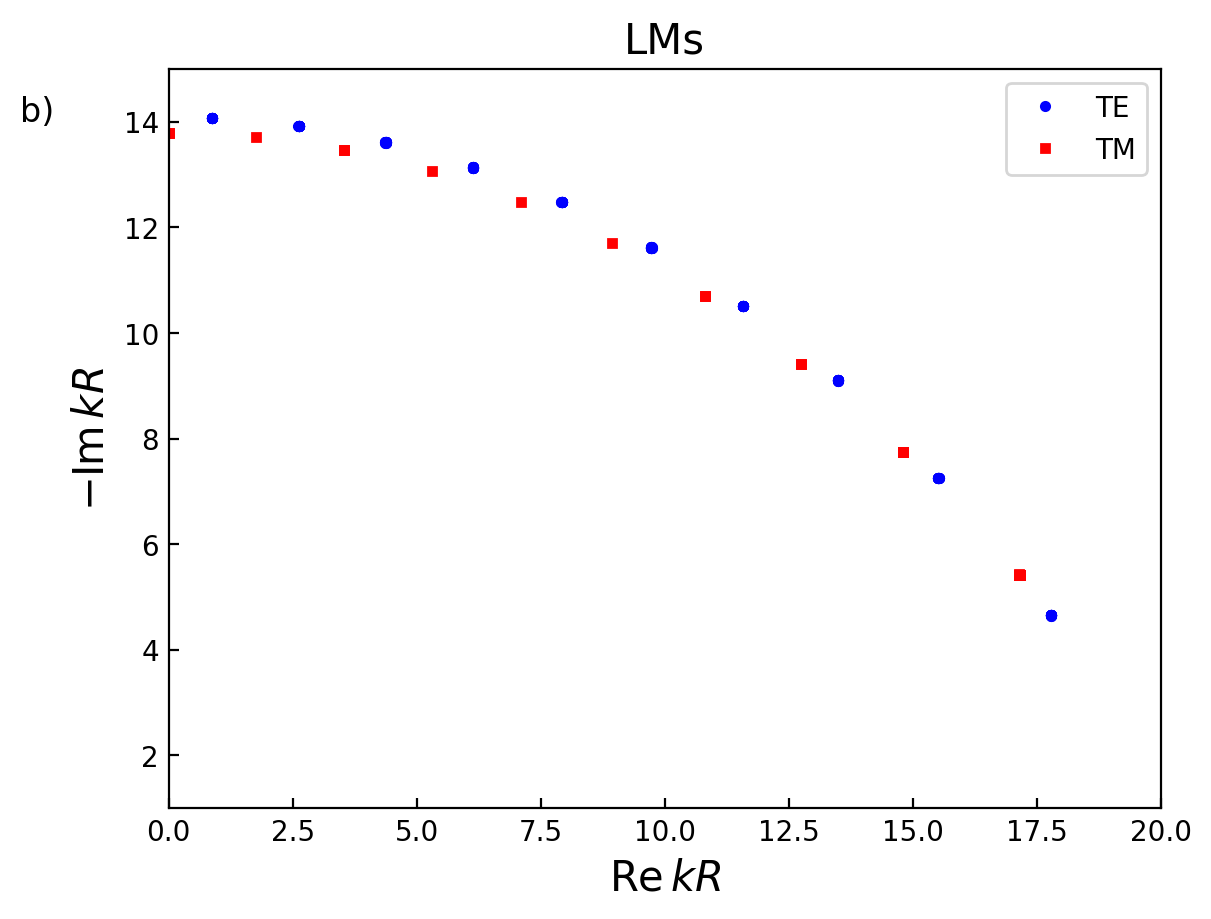}
\includegraphics[width=0.49\linewidth]{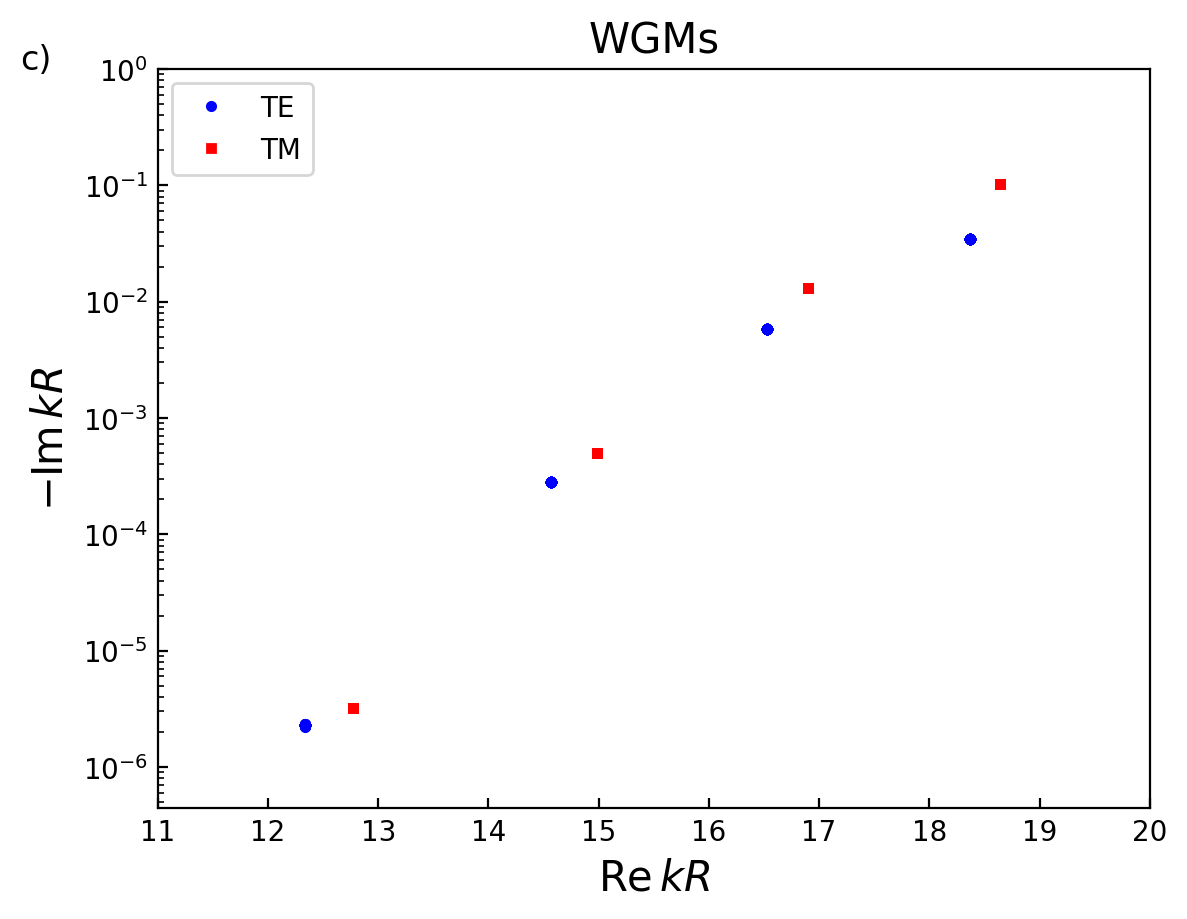}
\includegraphics[width=0.49\linewidth]{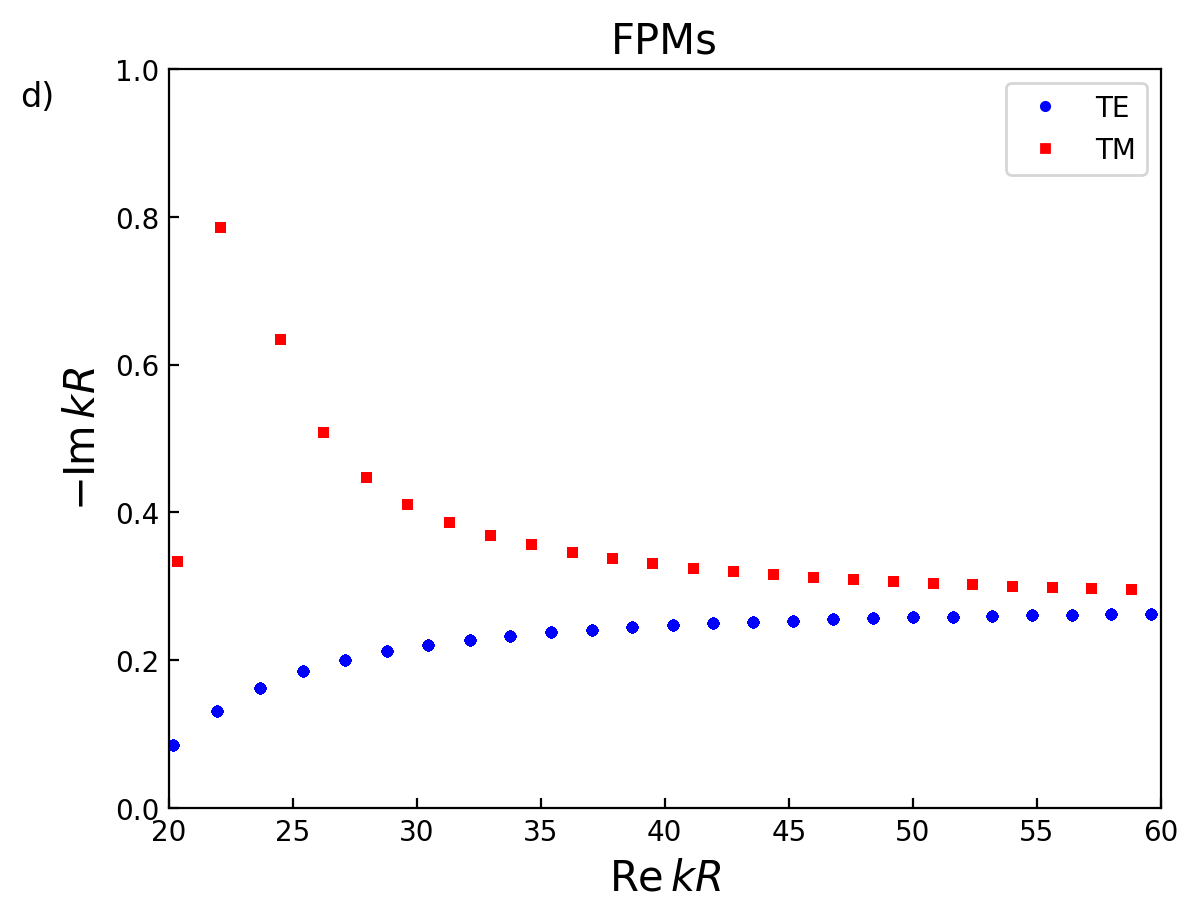}
\caption{(Color online) Wavenumbers of the TE (blue circles) and TM (red squares) $2l+1$ degenerate modes of a homogeneous dielectric sphere in a vacuum with refractive index $n_r=2$ and angular momentum number $l=20$. (a) All types of optical modes, (b) LMs, (c) WGMs (using a $\log_{10}$ scale for the imaginary part), and (d) FPMs. Reproduction of Fig.1 in \cite{sztranyovszky2022optical}.}
\label{l=20 modes}
\end{figure*}

Figures \ref{l=1 modes} and \ref{l=20 modes} show the numerical solutions to Eq.(\ref{secular}) for both TE and TM polarizations, with Fig.\ref{l=1 modes} using the parameters $l=1$ and $n_r=4$, and Fig.\ref{l=20 modes} using the parameters $l=20$ and $n_r=2$. Note that Fig.\ref{l=20 modes} reproduces Fig.1 in \cite{sztranyovszky2022optical}. The refractive index $n_r=4$ was chosen for small $l$ to ensure the system would exhibit WGMs because the quality-factor and number of WGMs increase with both $l$ and $n_r$. The $\text{Re}\,k<0$ domain of the spectrum is not shown because it is a mirror image of the $\text{Re}\,k > 0$ domain with respect to the imaginary axis, due to the fact that optical modes in any finite system are paired. Namely, for every RS with the wavenumber $k$ (with $\text{Re}\,k\neq0$) and electric field $\mathbf{E}$, there is always a conjugate RS with the wavenumber $-k^\ast$ and electric field $\mathbf{E}^\ast$.


In the spectra in Figs.\ref{l=1 modes} and \ref{l=20 modes}, the TE and TM modes alternate  as $\text{Re}\,k$ increases or decreases. The WGMs are the modes with a small imaginary component of $k$ and the real component $\lvert\text{Re}\,k\rvert <k_c$ where $k_c=l/R$ is the critical wavenumber of the total internal reflection. The $l=1$ fundamental mode and the eight $l=20$ modes in Fig.\ref{l=20 modes}(c) are WGMs. The modes with $\lvert\text{Re}\,k\rvert >k_c$ that are not leaky modes (LMs) are called Fabry-P\'erot modes (FPMs) of which there are a countable infinite number. LMs are the modes with a rather large imaginary component of $k$. For any $l\geqslant1$, there are $l$ TE and $l-1$ TM LMs  (each with $2l+1$ degeneracy). This is shown clearly for $l=20$, as there are 10 distinct LMs of each polarization, with each mode having a pair in the negative real domain except for the TM LM having $\text{Re}\,k=0$. The number of LMs is less clear for $l=1$ because of a TM mode with wavenumber $kR=1.039-0.501i$ which is the Brewster-peak mode considered to be a hybrid LM-FPM \cite{sztranyovszky2022optical}.

The physical interpretations of these optical modes are based on their reflections from the spherical surface  \cite{sztranyovszky2022optical}.  LMs, sometimes called external resonances \cite{meeten2019flow}, are such named because their electromagnetic field is distributed mostly outside the sphere and therefore these modes have large radiative losses. FPMs are due to reflections between two opposite sides of the sphere surface, similar to a parallel-mirror Fabry-P\'erot resonator \cite{perot1899application}. WGMs form by imperfect total internal reflection at the curved inner surface of the sphere \cite{oraevsky2002whispering}. These waves thus travel in great circles along the sphere surface, hence why their propagation can be clockwise (CW) or counterclockwise (CCW).

Excluding the case of $m=0$, WGMs in spheres exist in pairs with one CW and the other CCW. This is when $m<0$ and $m>0$, respectively, due to the often used  \cite{jackson1999classical} azimuthal function $e^{im\varphi}$ as a phasor in the complex plane. Here, chirality is the property of how much a mode is CW or CCW propagating. Replacing the usual $e^{im\varphi}$ with the real trigonometric functions $\chi_m(\varphi)$ in Eq.(\ref{chi}) means that the chirality is no longer given simply by the sign of $m$, but rather a combination of the functions $\chi_m(\varphi)$ with positive and negative $m$, for the same $|m|$. Spherically symmetric systems always have equal chirality contributions from CW and CCW propagation. Introducing a perturbation that breaks the spherical symmetry can lead to a chirality imbalance of the WGMs. Since eigenfunctions become equal (up to a factor of $i$) at an EP, it has been suggested that a pair of usually opposite chirality modes may have equal chirality at an EP \cite{wiersig2011structure}. This will be analyzed and discussed in depth in Secs.\ref{Purcell} and \ref{l=20}.

\section{Homogeneous sphere perturbed by point-like defects}\label{3}
In order to break the spherical symmetry, and thus reveal an EP, we introduce a finite number of point-like defects inside and/or outside the sphere. The perturbation of the permittivity due to these point-like defects is given by
\begin{equation}\label{DV}
\Delta \varepsilon (\mathbf{r}) = \sum_j \alpha_j \delta(\mathbf{r}-\mathbf{r}_j)
\end{equation}
where $\alpha_j$ is the strength of the $j$th perturber which is qualitatively the product of its electric permittivity and volume, $\mathbf{r}_j$ is the position vector of the $j$th perturber, and $\delta(\mathbf{r})$ is the three-dimensional Dirac delta function.

\subsection{The resonant-state expansion}
\label{3.1}
To solve this perturbation problem, we use the RSE \cite{muljarov2011brillouin}, which is a rigorous non-perturbative method for open optical systems, that maps Maxwell's equations onto a linear matrix eigenvalue problem and is capable of treating perturbations of arbitrary strength and shape. In simplified versions, it is also suitable for treating the case of degenerate states, and perturbation theory corrections to first, second \cite{doost2014resonant}, or higher orders can also be extracted.

We remark that the RSE has also been  applied to non-relativistic \cite{tanimu2018resonant} and relativistic open quantum systems, wherein one solves Schr\"odinger's or Dirac's wave equation with outgoing wave boundary conditions in an analogous way to solving Maxwell's equations. The RSE therefore provides a rigorous description of EPs in  those systems as well, however most of these results remain unpublished to date.

With an entirely internal perturbation, an RSE matrix with a complete basis of optical modes yields an exact solution for the perturbed system. However, outside the basis system, the resonant states are lacking  completeness due to the open nature of the system. As a result, the RSE does not presently work for external perturbations, even if all optical modes are included in the basis. Nevertheless, the RSE equation can be used in the same form for any external perturbations as soon as the corrections to the RS wavenumbers are limited to first-order in the perturbation strength or perturbation volume.  In fact, the RSE was shown to be correct to first-order for volume and/or boundary perturbations \cite{sztranyovszky2023first} or even for perturbations of the medium surrounding the optical system, although homogeneous medium perturbations can be treated rigorously by the RSE in any order \cite{almousa2021varying}. For non-degenerate modes, using the RSE in first-order implies a single-mode approximation, neglecting any off-diagonal elements of the perturbation matrix. For a subset of degenerate modes, all the matrix elements of the perturbation within such a subset must be kept as all of them contribute in first-order. This is exactly the case treated in this work, in which the optical modes of the dielectric spheres found in Sec.\ref{degenerate} are used as a basis for the RSE. Section \ref{l=1} considers both internal and external perturbations of the basis sphere, for $l=1$ (3 degenerate modes), Sec.\ref{Purcell} considers only internal perturbations for $l=1$, and Sec.\ref{l=20} considers only external perturbations for $l=20$ (41 degenerate modes).

The RSE matrix equation has the form \cite{muljarov2011brillouin}
\begin{equation}\label{RSE}
\sum_{n^\prime} H_{nn^\prime} C_{n^\prime\nu} = \frac{1}{\varkappa_\nu}C_{n\nu}\,,
\end{equation}
where $n$ is the matrix index counting the unperturbed modes, $\nu$ is the index that counts the perturbed modes, $C_{n\nu}$ is a square matrix whose columns are the eigenvectors, and $\varkappa_\nu$ is the wavenumber of the perturbed state $\nu$. The RSE matrix in the eigenvalue problem  Eq.(\ref{RSE}) is given by
\begin{equation}\label{H}
H_{nn^\prime} = \frac{\delta_{nn^\prime}}{k_n} + \frac{V_{nn^\prime}}{\sqrt{k_n}\sqrt{k_{n^\prime}}}\,,
\end{equation}
where $\delta_{nn^\prime}$ is the Kronecker delta and
\begin{equation}\label{V int}
V_{nn^\prime} = \int d\mathbf{r}\, \mathbf{E}_n(\mathbf{r}) \cdot \Delta \varepsilon(\mathbf{r}) \mathbf{E}_{n^\prime}(\mathbf{r})
\end{equation}
is the perturbation matrix, with
$\mathbf{E}_n$ being the electric field of mode $n$. It should be noted that Eq.(\ref{H}) is missing a factor of $1/2$ in the second term compared to \cite{doost2014resonant,muljarov2011brillouin} which is accounted for by $A_l^{\text{TE}}$ in Eq.(\ref{Anorm}) losing a factor of $\sqrt{2}$ compared to \cite{doost2014resonant}.

As well as the wavenumbers, the electric fields are also changed by the perturbation. These perturbed electric fields, which are the eigenfunctions of the perturbed Maxwell's equations, are expanded into the states of the basis system \cite{muljarov2011brillouin}
\begin{equation}\label{Ecal}
\mathbfcal{E}_\nu(\mathbf{r}) = \sqrt{\varkappa_\nu} \sum_n \frac{1}{\sqrt{k_n}} C_{n\nu} \mathbf{E}_{n}(\mathbf{r})\,.
\end{equation}
This eigenfunction expansion is fundamental to the derivation of the RSE and is why it is so named. The eigenvectors of the RSE matrix equation satisfy the orthonormality relation \cite{lobanov2018resonant}
\begin{equation}\label{norm}
\sum_n C_{n\nu} C_{n\nu^\prime} = \delta_{\nu\nu^\prime}\,.
\end{equation}
It should be noted that while the eigenvectors of a Hermitian matrix are normalized using the square modulus, and their orthogonality relation contains the complex conjugate of one of the eigenvectors, this is not the case for non-Hermitian matrices [as it is clear from Eq.(\ref{norm})] which are the primary consideration of this paper.

To bring the RSE from the general case to one that treats a basis system with point-like defects, we substitute Eq.(\ref{DV}) into Eq.(\ref{V int}) to get the perturbation matrix
\begin{equation}\label{V}
V_{nn^\prime} = \sum_j \alpha_j \mathbf{E}_n(r_j,\theta_j,\varphi_j) \cdot \mathbf{E}_{n^\prime}(r_j,\theta_j,\varphi_j)\,,
\end{equation}
where $r_j$, $\theta_j$, and $\varphi_j$ are the radial, polar, and azimuthal positions of the $j$th perturber, respectively.
In Secs.\ref{l=1}-\ref{l=20}, using Eq.(\ref{basis}) as the basis system and Eq.(\ref{DV}) as the perturbation, the matrix $H_{nn^\prime}$ is diagonalized to find the perturbed wavenumbers $\varkappa_\nu$, which are the reciprocal of the eigenvalues of $H_{nn^\prime}$, and the expansion coefficients $C_{n\nu}$ which are the elements of the eigenvector blocks and contribute to the perturbed electric fields $\mathbfcal{E}_\nu$ via Eq.(\ref{Ecal}). The RSE matrix is symmetric due to the lack of conjugation in Eq.(\ref{V int}). However, if  instead of the real azimuthal functions in Eq.(\ref{chi}), the complex functions $e^{im\varphi}$ are used, Eq.(\ref{V int}) requires a conjugation of the angular part of $\mathbf{E}_n(\mathbf{r})$. This thus leads to a non-symmetric matrix which is equally valid but can complicate analysis and is thus avoided in this work.

While this paper focuses on dielectric spheres perturbed by point-like defects, other geometries and dimensionalities can be rigorously described by the RSE, as demonstrated in \cite{doost2012resonant,doost2014resonant,armitage2014resonant,neale2020resonant,neale2021accidental,almousa2021varying}. We expect that the approach used here to study EPs can be applied to other complicated systems by changing the basis states and/or perturbation.

\subsection{Resonant-state expansion matrix truncation}\label{truncation}
Since there are a countable infinite number of the modes of an optical system, and the RSE matrix index $n$ counts all of them, $H_{nn^\prime}$ is an infinite matrix. While matrices of this size can be treated analytically in a handful of cases, many problems have only numerical solutions which require matrices of finite size. So, in practice, a basis with a finite number of modes is normally used for the RSE. Usually, the size of the basis is determined by the required accuracy of the RSE calculation; in this way one keeps the method numerically exact \cite{doost2012resonant}.

An essential technique for truncating the RSE matrix to finite size exploits the fact that optical modes with very different values of $\lvert k_n \rvert$ do not interact as much as closer ones. This is due to the second-order Rayleigh-Schr\"odinger perturbation theory corrections being inversely proportional to the wavenumber difference \cite{doost2014resonant}. As soon as only first-order corrections to the RS wavenumbers are of interest,  the exclusion of all non-degenerate modes from the basis is justified, although nearby modes can still be included and their effect is investigated in Sec.\ref{l=1}.
In a basis consisting of only degenerate modes of a spherical system, all having the same wavenumber, the index $n$ labeling the basis RSs within such a degenerate block of matrix $H_{nn^\prime}$ can be identified as the magnetic quantum number $m$.

Further truncation can be achieved when all of the perturbers are on the same plane. This allows all perturbers to be considered on the equatorial plane $\theta_j=\pi/2$ where the inner products of the electric fields of some modes vanish. With all perturbers on one plane, modes with equal $l$, same polarization, and different $m$ values of opposite parity become orthogonal. A significant consequence of this is that each degenerate block of matrix $H_{nn^\prime}$  is further split into two independent blocks, one for even-$m$ and the other for odd-$m$ modes which can be solved separately. Furthermore, TE electric fields with $m=0$ and even $l$ vanish in this case. The proofs for these statements are provided in Appendix \ref{3.2}. %

\subsection{Two point-like defects}\label{2 perturb}
Mathematically, $H_{nn^\prime}$ can feature EPs due to being non-Hermitian. In fact, the eigen-wavenumbers of the unperturbed system $k_n$  are complex. They contribute not only to the diagonal part of  $H_{nn^\prime}$, according to Eq.(\ref{H}), but also to ${\cal R}_l(r)$, making them, and consequently the perturbation matrix $V_{nn^\prime}$, complex, see Eqs.(\ref{Rl}) and (\ref{V}). However, in TE polarization, at least two perturbers are required for the system to exhibit an EP. If there is only one perturber, the radial function can be factored out of the matrix Eq.(\ref{V}) leaving a Hermitian matrix multiplied by a complex factor, and Hermitian matrices by no means can feature EPs. The same is also true for multiple perturbers all sharing the same radial position, since the angular part of the basis RS wave function is real, see Eq.(\ref{ETE}).

Sections \ref{l=1}--\ref{l=20} of this work consider only the two perturber case with $j=1,2$ in Eq.(\ref{DV}). For degenerate basis modes, changing both $\alpha_1$ and $\alpha_2$ while maintaining their ratio $\alpha=\alpha_2/\alpha_1$ simply scales the eigenvalues and eigenvectors, so only the ratio matters for locating an EP. Due to the spherical symmetry, the absolute azimuthal position of the perturber $\varphi_j$ is not significant, but the angle between the two perturbers $\Delta \varphi=\varphi_2-\varphi_1$ is. The polar angles of the perturbers $\theta_j$ and their difference are not free parameters because two points can always be considered to be on the same plane, and consequently, the parity selection rules in Appendix \ref{3.2} can always be used in this case. The free parameters that are used for searching for EPs are thus $\alpha$, $r_1$, $r_2$, and $\Delta \varphi$.

\section{Dipolar modes}\label{l=1}
With two perturbers, $j=1,2$, we consider only the $l=1$ triply degenerate fundamental TE mode (the WGM) of the sphere with $n_r=4$, shown in Fig.\ref{l=1 modes}, which has the unperturbed wavenumber $k_0R=0.754 - 0.024i$. The full RSE matrix $H_{nn^\prime}$ is then reduced to a $3\times3$ matrix. We can truncate this matrix further to $2\times2$ by considering only $m=\pm1$ states, owing to the parity selection rules derived in Appendix \ref{3.2} and summarized in Sec.\ref{truncation}. Then the EP condition can be treated entirely analytically, resulting in simple explicit expressions.

\begin{figure}
\includegraphics[width=1.0\linewidth]{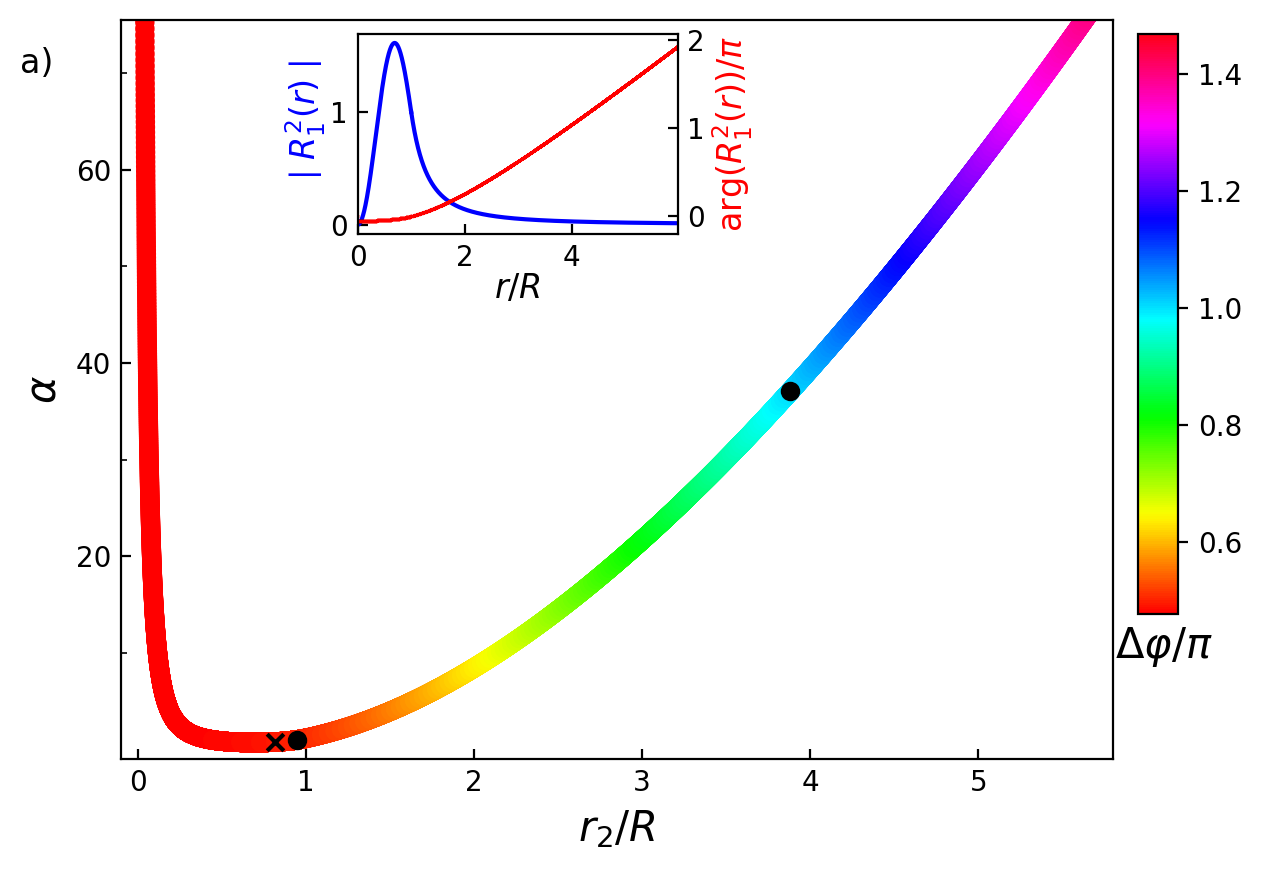}
\includegraphics[width=1.0\linewidth]{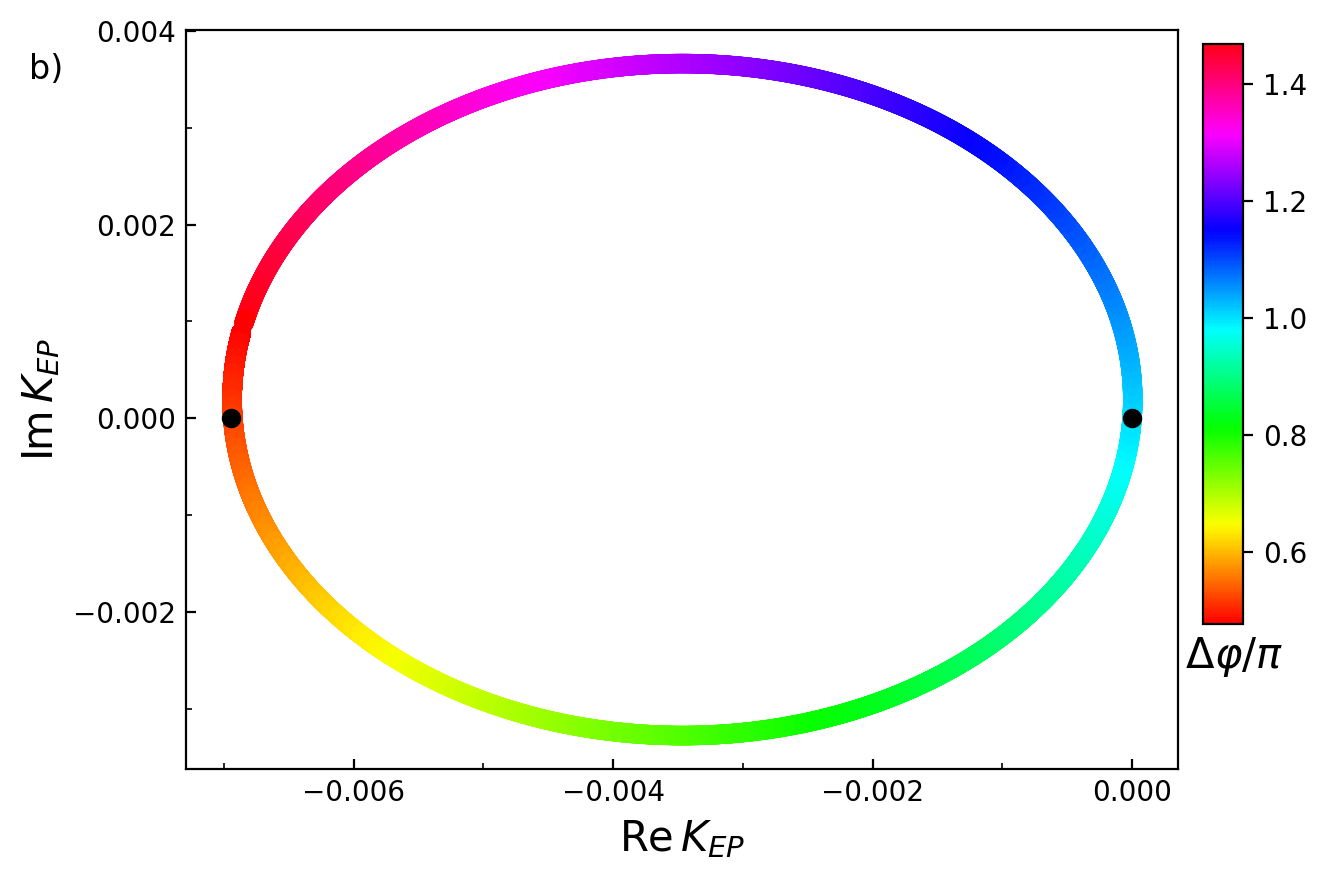}
\includegraphics[width=0.75\linewidth]{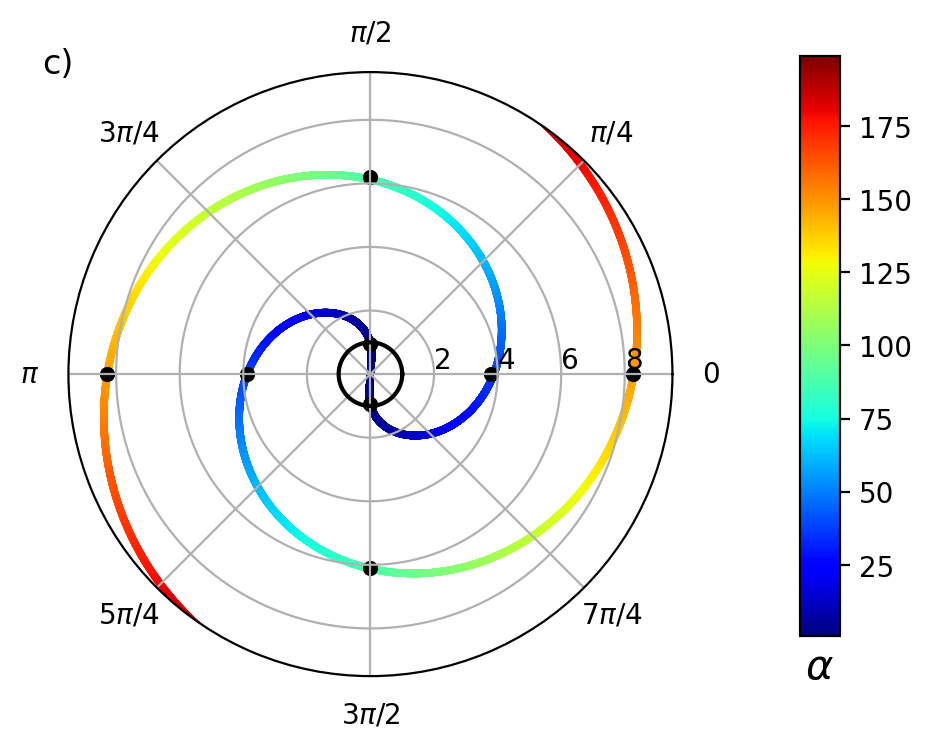}
\caption{(Color online) Solutions of Eq.(\ref{H condition}), the condition where the eigenvalues of $H_{nn^\prime}$ (with an $m=\pm1$ two-mode TE basis of the fundamental WGM in Fig.\ref{l=1 modes}) are degenerate, in the parameter space of the system (dielectric microsphere in vacuum with two perturbers).  Black circles denote the parameters of DPs. The radial position of the first perturber is fixed at $r_1/R=0.95$. (a) Parameters satisfying Eq.(\ref{H condition}): $\alpha$, $r_2/R$, and $\Delta \varphi/\pi$ with the latter represented by color shown in the range $\Delta \varphi/\pi=0.477$ to $\Delta \varphi/\pi = 1.477$. Inset plots the modulus of the squared radial function $\lvert {\cal R}_1^2(r)\rvert$ in the left axis and its phase $\text{arg}{\,\cal R}_1^2(r)/\pi$ in the right axis against radial position. Black cross denotes the parameters of the EP shown in Fig.\ref{l=1 EP}. (b) $K_{\rm EP}$ for the same degeneracies as (a) with color representing the same range of $\Delta\varphi/\pi$. (c) Positions of the second perturber $r_2/R$ and $\varphi_2=\Delta\varphi$ that lead to EPs in polar coordinates (with $\varphi_1=0$ and $\theta_1=\theta_2=\pi/2$) with the corresponding $\alpha$ shown by color. Black thick hollow circle shows the surface at $r=R$ of the microsphere.
}
\label{EA}
\end{figure}

In fact, the truncation of the RSE matrix to this size allows us to use the degeneracy condition Eq.(\ref{EP2 condition}),
\begin{equation}\label{H condition}
H_{11}-H_{22} \pm 2i H_{12} = 0\,,
\end{equation}
where the indices $n=1$ and $n=2$ denote the modes with $m=-1$ and $m=1$, respectively. Equation (\ref{H condition}) is solved explicitly in Appendix C leading to the following EP condition:
\begin{equation}\label{degen condition}
e^{\pm 2i\Delta\varphi} = -\alpha \frac{{\cal R}_1^2(r_2)}{{\cal R}_1^2(r_1)} \,.
\end{equation}
To calculate the parameters that result in an EP degeneracy, this complex equation is split into its magnitude
\begin{equation}\label{alpha}
\frac{\alpha_2}{\alpha_1} = \alpha = \left\lvert \frac{{\cal R}_1(r_1)}{{\cal R}_1(r_2)} \right\rvert^2
\end{equation}
and phase
\begin{equation}\label{phi}
\varphi_2 - \varphi_1 = \Delta\varphi = \text{arg}\frac{{\cal R}_1(r_2)}{{\cal R}_1(r_1)} \pm \frac{\pi}{2}
\end{equation}
for any given values of $r_1$ and $r_2$. Note that the $\pm$ sign in Eq.(\ref{degen condition}) arises from the mirror symmetry of the perturbed system, making the phases $\Delta\varphi$ and $-\Delta\varphi$ indistinguishable.  We therefore choose the positive sign in $e^{\pm2i\Delta\varphi}$ without loss of generality. The $\pm$ sign in Eq.(\ref{phi}) is of different nature, which has nothing to do with the mirror symmetry of the system but is rather related to the fact that the coupled modes with $m=-1$ and $m=1$ have the difference $\Delta m=2$ in the magnetic quantum number. This results in the rotational symmetry (with the rotation angle of $\pi$) of the EP diagram in Fig.\ref{EA}(c), see below for a detailed discussion. Note also that if $\alpha_1$ and $\alpha_2$ have opposite signs, so that $\alpha<0$, the left-hand side of Eq.(\ref{alpha}) gets an extra factor of $-1$, and a phase of $\pi/2$ is added to Eq.(\ref{phi}).

Omitting the index $\nu$ for brevity, this paper uses a scaled dimensionless wavenumber
\begin{equation}\label{K}
K = \frac{\varkappa - k_0}{\alpha_1} R^4
\end{equation}
such that it is zero at the unperturbed wavenumber ($\varkappa= k_0$) and scaled by perturbation strength. In fact, for a single perturber, $K$ does not depend on the perturber strength $\alpha_1$ in first-order. For two perturbers, $K$ depends only on the ratio $\alpha$, (again, to first-order in $\alpha_1$ and $\alpha_2$). In general, there are two values of $K$ which are different and coalesce only at EPs or DPs.
As derived in Appendix \ref{explicit}, the value of $K=K_{\rm EP}$ for an EP is given by
\begin{equation}\label{Kep}
K_{\rm EP} \!=\! \frac{k_0 R^4}{\alpha_1} \!\left(\! \left[ 1 + \frac{3(A_1^{\text{TE}})^2}{8\pi} \alpha_1 {\cal R}_1^2(r_1) \left( 1\! -\! e^{2i\Delta\varphi} \right) \right]^{-1} \!\!\!\!  - 1  \!\right)\!,
\end{equation}
which in first-order can be approximated as
\begin{equation}\label{K1st}
K_{\rm EP} \approx - {k_0}{R^4} \frac{3(A_1^{\text{TE}})^2}{8\pi} {\cal R}_1^2(r_1) \left( 1 - e^{2i\Delta\varphi} \right)
\end{equation}
and is clearly independent of $\alpha_1$, $\alpha_2$, and $r_2$. Here, for instance, the parameters $\alpha$ and $r_2$ were varied to achieve an EP for the given $r_1$ and $\Delta\varphi$. Note that $K_{\rm EP}$ given by Eqs.(\ref{Kep}) or  (\ref{K1st}) takes the same values if a phase $p\pi$, where $p$ is any integer, is added to $\Delta\varphi$. Furthermore, $K_{\rm EP}$ given by Eqs.(\ref{Kep}) or  (\ref{K1st}), reaches DPs at $\Delta\varphi=p\pi/2$, returning exactly to the unperturbed values ($K=0$) at $\Delta\varphi=p\pi$, where $p$ is an integer.

With the radial position of the first perturber fixed at $r_1/R=0.95$, the parameters $\alpha$, $r_2$, and $\Delta\varphi$, satisfying the conditions Eqs.(\ref{alpha}) and (\ref{phi}) for EPs, are plotted in Fig.\ref{EA}(a), with the corresponding $K_{\rm EP}$ values from Eq.(\ref{Kep}) plotted in Fig.\ref{EA}(b). Due to the periodicity  in $\Delta\varphi$ of the expression in Eq.(\ref{Kep}), Fig.\ref{EA}(b) shows all possible values of $K_{\rm EP}$, although other parameters at EPs, such as $r_2$, are not periodic functions of $\Delta\varphi$, see Fig.\ref{EA}(c). While $K_{\rm EP}$ is independent of $\alpha_1$ to first-order, as explained above, it takes rather small values. This is due to a small factor $3(A_1^{\text{TE}})^2 R^{3} /(8\pi)\approx 0.004 $ in Eq.(\ref{Kep}), resulting from the rather high refractive index of the sphere.  The set of EPs shown in Fig.\ref{EA}(a) is then extended to large $r_2$ and $\alpha$ in polar coordinates in Fig.\ref{EA}(c). Multiple adjacent EPs in a line are called an exceptional arc (EA) \cite{zhou2018observation}; EPs and EAs are, respectively, zero- and one-dimensional objects \cite{bai2022nonlinearity}. Fig.\ref{EA} therefore shows an example of an EA. This EA takes the form of a bilateral spiral in the parameter space and an ellipse in the complex $k$-space. It has been suggested that EPs of $N$th-order require $2N-2$ parameters to be simultaneously adjusted \cite{mandal2021symmetry}. The treatment of Eq.(\ref{degen condition}) as a phasor in the complex plane, alongside Figs.\ref{EA}(a) and \ref{EA}(c), supports this for the case of second-order EPs.

In Fig.\ref{EA}(a), for the chosen $r_1$, the angle between perturbers is close to $\pi/2$ for internal perturbations since the WGMs, having a high quality factor, are described by an almost real wave function. While the argument of the angular function $\text{arg}\,{\cal R}_1^2(r)$, plotted in the inset of Fig.\ref{EA}(a), is close to $0$ for small radii, Eq.(\ref{degen condition}) also has a factor of $-1$, resulting in $\Delta\varphi\approx\pi/2$ for EPs inside the sphere. On the other hand, the perturber strength ratio at EPs diverges for smaller $r_2$. This is because $\alpha$ in Eq.(\ref{alpha}) is inversely proportional (in the case of varied $r_2$) to the radial function $\lvert {\cal R}_1^2(r) \rvert$, also plotted in the inset of Fig.\ref{EA}(a), which is small at small values of the radial position.  The radial function also has a maximum at $r/R=0.69$ which causes the minimum of $\alpha$ at around the same radius of the second perturber in Fig.\ref{EA}(a). With the second perturber outside the sphere, $\Delta \varphi$ becomes much more sensitive to $r_2$ and $\alpha$ because the imaginary part of the wave functions of the WGMs gradually increases outside the sphere. Beyond this, the WGM wave function exponentially grows at large distances from the open optical system \cite{weinstein1969open}, due to the complexity of the RS wavenumbers. Because of the periodicity of Eq.(\ref{phi}), for a given $\Delta \varphi$, the presented solution generates an infinite number of possible EPs with the same wavenumber at different values of $r_2$ and $\alpha$. However, the fact that this solution is valid only in first order imposes a restriction for its use for very large values of $\alpha$.  Fig.\ref{EA}(c) shows this resultant bilateral spiral of second perturber positions that satisfy the degeneracy condition. Not all of the degeneracies plotted are EPs; since $H_{12}$ vanishes at $\Delta \varphi = p\pi /2$, where $p$ is any integer, these points are DPs, due to the symmetry of the system not being sufficiently broken by the perturbation. The DP locations are derived in Appendix \ref{explicit} and shown by black circles in Fig.\ref{EA}.

\begin{figure}[!ht]
\centering
\includegraphics[width=1\linewidth]{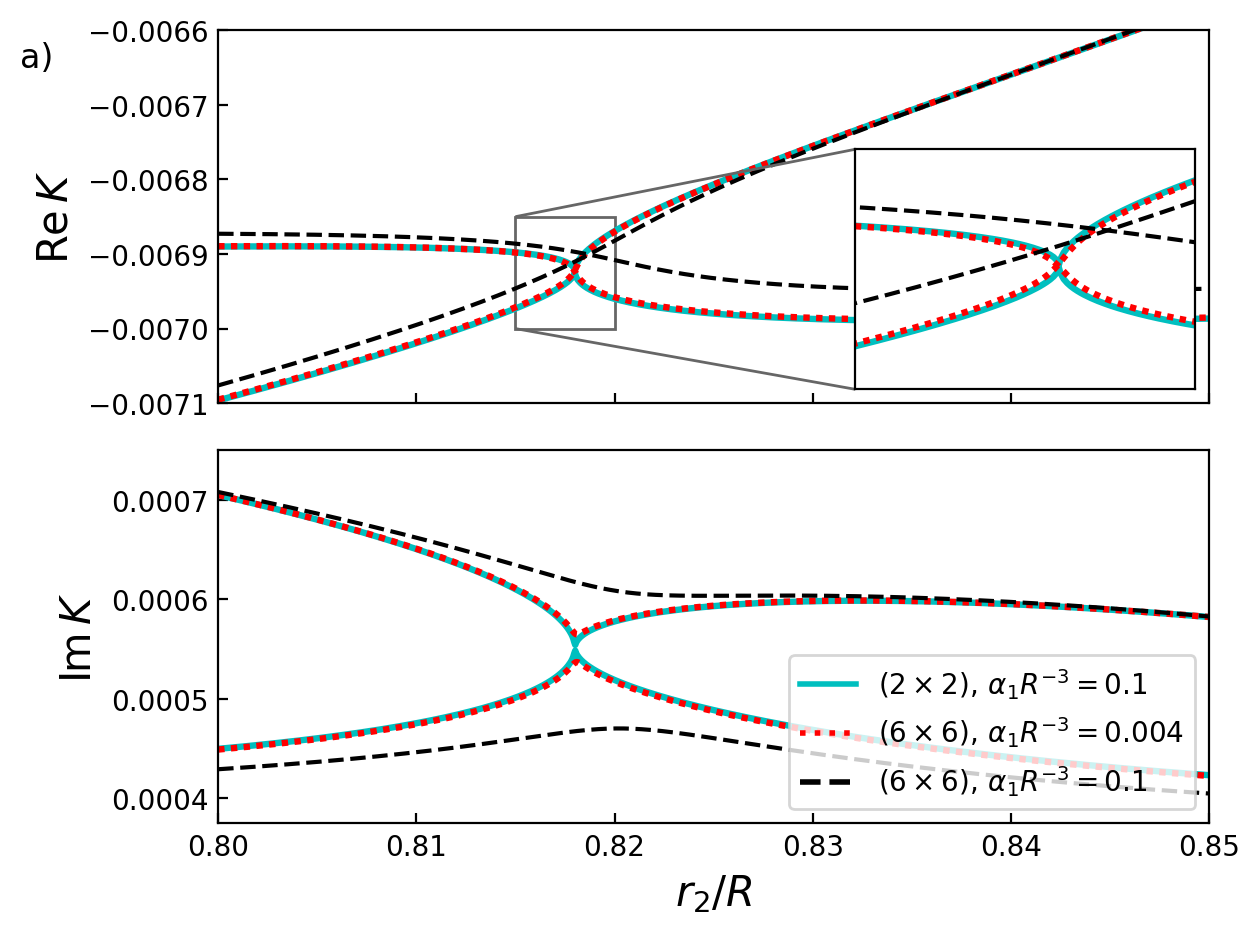}
\includegraphics[width=1\linewidth]{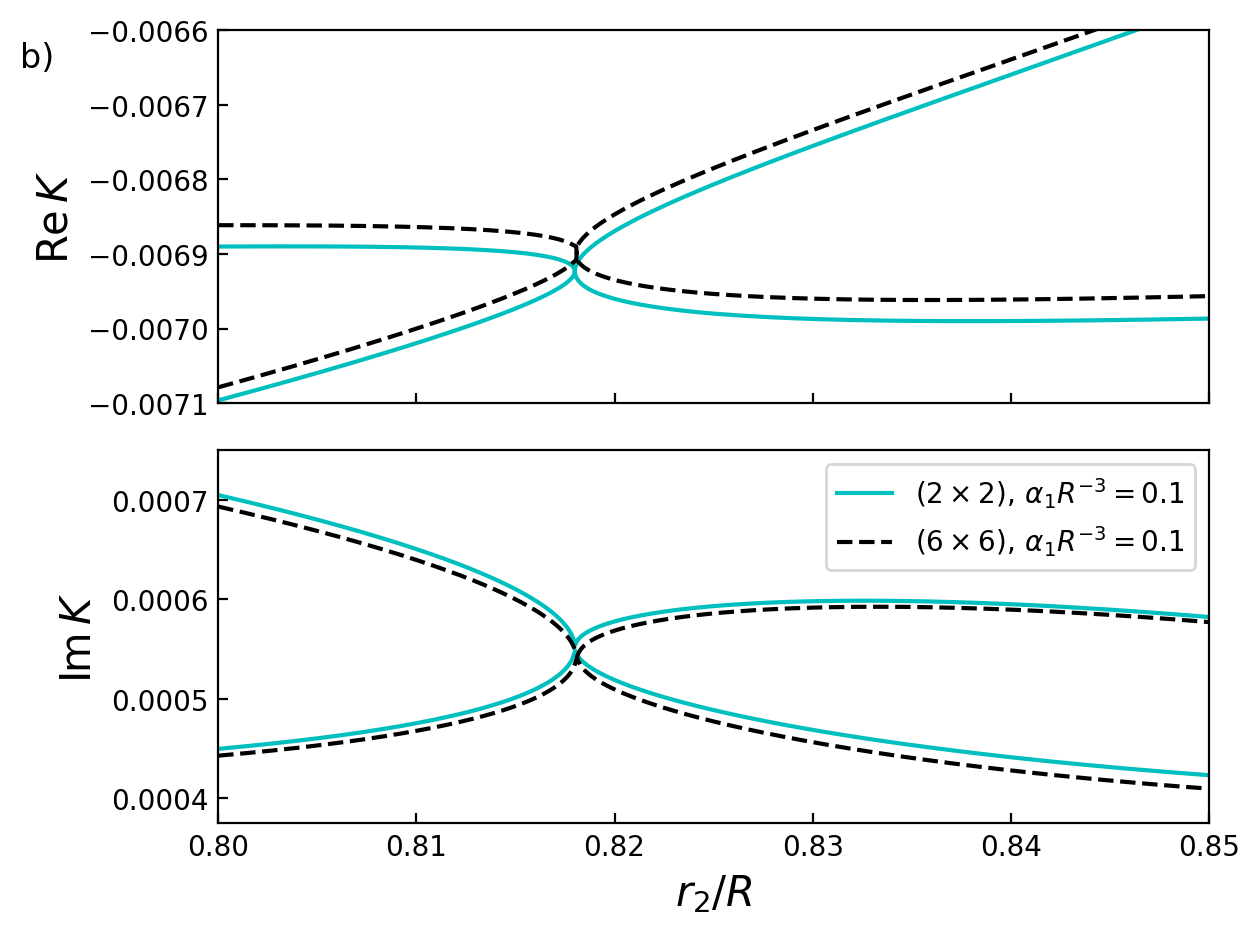}
\caption{(Color online) Dimensionless wavenumbers $K$ of perturbed RSs calculated from $H_{nn^\prime}$, with $l=1$, against $r_2/R$, for a microsphere with $n_r=4$ and two perturbers. (a) For the cyan solid lines, $H_{nn^\prime}$ has the mode basis of a triply degenerate TE WGM only accounting for $m=\pm1$. For the red dotted and black dashed lines, $H_{nn^\prime}$ has the mode basis of a triply degenerate TE WGM and its closest triply degenerate TM FPM, both with $m=0,\pm1$. Red dotted (black dashed) lines have perturber strengths $\alpha_1 R^{-3} =0.004$ and $\alpha_2 R^{-3} =0.003107$ ($\alpha_ 1 R^{-3} =0.1$ and $\alpha_2 R^{-3} =0.0777$). Cyan solid lines are the $K$-values of the $2\times2$ matrix which are equal for either factor of perturber strength, to first order. Other perturbation parameters are fixed at $r_1/R=0.95$ and $\Delta \varphi=1.547$. (b) As the cyan solid lines and black dashed lines in (a) but the matrix with $\alpha_ 1 R^{-3} =0.1$ including the $l=1$ TM mode (black dashed lines) is fine-tuned to restore the EP by changing the angle between perturbers to $\Delta \varphi=1.5494$.}
\label{l=1 EP}
\end{figure}

Looking closer at an example of one of these EPs inside the sphere (the black cross in Fig.\ref{EA}(a)), the RSE matrix is diagonalized numerically to find the nearby perturbed wavenumbers.  $K$-values for the two perturbed RSs are plotted in Fig.\ref{l=1 EP}(a) against a varied radial position of the second perturber around an EP. The other perturbation parameters are fixed at $\alpha_1 R^{-3}=0.004 $, $\alpha_2 R^{-3} =0.003107$, $\Delta \varphi=1.547$, and $r_1/R=0.95$. At the EP, $r_2/R=0.818$. The characteristic shape of an EP, which was demonstrated in Fig.\ref{ex EP}, is present in Fig.\ref{l=1 EP}(a).

To show how accurate approximating the RSE to a basis with only degenerate modes is, we now include in the basis the closest modes to the earlier selected fundamental TE WGM. This is a dipolar triply degenerate TM FPM at $kR= 1.053 - 0.072i$ as shown by Fig.\ref{l=1 modes}. By including TM modes, the $m=0$ states are no longer orthogonal with all other modes, so $m=0,\pm1$ is the required basis for each polarization. The resulting $6\times6$ RSE matrix is diagonalized with the same parameters as the $2\times2$ matrix, demonstrating little visual difference in the plotted wavenumbers between the two models (compare cyan solid and red dotted lines). To compare different perturbation strengths of the system, the same matrices are diagonalized again but with both $\alpha_1$ and $\alpha_2$ multiplied by $25$. It is worth noting that while other modes contribute (including those with different $l$) in second and higher orders, they are much less significant than the closest mode due to the larger difference in wavenumber.

Comparing the two perturbation strengths which differ by the factor of 25, it is clear that the different size matrices are in closer agreement when the perturbation is weaker because stronger perturbations have greater effect on the modes and second-order perturbation theory terms become more significant. The consequence of this is that truncation to a basis of only degenerate modes is accurate only for weak perturbation strengths since it is a first-order approximation. Further modes not included in this calculation have a similar effect on the EP when considered in the truncated basis, although the effect is smaller as the difference in wavenumber between the modes increases. This article is dealing with small matrices for illustrative purposes but many resonant states (often up to order $10^3$) should be accounted for in an accurate calculation for strong perturbations. Note that Refs.\cite{muljarov2011brillouin,doost2014resonant,armitage2014resonant,lobanov2019resonant,muljarov2020full,
sztranyovszky2022optical,almousa2021varying} provide a convergence analysis for the RSE treating various spherical and non-spherical systems using the modes of a dielectric sphere as a basis.

While the EP is lost after accounting for more modes, it can be easily restored by fine-tuning the parameters again. In fact, while second-order EPs are found by fine-tuning two parameters, only one parameter needs to be fine-tuned to restore the EP in this case. To demonstrate this, we take the matrix including in its basis the $l=1$ fundamental TM modes and adjust $\Delta \varphi$ by $-0.0024$ [compared to the value for the $2\times 2$ matrix, calculated via Eq.(\ref{phi})] to $\Delta \varphi=1.5494$. The $K$-values for this $6\times 6$ fine-tuned matrix are plotted in Fig.\ref{l=1 EP}(b) which shows that the EP has been restored but slightly shifted in $K$.

\section{Squared-Lorentzian spectrum near exceptional points}\label{Purcell}

Using the basis of the TE dipolar RSs considered in Sec.\,\ref{l=1}, we now focus on the optical spectrum of the system, in order to see how the normal Lorentzian line shape is modified when the system approaches an EP, developing an additional squared-Lorentzian component \cite{yoo2011quantum,takata2021observing,SchomerusPRA22}. We focus here on a simple but experimentally relevant spectral function, the Purcell factor (PF), which is a measure of local density of states of the optical system for a given polarization of the electric field, and is therefore comparable to a measurable micro-photoluminescence~\cite{PellegrinoPRL20}. Note, however, that the often studied Petermann factor which characterizes the behavior of non-Hermitian systems near EPs \cite{wiersig2020prospects} is not analyzed here.

According to the exact theory of the Purcell effect~\cite{muljarov2016exact}, the PF is given by
\begin{equation}
F(q)=\frac{3\pi}{q}\sum _n {\rm Im}\,\frac{1}{\mathcal{V}_n k_n(k_n-q)}
\label{eq-Purcell}
\end{equation}
where
\begin{equation}
\frac{1}{\mathcal{V}_n}=\left[{\bf e}\cdot {\bf E}_n({\bf r}_d)\right]^2
\end{equation}
is the inverse mode volume, ${\bf r}_d=(r_d,\theta_d,\varphi_d)$ is the position of a point-like emitter (not to be confused with the point-like perturbers treated in this work), ${\bf e}$ is the unit vector of its polarization, which is chosen below in the $z$-direction, and $q$ is the {\it real} light wavenumber of the emitter. Clearly, all the RSs of the system contribute to its PF as individual complex Lorentzian lines. Truncating the infinite sum in Eq.\,(\ref{eq-Purcell}) and keeping only the contribution of the dipolar WGMs, we obtain
\begin{equation}
F(q)=\frac{3\pi}{q}\sin^2\theta_d\, {\rm Im} \left[ \frac{\tilde{\cal R}^2(r_d)}{k_0} f(q,\varphi_d) \right]
\label{eq-Purcell1}
\end{equation}
where the function
\begin{equation}
f(q,\varphi)=\frac{1}{k_0-q}
\label{f-sym}
\end{equation}
is purely Lorentzian and independent of $\varphi$ for the unperturbed, i.e. spherically-symmetric, system. In fact, in this case (and also in the perturbed system), the $m=0$ state does not contribute. The $m=-1$ and $m=1$ unperturbed modes, denoted below (as in Sec.\,\ref{l=1}) with the indices $n=1$ and $n=2$, respectively, have the wavenumbers $k_1=k_2=k_0$ and the electric fields
\begin{equation}
{\bf E}_1({\bf r})= \tilde{\cal R}(r)
\begin{pmatrix}
0\\
\cos\varphi\\
-\cos\theta \sin\varphi
\end{pmatrix}
\end{equation}
and
\begin{equation}
{\bf E}_2({\bf r})= \tilde{\cal R}(r)
\begin{pmatrix}
0\\
-\sin\varphi\\
-\cos\theta \cos\varphi
\end{pmatrix},
\end{equation}
in accordance with Eq.\,(\ref{ETE}), where
\begin{equation}
\tilde{\cal R}(r)=\sqrt{\frac{3}{4\pi}} A_1^{\rm TE} {\cal R}_1(r)\,.
\end{equation}
Therefore, the sum of their inverse mode volumes is given by
\begin{equation}
\frac{1}{\mathcal{V}_1}+\frac{1}{\mathcal{V}_2}= \sin^2\theta_d \tilde{\cal R}^2(r_d)\,.
\end{equation}

For the perturbed RSs, calculated in the two-state basis,
\begin{equation}
\mathbfcal{E}_\nu ({\bf r})=\sqrt{\frac{\varkappa_\nu}{k_0}}
\left[ C_{1\nu} {\bf E}_1({\bf r}) + C_{2\nu} {\bf E}_2({\bf r})\right]\,,
\end{equation}
in accordance with Eq.\,(\ref{Ecal}), so that
\begin{equation}
f(q,\varphi)=\sum_{\nu=\pm} \frac{(C_{1\nu} \cos\varphi- C_{2\nu} \sin\varphi )^2}{\varkappa_\nu-q}
\label{f-general}
\end{equation}
is a sum of two complex Lorentzian functions.
To evaluate this expression in general and near an EP, we solve the $2\times 2$ RSE matrix eigenvalue problem,
\begin{equation}
\begin{pmatrix}
H_{11} & H_{12} \\
H_{12} &H_{22}
\end{pmatrix}
\begin{pmatrix}
C_{1\nu} \\
C_{2\nu}
\end{pmatrix}
=
\frac{1}{\varkappa_\nu}
\begin{pmatrix}
C_{1\nu} \\
C_{2\nu}
\end{pmatrix}\,,
\label{matrix-problem}
\end{equation}
determining  the eigenvalues
\begin{equation}
\frac{1}{\varkappa_\nu} = \frac{H_{11}+H_{22}+\nu\Delta}{2}
\end{equation}
and the properly normalized eigenvectors
\begin{equation}
\begin{pmatrix}
C_{1\nu} \\
C_{2\nu}
\end{pmatrix}
={\cal N}_\nu
\begin{pmatrix}
1 \\
\frac{H_{22}-H_{11}+\nu\Delta}{2H_{12}}
\end{pmatrix}
\end{equation}
of the two perturbed RSs $\nu=\pm$, where
\begin{equation}\label{Delta}
\Delta=\sqrt{(H_{22}-H_{11})^2+4H_{12}^2} \, ,
\end{equation}
and
\begin{equation}
{\cal N}_\nu = \sqrt{ \frac{2H_{12}^2}{\Delta[\Delta+\nu(H_{22}-H_{11})]}}
\end{equation}
are the normalization constants. This is the exact solution of the truncated $2\times 2$ RSE matrix problem, so that the use of the eigenvalues $\varkappa_\nu$ and the expansion coefficients $C_{1\nu}$ and  $C_{2\nu}$ in Eq.\,(\ref{f-general}) determines the spectrum for any sufficiently small perturbation for which the truncation works well. Keeping only the WGMs in the sum is in turn well justified by their dominant contribution to the spectrum \cite{muljarov2016exact}.

Now focusing on the spectral properties near the EP, we take the limit $\Delta \to 0$, expanding all the  expressions (depending on $\Delta$) in powers of $\Delta$ and keeping terms to first-order in $\Delta$. In this approximation, Eq.\,(\ref{Delta}) coincides with the EP condition Eq.\,(\ref{H condition})
\begin{equation}
\frac{H_{22}-H_{11}}{2H_{12}} = \pm i
\end{equation}
(in fact, corrections are proportional to $\Delta^2$), the eigenvectors simplify  to
\begin{equation}
\begin{pmatrix}
C_{1\nu} \\
C_{2\nu}
\end{pmatrix}
=\sqrt{\mp \nu \frac{iH_{12}}{\Delta}}
\begin{pmatrix}
1 \pm \nu i\frac{ \Delta}{4H_{12}} \\
\pm i +\nu \frac{\Delta}{4H_{12}}
\end{pmatrix}
\label{c12}
\end{equation}
and the eigenvalues to
\begin{equation}
\varkappa_\nu= \bar{\varkappa}-\nu\frac{\bar{\varkappa}^2\Delta}{2}\,,\ \ \ \ {\rm where} \ \ \ \bar{\varkappa}=\frac{2}{H_{11}+H_{22}}\, ,
\label{kappa12}
\end{equation}
giving the value $\varkappa_+=\varkappa_-=\bar{\varkappa}$ at the EP. Substituting
Eqs.\,(\ref{c12}) and (\ref{kappa12}) into (\ref{f-general}), we arrive after simple algebra at
\begin{equation}
f(q,\varphi)=\frac{1}{\bar{\varkappa}-q} \mp iH_{12} e^{\mp2i\varphi }\frac{\bar{\varkappa}^2}{(\bar{\varkappa}-q)^2}\,,
\label{f-EP}
\end{equation}
where in addition to the Lorentzian term having the same form as in Eq.\,(\ref{f-sym}), a new, squared-Lorenzian term appears, which strongly depends on the angular position of the emitter, thus emphasizing on the strong chirality of the modes at the EP. This term is however rather small as it has a factor of $H_{12}$, proportional to the strengths of the perturbers, so it is a first-order correction to the Lorentzian spectrum. Note also that Eq.\,(\ref{f-EP}) is valid in the vicinity of the EP for which $|\Delta/H_{12}|\ll 1$.

\begin{figure*}
$\varphi_d=\pi/8$ \hskip 7cm $\varphi_d=\pi/4$ \\
\includegraphics[width = 0.45\linewidth]{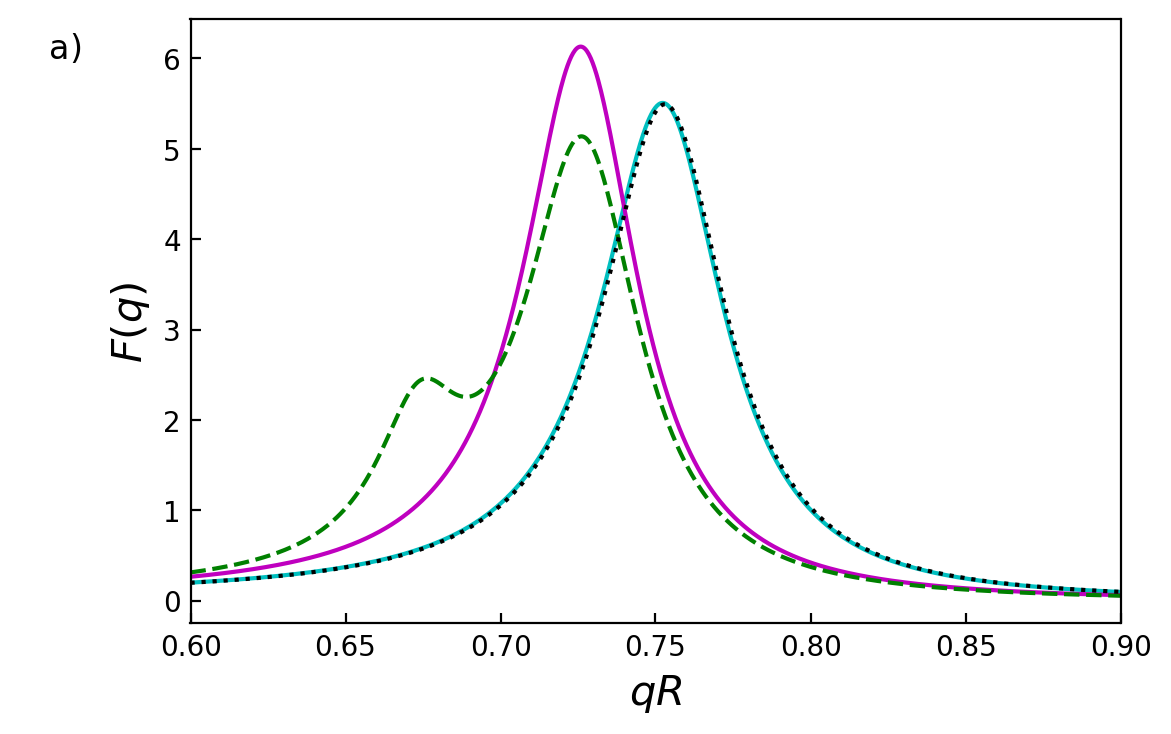}
\includegraphics[width = 0.45\linewidth]{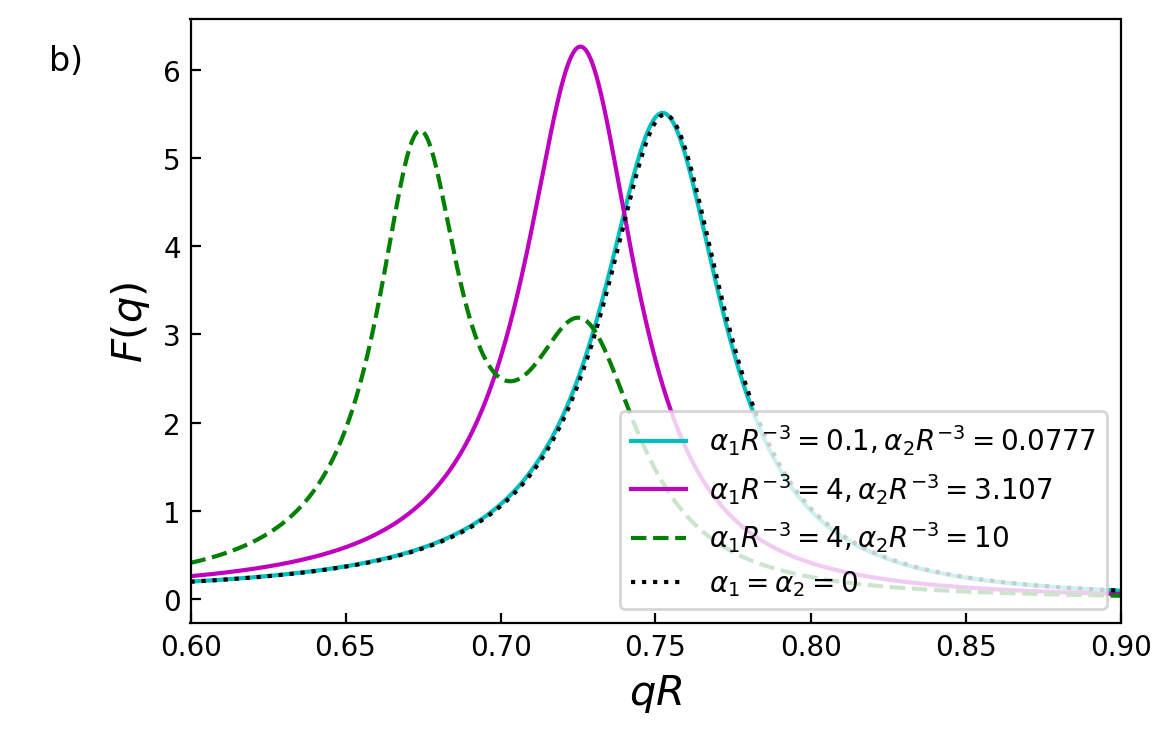}
\includegraphics[width = 0.45\linewidth]{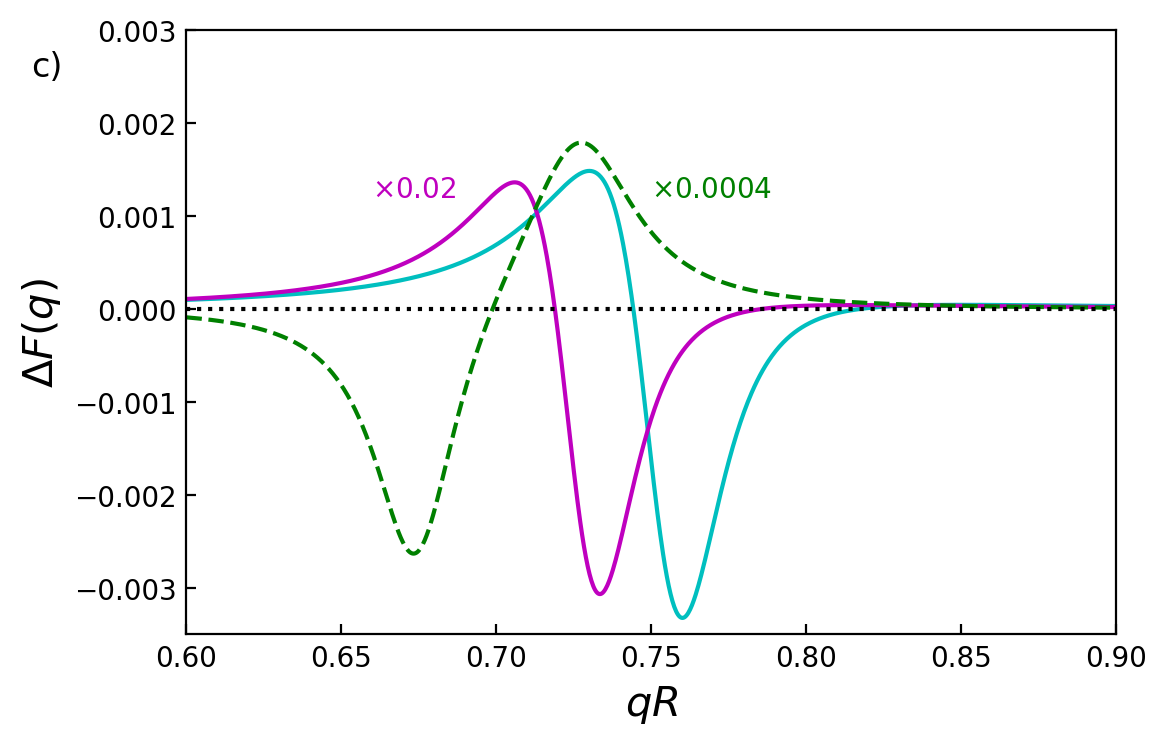}
\includegraphics[width = 0.45\linewidth]{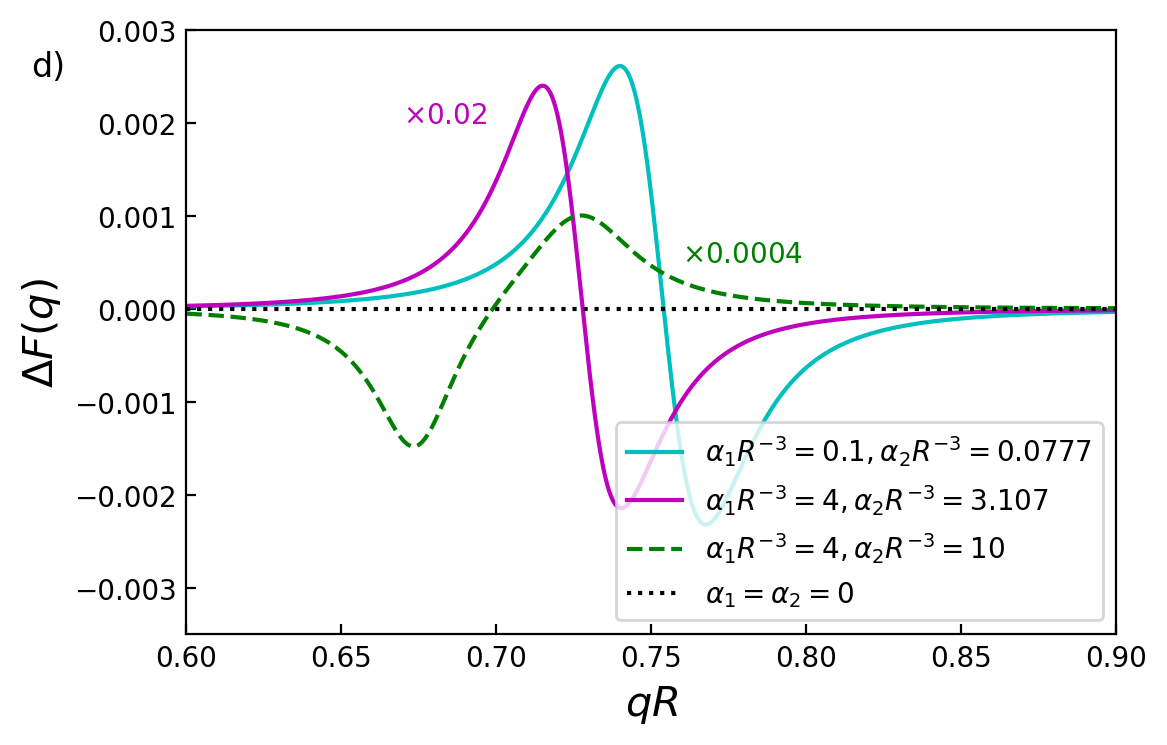}
\includegraphics[width = 0.45\linewidth]{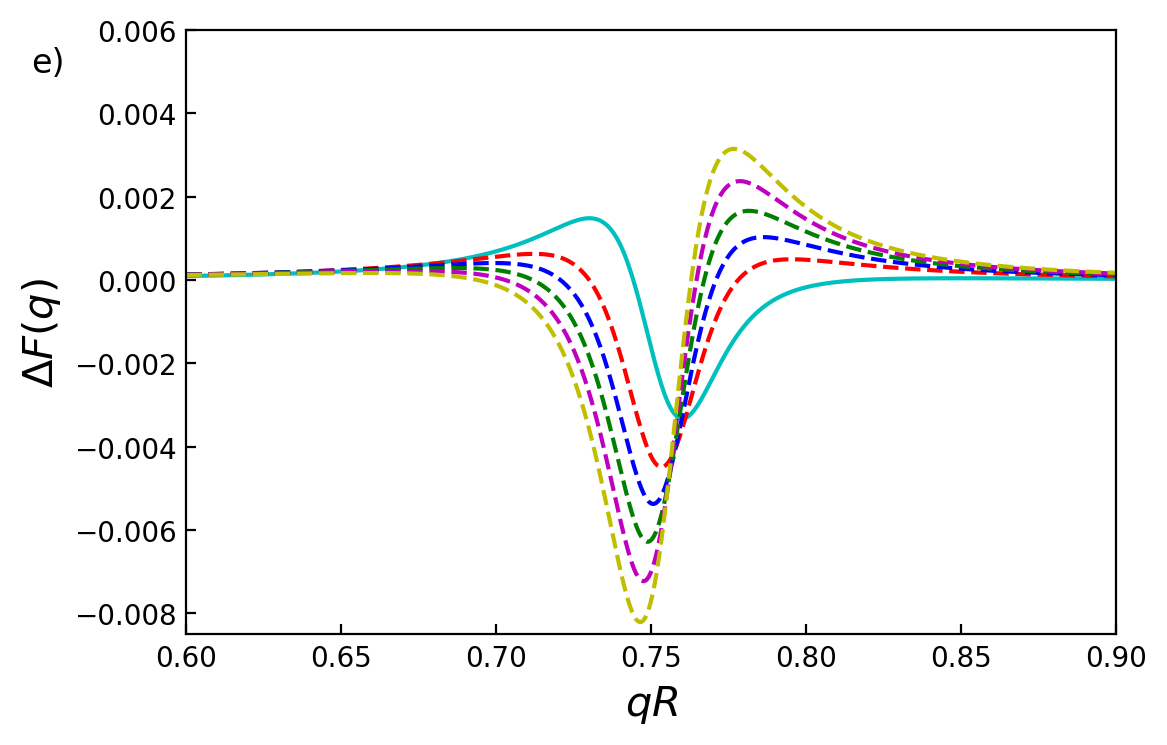}
\includegraphics[width = 0.45\linewidth]{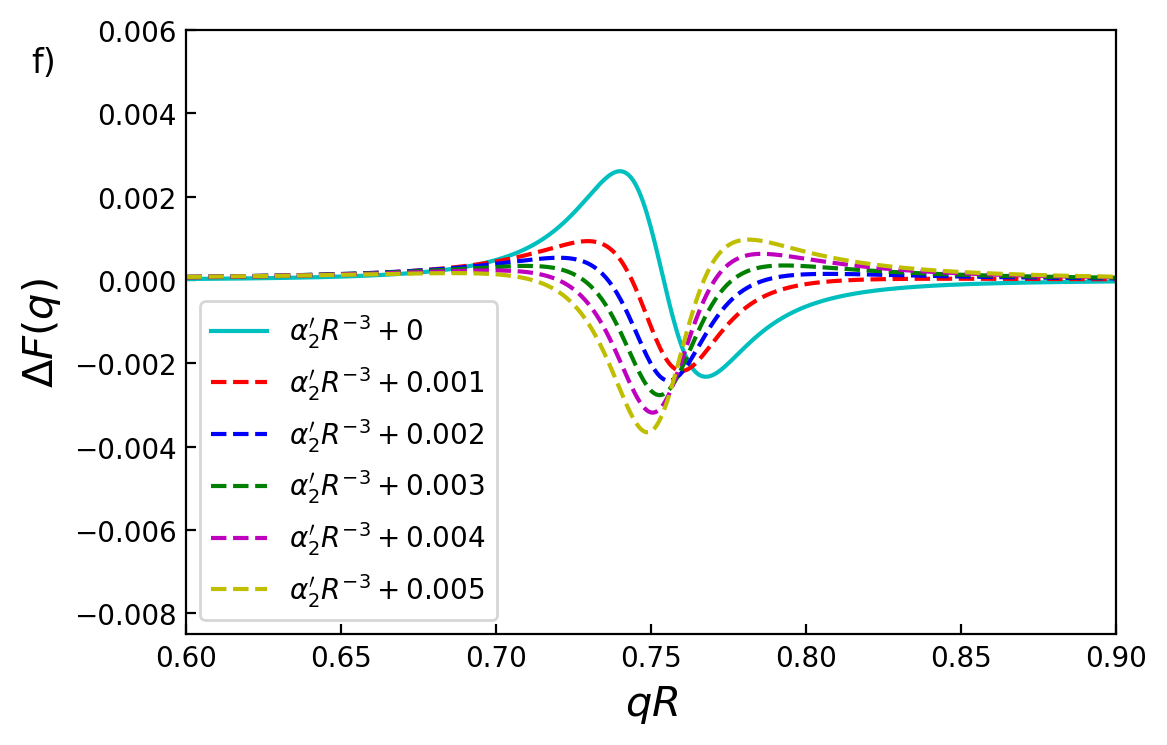}
\includegraphics[width = 0.45\linewidth]{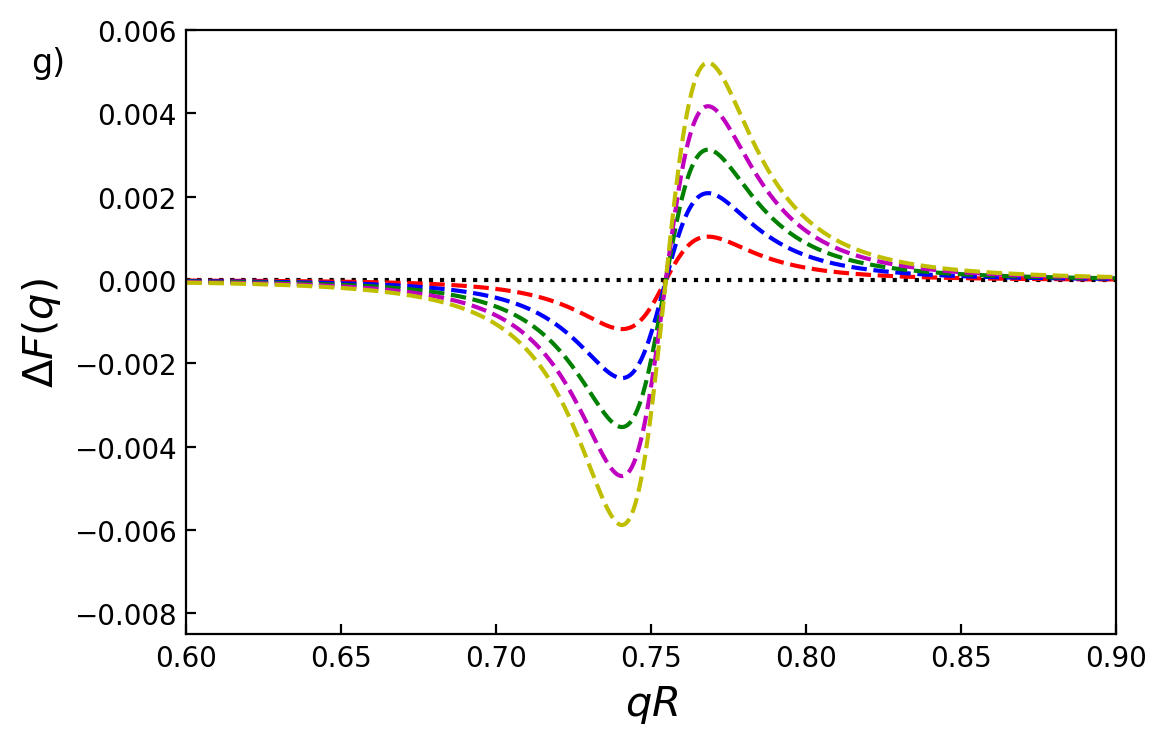}
\includegraphics[width = 0.45\linewidth]{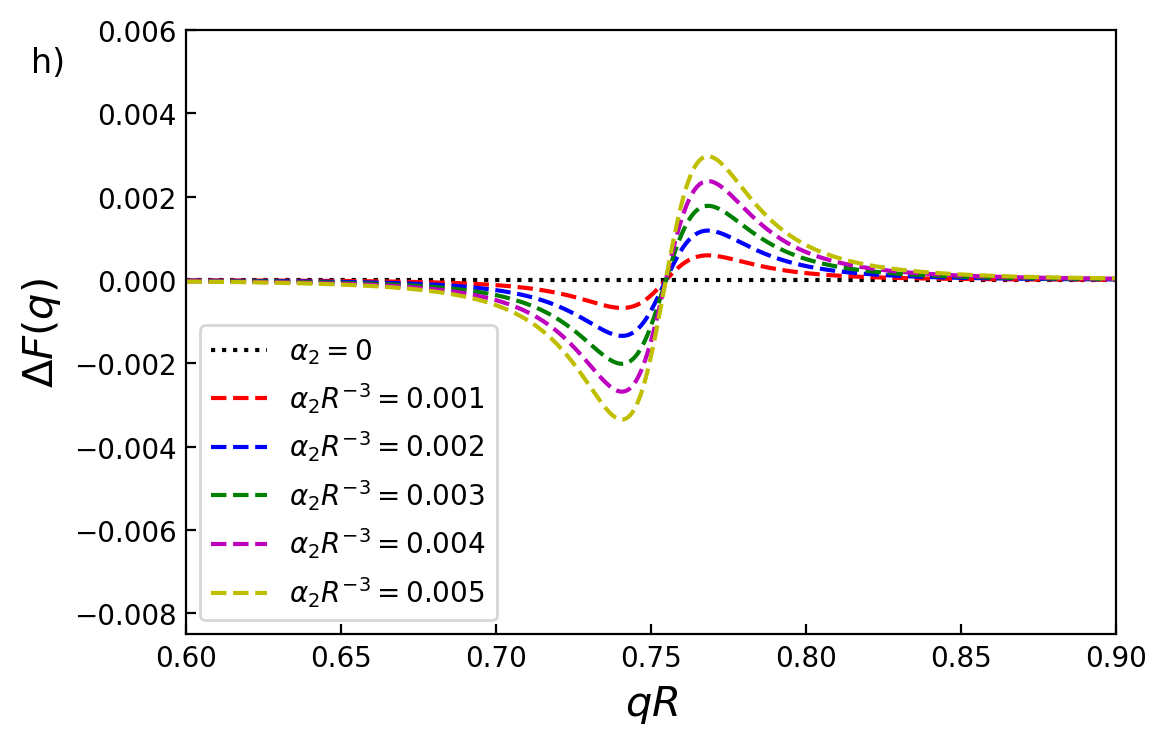}
\caption{(Color online) Purcell factor $F(q)$ and the non-Lorentzian part of the Purcell factor $\Delta F(q) $ as functions of light emitter wavenumber $qR$ for a microsphere of refractive index $n_r=4$  with two perturbers, calculated for the perturbed fundamental TE WGM with $l=1$. The perturbation parameters are: $r_1/R=0.95$, $r_2/R=0.818$, and $\Delta \varphi=1.547$, with $\alpha=0.777$ at EPs. Emitter is placed at $r_d/R=1$, $\theta_d=\pi/2$, and $\varphi_d = \pi/8$ for the left panels and $\varphi_d = \pi/4$ for the right panels. (a) and (b): PF $F(q)$ calculated using Eq.(\ref{f-general}) at an EP with $\alpha_1 R^{-3} = 0.1$ and $\alpha_2 R^{-3} = 0.0777$ (cyan/light gray solid line), at an EP with $\alpha_1 R^{-3} = 4$ and $\alpha_2 R^{-3} = 3.107$ (magenta/dark gray solid line), away from the EP with $\alpha_1 R^{-3} = 4$ and $\alpha_2 R^{-3} = 10$ (green dashed line), and for the unperturbed system (black dotted line). (c) and (d): As (a) and (b) but instead plotting only the differential Purcell factor $\Delta F(q) $. The actual results for the green dashed and magenta/dark gray solid lines can be obtained by multiplying by factors of $4\times 10^{-4}$ and $0.02$, respectively. (e) and (f): Differential PF for the EP with $\alpha_1 R^{-3} = 0.1$ and $\alpha_2^\prime R^{-3} = 0.0777$ (cyan solid line) and for small variations from $\alpha_2^\prime$ away from the EP (colored dashed lines). (g) and (h): Differential PF for the unperturbed system $\alpha_1=\alpha_2=0$ (black dotted line) and at small variations away from the DP (colored dashed lines). }
\label{Purcell_fig1}
\end{figure*}

To see more clearly the chirality of the modes at the EP, we write the explicit form of the perturbed fields to first-order in $\Delta$:
\begin{eqnarray}\label{E-chiral}
\mathbfcal{E}_\nu ({\bf r})= &
\displaystyle
\sqrt{\mp\nu \frac{iH_{12}}{\Delta}\frac{\bar{\varkappa}}{k_0}} \tilde{\cal R}(r)
\left[
\left(1-\nu\frac{\varkappa\Delta}{4}\right) e^{\mp i\varphi}
\begin{pmatrix}
0\\
1\\
\mp i\cos\theta
\end{pmatrix}
\right.
\nonumber\\
&
\displaystyle
\left. \pm \nu i \frac{\Delta}{4 H_{12}} e^{\pm i\varphi}
\begin{pmatrix}
0\\
1\\
\pm i\cos\theta
\end{pmatrix}
\right].
\end{eqnarray}
While both modes are almost fully chiral with the degenerate chirality owing to the dominant and divergent CW term (proportional to $e^{-i\varphi}$), if we choose the upper sign in Eq.\,(\ref{E-chiral}), this is likely to be a small effect in any observable, since the major contribution comes from an interference of the divergent CW and the vanishing CCW term (proportional to $e^{i\varphi}$), leading to a finite Lorentzian part of the spectrum with no phase dependence, as is clear from Eq.\,(\ref{f-EP}).

We show in Fig.\ref{Purcell_fig1} the optical spectra for the Purcell factor of this system and the effect EPs have on them, using  Eqs.(\ref{eq-Purcell}) and (\ref{f-general}), the latter being equivalent to Eq.(\ref{f-EP}) at EPs. The emitter position is chosen to be $r_d/R = 1$, $\theta_d = \pi/2$, and both $\varphi_d=\pi/8$ and $\varphi_d=\pi/4$ are considered separately in the left and right panels, respectively. Figures \ref{Purcell_fig1}(a) and \ref{Purcell_fig1}(b) plot $F(q)$  for (i) the unperturbed system, (ii) the EP shown in Fig.\ref{l=1 EP} with $\alpha_1 R^{-3} = 0.1$ and $\alpha_2 R^{-3} = 0.0777$, (iii) the EP in Fig.\ref{l=1 EP} for the same $\alpha=0.777$ but with $\alpha_1 R^{-3} = 4$ and $\alpha_2 R^{-3} = 3.107$, and (iv) away from the EP at $\alpha_1 R^{-3} = 4$ and $\alpha_2 R^{-3} = 10$. These plots demonstrate that the Lorentzian originating from the basis system dominates for weak perturbations where the truncation approximation described in Sec.\ref{l=1} is valid. Observation of the double Lorentzian for non-degenerate modes requires a stronger perturbation, which may go beyond first-order limits. However, these plots show that even with a strong perturbation, the EP exhibits only one peak in the spectrum due to the degeneracy of the eigenstates. %

Since the non-Lorentzian part of the spectrum is small, we introduce a differential PF defined as
\begin{equation}
\Delta F(q) = F(q) - {\rm Im} \, \frac{B}{\bar{\varkappa} - q} \, ,
\label{differential PF}
\end{equation}
with the appropriate constant $B$, as follows from Eqs.(\ref{eq-Purcell}) and (\ref{f-EP}). In the differential PF Eq.(\ref{differential PF}), the Lorentzian part is removed, making it equivalent at an EP to the second, squared-Lorentzian  term in Eq.(\ref{f-EP}). These non-Lorentzian spectra are plotted in Figs.\ref{Purcell_fig1}(c) and \ref{Purcell_fig1}(d) with the same parameters as before. The unperturbed system has an entirely Lorentzian spectrum so its $\Delta F(q)$ vanishes. The amplitudes of the squared-Lorentzians at EPs are proportional to perturbation strength (see Eq.(\ref{f-EP})) so the spectra with strong perturbations are scaled down for visibility. These squared-Lorentzians have almost the same shape and are shifted in frequency with respect to each other owing to the perturbation-induced frequency shift of the modes.   The non-Lorentzian spectrum away from the EP has a minimum at one of its Lorentzian peaks and a maximum at the other. Meanwhile, the EPs' non-Lorentzians have a minimum-maximum pair centered around their single Lorentzian peaks, in accordance with Eq.(\ref{f-EP}). It is notable that $\Delta F(q)$ near an EP resembles the change in circular-dichroism (the absorption difference between modes of opposite chirality) observed for $\bf \Omega$-shaped plasmonic nanoantennas studied in \cite{both2022nanophotonic}. This is because the chirality of each mode are sensitive in an EP's vicinity, see \cite{wiersig2011structure} and  Sec.\ref{20.1} below. While the squared-Lorentzian emission spectrum may be unresolvable at small perturbations due to the dominating spectrum of the basis system, measuring the circular dichroism \cite{both2022nanophotonic} may provide a way to spectrally observe EPs in this system.

Spectra in the vicinity of an EP (within $|\Delta/H_{12}|\ll 1$) show similar features as the spectrum at the EP (which can never be exactly reached). To show this, we plot the non-Lorentzian part of the spectrum, at the EP with $\alpha_1 R^{-3} = 0.1$ and $\alpha_2 R^{-3} = 0.0777$, alongside small variations from it in Figs.\ref{Purcell_fig1}(e) and \ref{Purcell_fig1}(f). We call $\alpha_2^\prime$ the value of $\alpha_2$ at the EP and make small variations of $\alpha_2$ from $\alpha_2^\prime$. We then compare this to the same deviations from the unperturbed DP plotted in Figs.\ref{Purcell_fig1}(g) and \ref{Purcell_fig1}(h). Clearly, around EPs, the non-Lorentzian lineshape turns out to be very sensitive to the parameters of the system and the emitter position, whereas it stays the same with small deviations from a DP.

\section{Whispering-gallery modes with high angular momentum}\label{l=20}
\subsection{Twenty-mode basis}\label{20.1}
Again with two perturbers, $j=1,2$, we now focus on a perturbation of the fundamental WGM of the sphere with $n_r=2$ shown in Fig.\ref{l=20 modes}(c). This mode has the wavenumber $k_0 R=12.33404942-2.27\times 10^{-6}i$ and angular momentum $l=20$, so it is 41-fold degenerate.  Using the same truncation approximation as in Sec.\ref{l=1}, we therefore start with a basis of 41 degenerate states. Owing to the selection rules in Appendix \ref{3.2}, the $41\times41$ RSE matrix is truncated further to a $20\times20$ matrix by excluding the odd $m$ and $m=0$ states from the basis. Still, Eq.(\ref{EP2 condition}) cannot be used to find degeneracies here before a transformation to a $2\times2$ matrix is applied in Sec.\ref{Orthogonal}. In this section, $\nu$ accordingly takes integer values from $1$ to $20$.

\begin{figure}
\centering
\includegraphics[width=\linewidth]{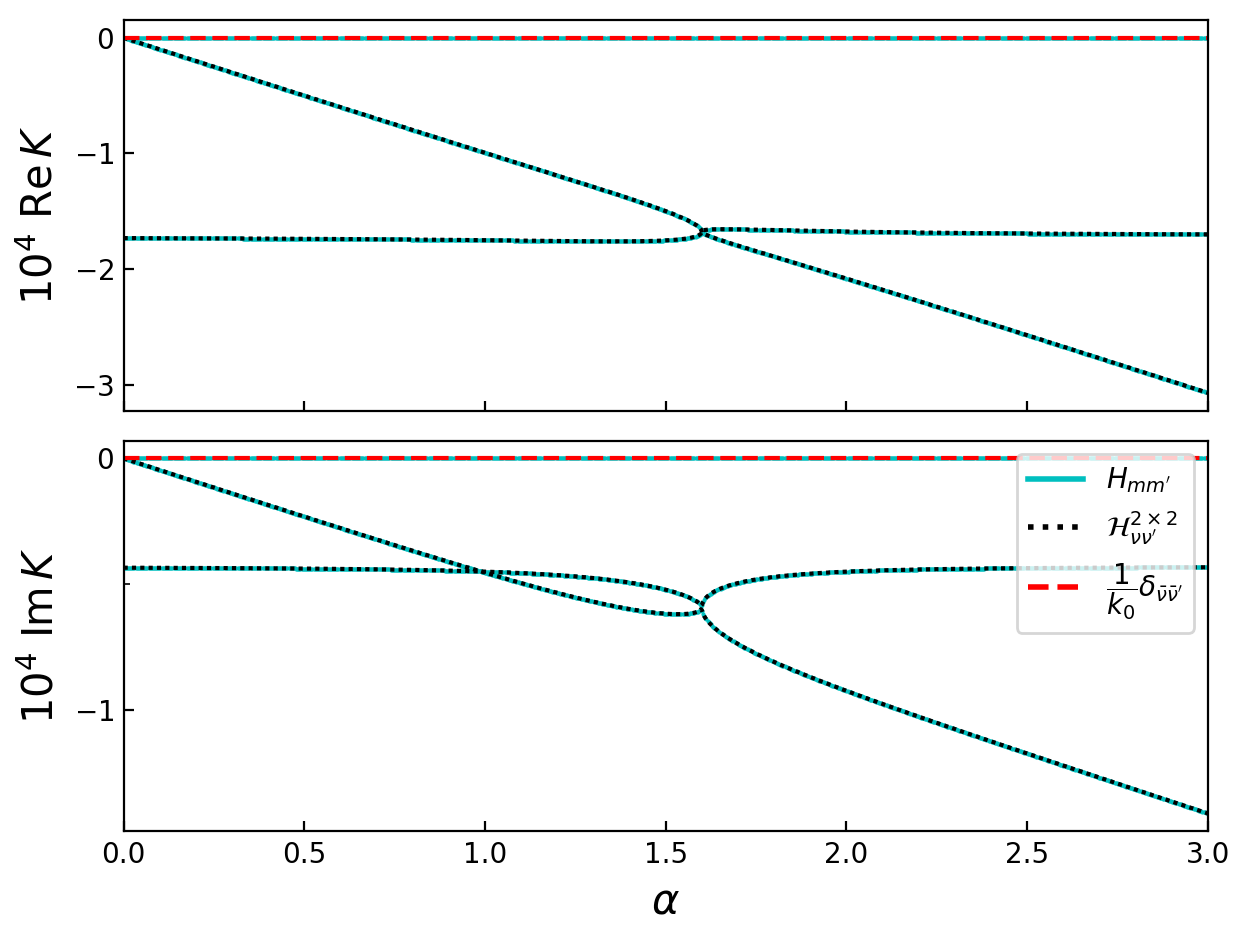}
\caption{(Color online) Scaled dimensionless wavenumbers $K$ around an EP for a microsphere with refractive index $n_r=2$, angular momentum $l=20$, and two perturbers, calculated using a $20\times20$ matrix $H_{mm^\prime}$ (cyan solid lines), $2\times2$ matrix $\mathcal{H}_{\nu\nu^\prime}^{2\times 2}$ (black dotted lines), and the diagonal $18\times18$ matrix  $k_0^{-1}\delta_{\bar{\nu}\bar{\nu}^\prime}$ (red dashed lines) against the perturber strength ratio $\alpha$, with $\sigma^\prime=\alpha^\prime=10$ used for the orthogonal transformation.
The other parameters are fixed at $r_1/R=1.5$, $r_2/R =1.5542 $, and $\Delta \varphi=1.199605$. The actual values of $K$ are obtained by multiplying the shown numbers by $10^{-4}$ as indicated.
}
\label{l=20 EP}
\end{figure}

The $20\times20$ RSE matrix is numerically diagonalized to find the perturbed wavenumbers and expansion coefficients.  The normalized perturbed wavenumbers $K$ [defined by Eq.(\ref{K})] are shown in Fig.\ref{l=20 EP} as functions of the perturber strength ratio $\alpha$. Note that the two perturbers are now placed outside the sphere to make the system more suitable for a potential experimental verification.
It is clear that only two states, which we label by $\nu=1,2$, are affected by the perturbation as their wavenumbers change with a variation of the perturbation parameter and show the characteristic shape of an EP
at $\alpha=1.6$, $\Delta \varphi=1.199605$, $r_1/R=1.5$, and $r_2/R=1.5542$.  We therefore from here onwards call them {\it affected} states. The other 18 states have the wavenumber $K=0$ (corresponding to $\varkappa_{\nu>2}=k_0$), independent of the perturbation parameters and are called here {\it unaffected} states, although their wave functions do depend on the perturbation, as discussed below.
More generally, using the selection rule in Appendix \ref{3.2}, it can be seen from Eq.(\ref{ETE}) that for $N$ perturbers, there are $2N$ affected modes in TE polarization for even $l$, half of them for even $m$ and half for odd $m$. This is supported by Fig.\ref{l=20 EP} where we can see that there is only one (even $m$) affected state for $\alpha=0$ which is the case for a single perturber.

\begin{figure}
\includegraphics[width=\linewidth]{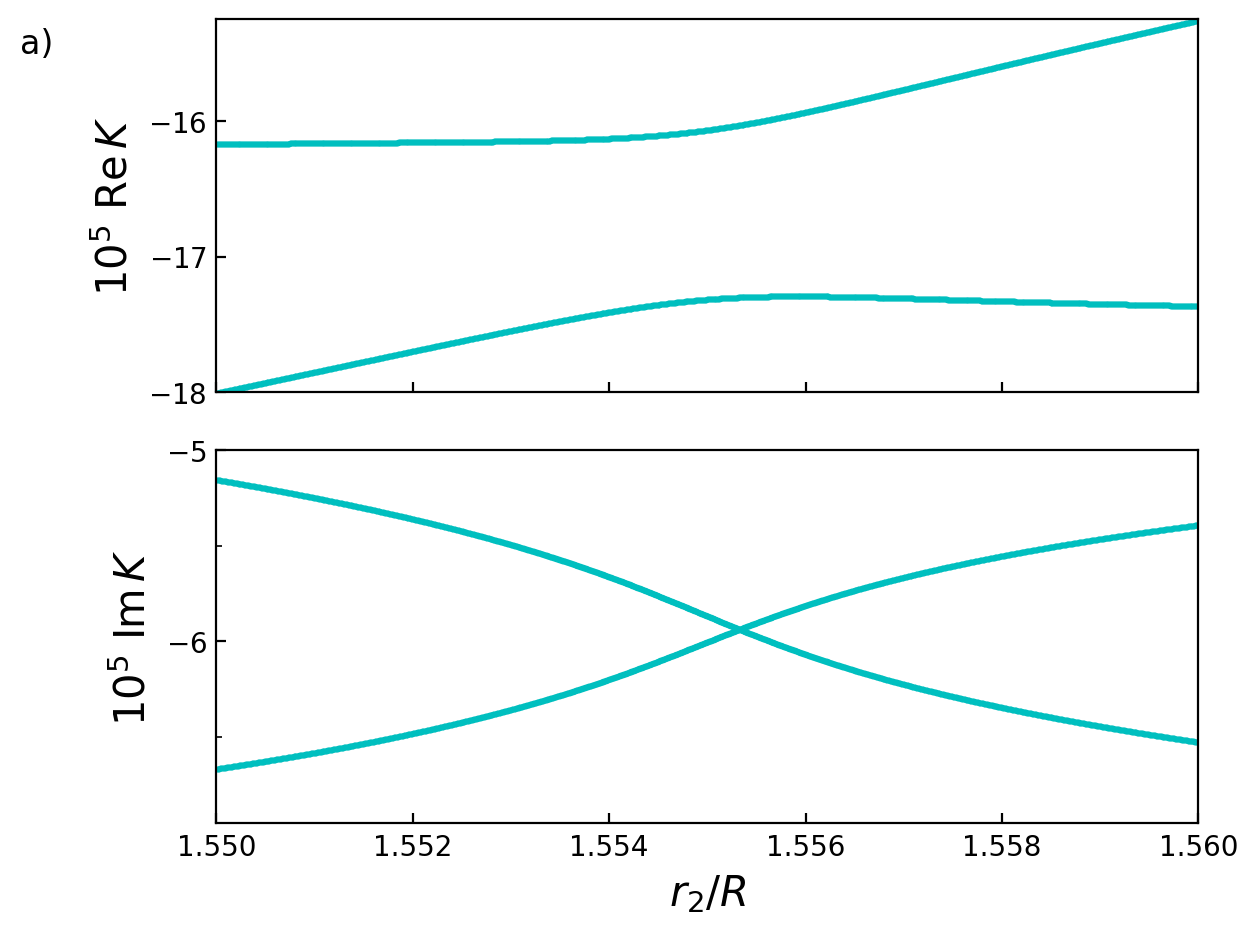}
\includegraphics[width=\linewidth]{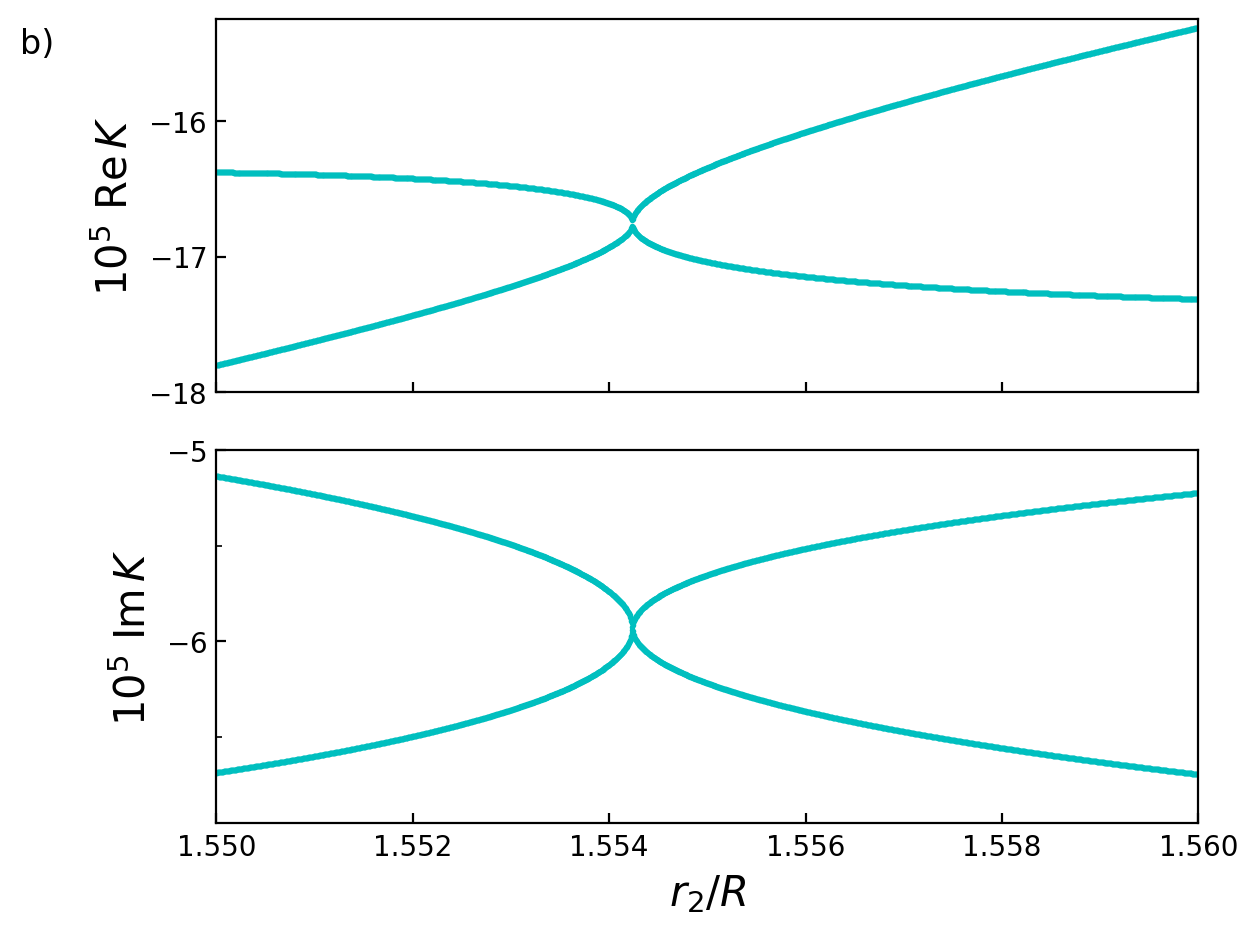}
\includegraphics[width=\linewidth]{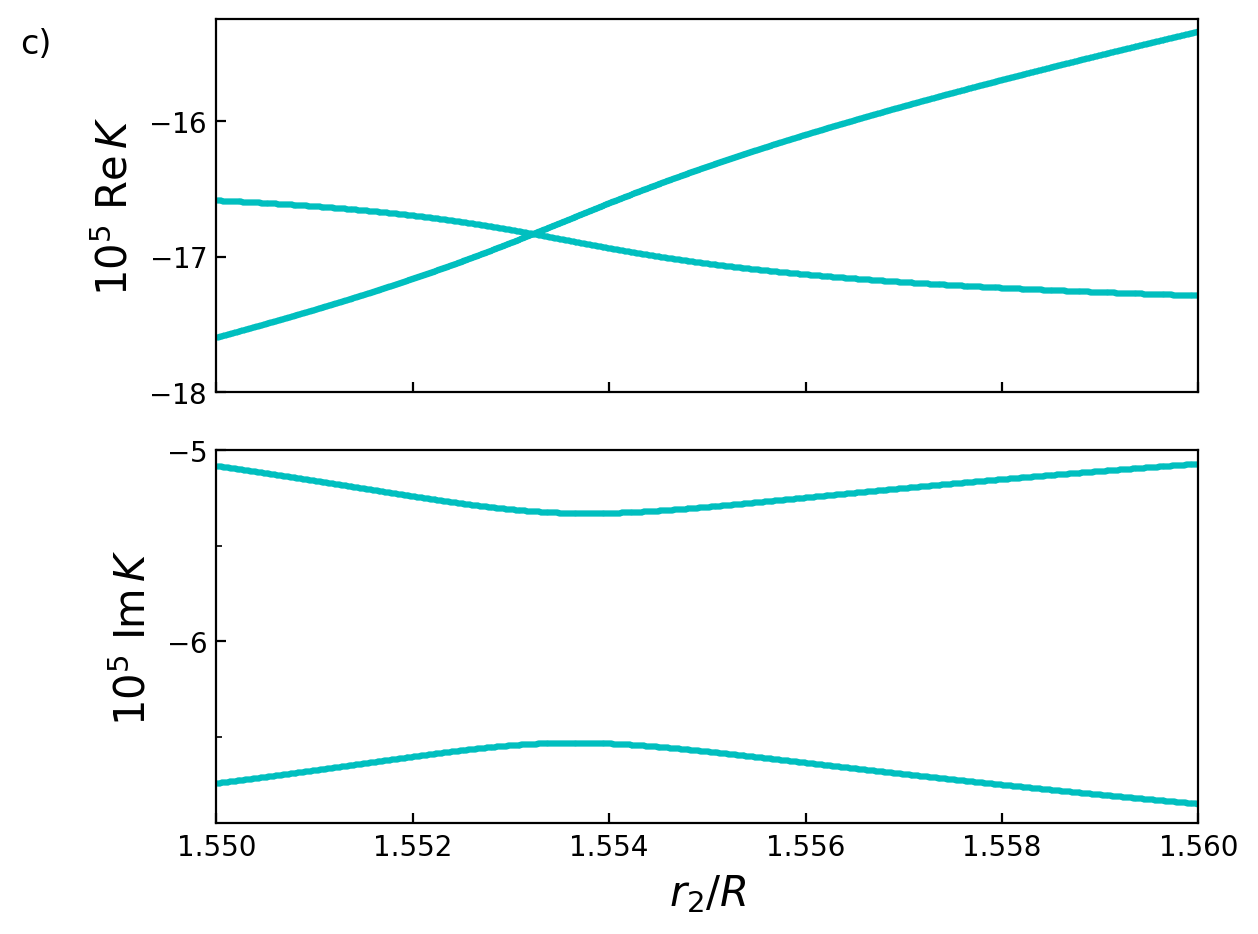}
\caption{As Fig.\ref{l=20 EP} but for a $20\times20$ matrix only, varying $r_2$, and fixing $\alpha=1.6$, $r_1/R=1.5$, and $\Delta \varphi$, with  (a) $\Delta \varphi = 1.200605$ (strong coupling), (b) $\Delta \varphi = 1.199605$ (EP), and (c) $\Delta \varphi = 1.198605$ (weak coupling). The actual values of $K$ are obtained by multiplying the shown numbers by $10^{-5}$ as indicated.}
\label{coupling}
\end{figure}

A transition from weak to strong coupling occurs at an EP which is also the point of critical coupling \cite{CaoRMP15}. The EP in Fig.\ref{l=20 EP} is replotted now with Fig.\ref{coupling}(b) instead varying the parameter $r_2/R$ and fixing $\alpha=1.6$, $r_1/R=1.5$, and $\Delta\varphi=1.199605$. The strong (weak) coupling regime is shown in Fig.\ref{coupling}(a) (\ref{coupling}(c)) by adding (subtracting) $10^{-3}$ to (from) $\Delta \varphi$. These figures clearly demonstrate the characteristics of the real and imaginary parts in the strong and weak coupling regimes. These are an avoided crossing of the real part and crossing of the imaginary part of the eigenvalues (here the RS wavenumbers) in the strong coupling, and vice versa in the weak coupling. Approaching the EP from the strong coupling regime, the Rabi splitting strictly vanishes at the EP and the linear crossing of the imaginary part of the wavenumber also gains the characteristic shape of the EP as shown by Fig.\ref{coupling}(b). Small Rabi splitting is thus an indication of proximity to a degeneracy in this system. Whether or not the system then becomes weakly coupled distinguishes an EP from a DP.\\ %

Now considering the wave functions of the perturbed WGMs, the expansion coefficients obtained from the diagonalization of the RSE matrix are used in Eq.(\ref{Ecal}) along with the TE electric fields from Eq.(\ref{ETE}). The perturbed electric fields are found by a summation over even $m$ from $m=-20$ to $m=20$ excluding $m=0$. The modulus of these fields on the sphere surface $r=R$ are color plotted in Fig.\ref{1st WF}, as well as in Figs.\ref{2nd WF}-\ref{last WF} in Appendix \ref{C}. In these plots, $\varphi$ is represented intuitively by the polar angle but $\theta$ is represented by the radial coordinate $\rho(\theta) = \theta$. %
As a result, the center of these diagrams is the south pole and the outer circle is the north pole of the sphere; this distortion should be considered when interpreting these graphs.

\begin{figure*}
\centering
\includegraphics[scale=0.59]{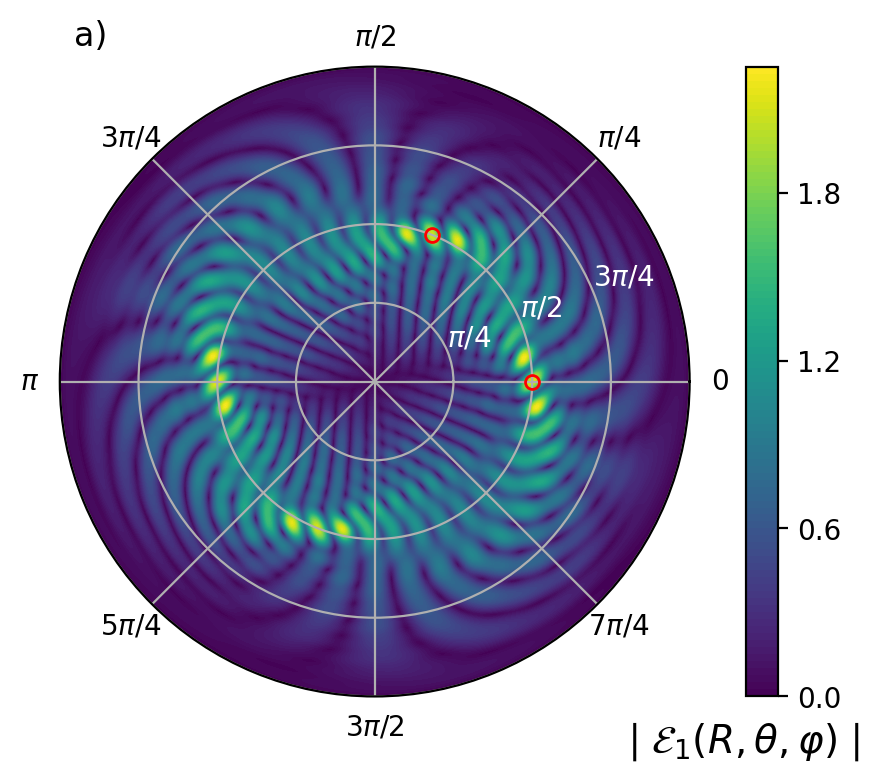}
\includegraphics[scale=0.59]{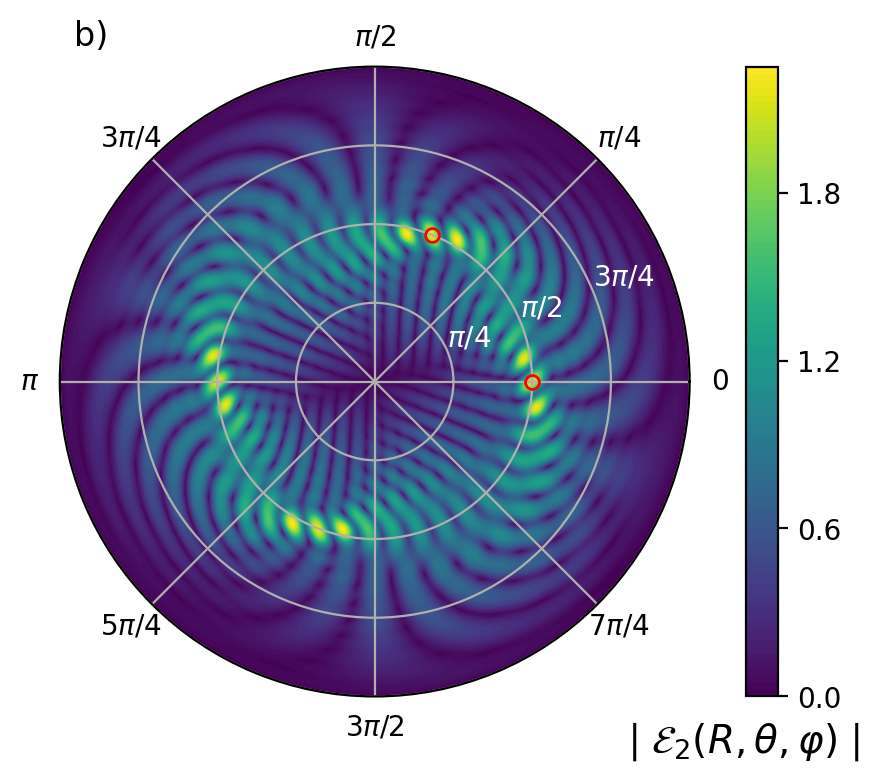}
\includegraphics[scale=0.59]{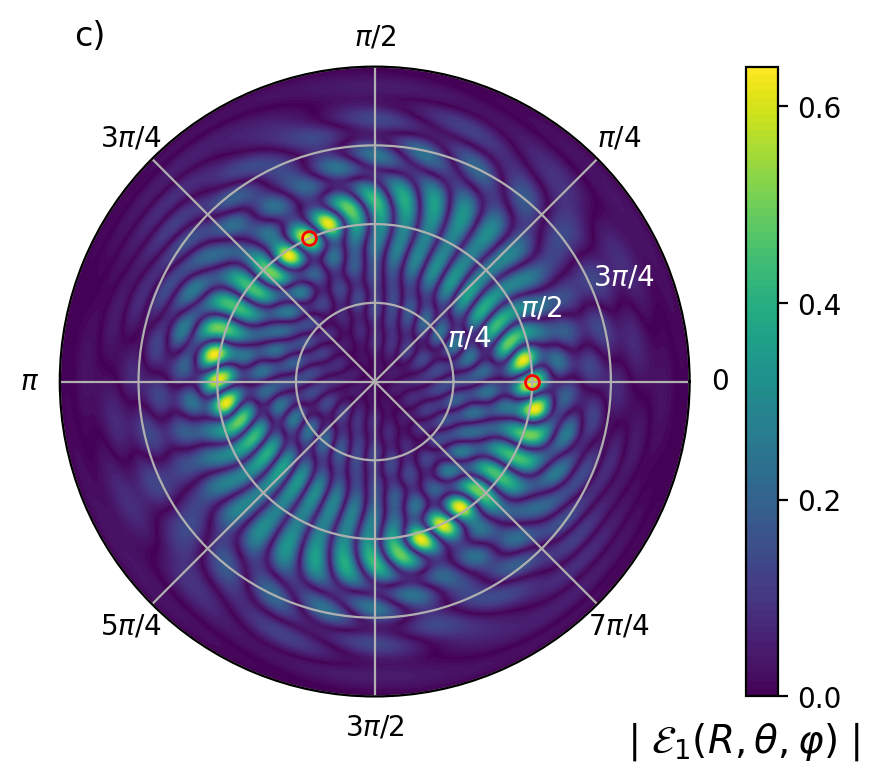}
\includegraphics[scale=0.59]{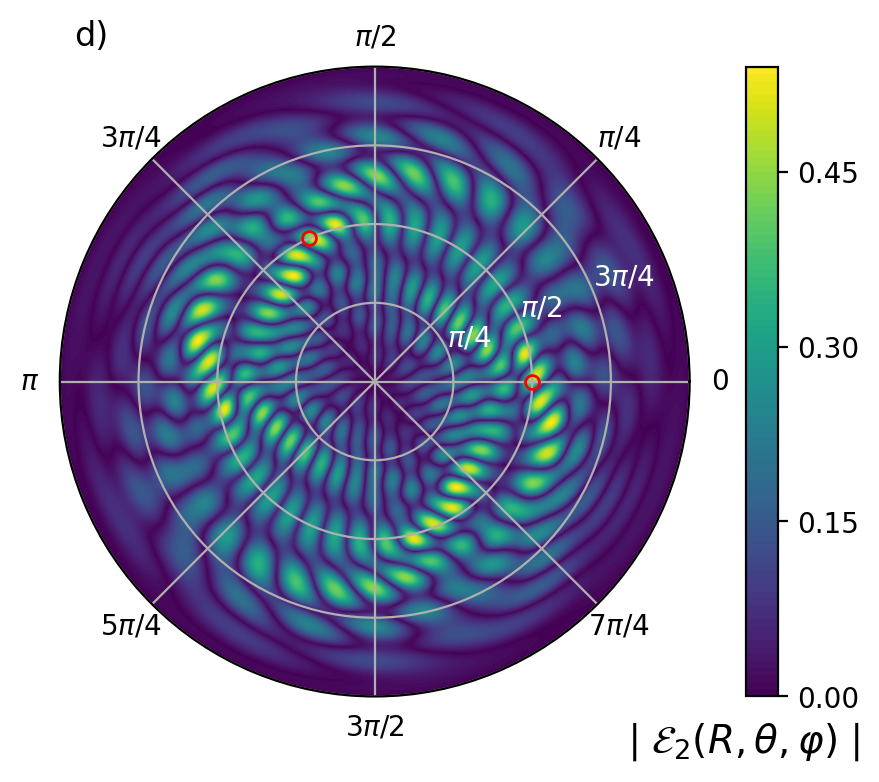}
\includegraphics[scale=0.59]{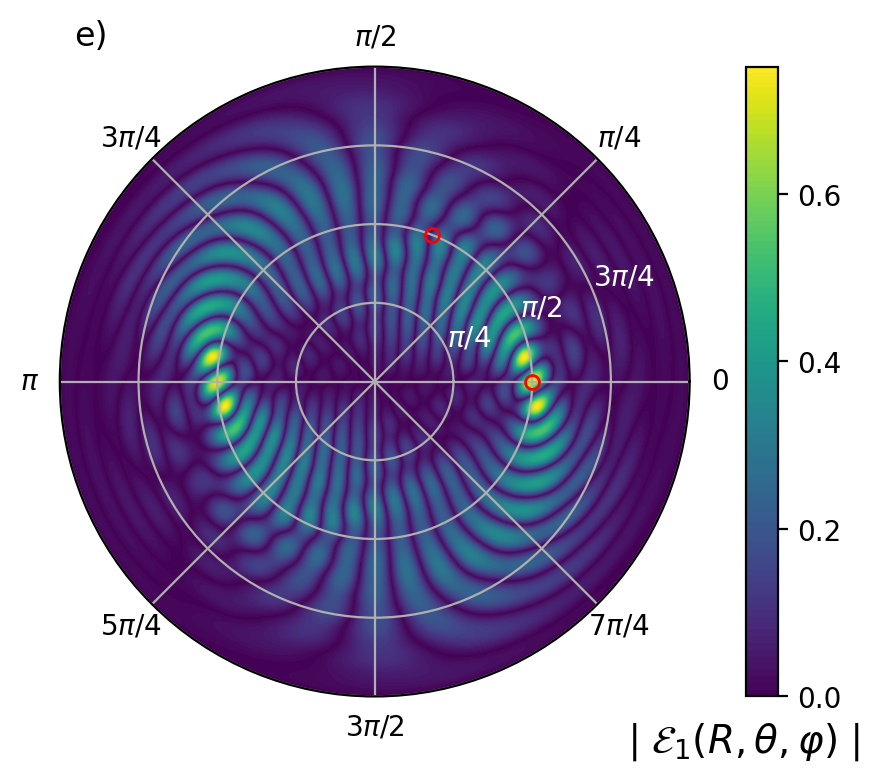}
\includegraphics[scale=0.59]{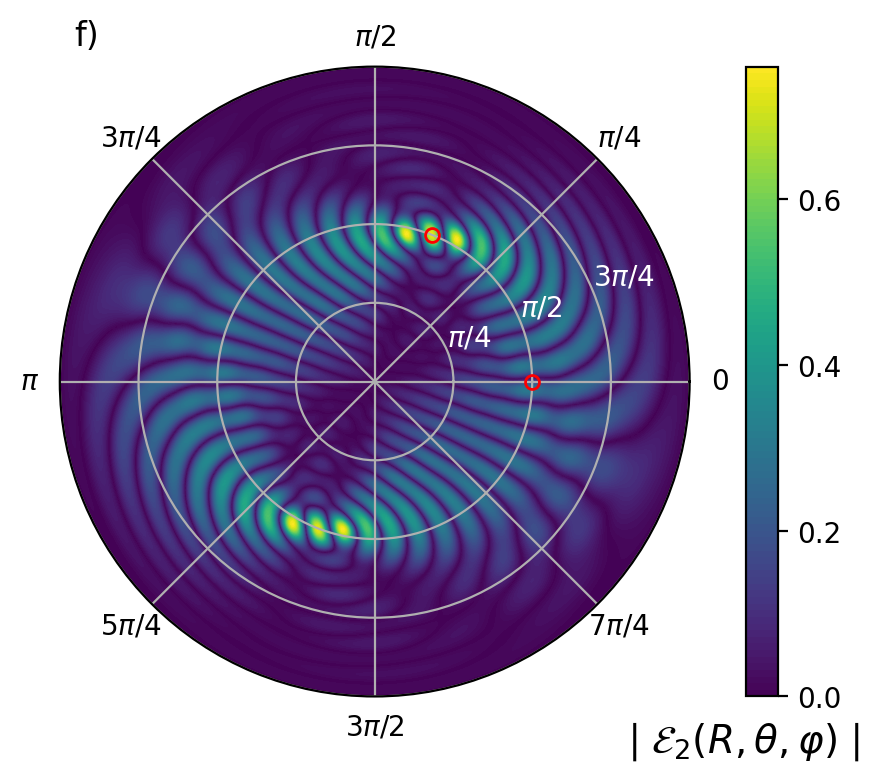}
\caption{(Color online) Modulus of perturbed $l=20$ electric fields at the sphere surface as a function of azimuthal and polar angles for $\nu=1$ (left panels) and $\nu=2$ (right panels). $\varphi$ is shown by the angular axis and $\theta$ is shown by the radial axis. Brighter colors represent a stronger field. Red hollow circles show the angular position of the perturbers. Panels (a) and (b) show the wave functions at the EP with $\Delta\varphi=1.199605$ and $\alpha=1.6$. Panels (c) and (d) show the wave functions away from the EP for $\Delta\varphi=2$ and $\alpha=1.6$. Panels (e) and (f) show the wave functions away from the EP for $\Delta\varphi=1.199605$ and $\alpha=3$. The perturber radii are $r_1/R=1.5$ and $r_2/R=1.55420$ for all panels.
}
\label{1st WF}
\vskip4cm
\end{figure*}

\begin{figure*}
\centering
\includegraphics[width=0.49\linewidth]{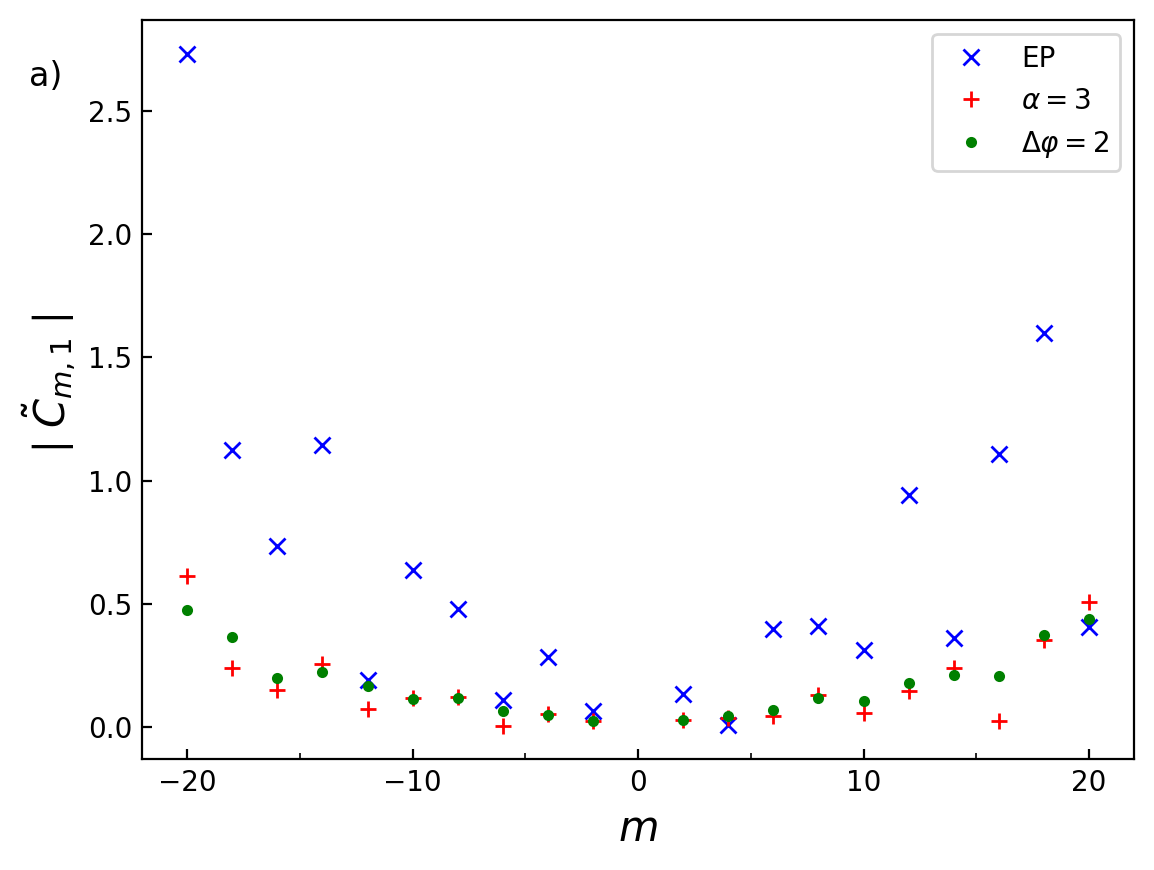}
\includegraphics[width=0.49\linewidth]{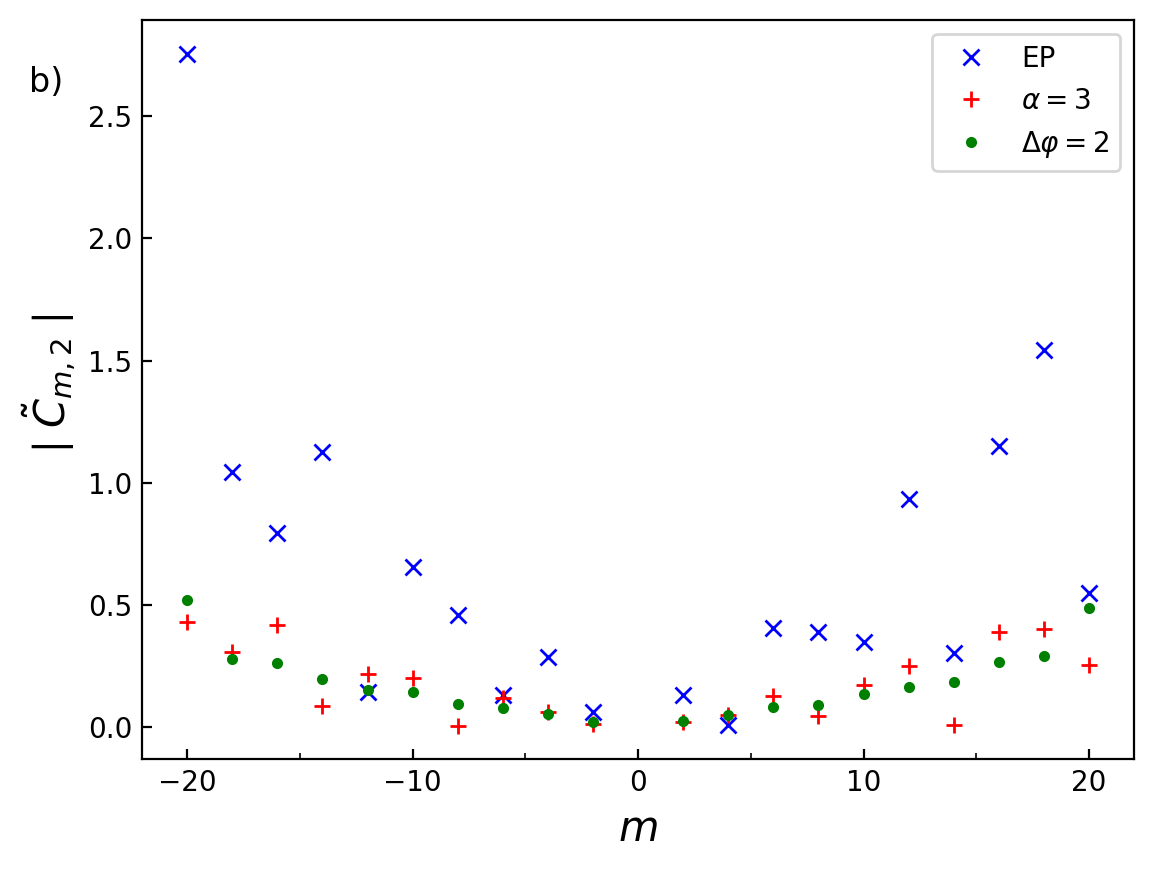}
\includegraphics[width=0.49\linewidth]{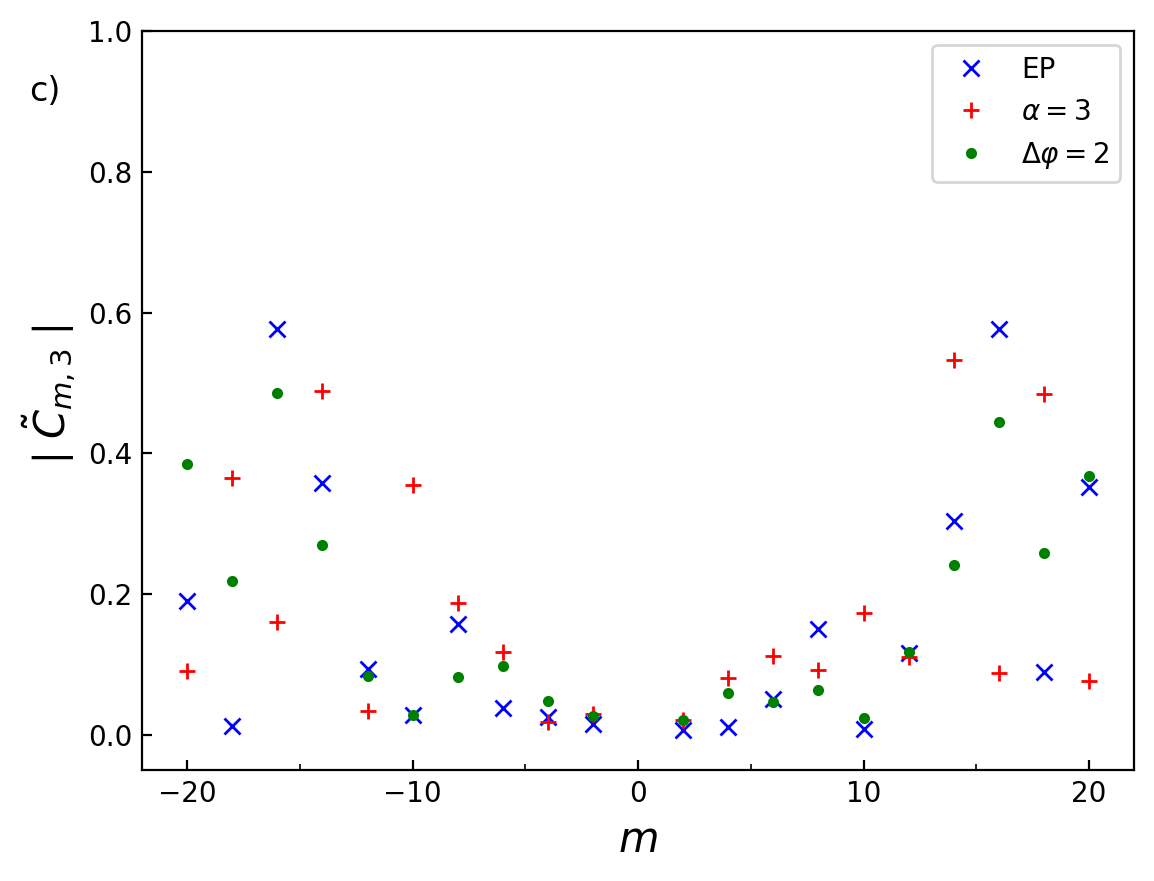}
\includegraphics[width=0.49\linewidth]{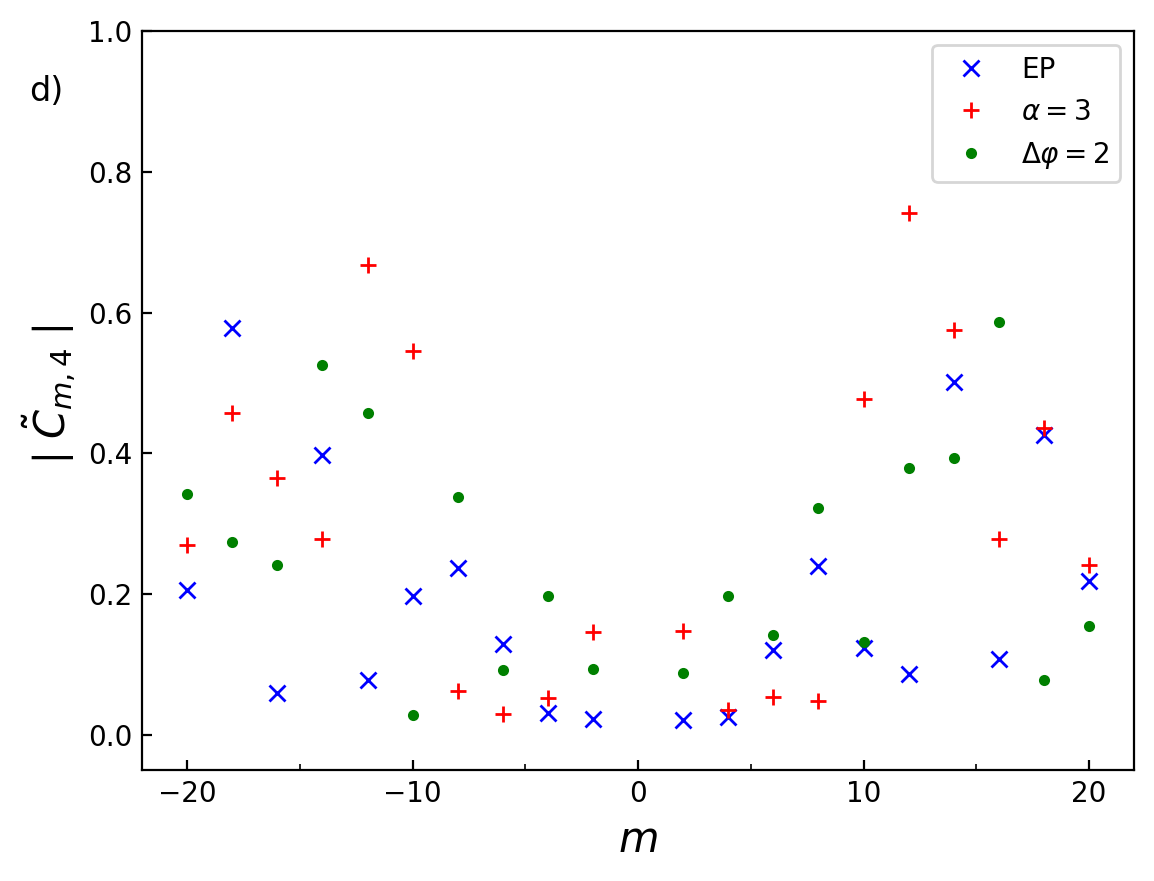}
\caption{(Color online) Modulus of RSE expansion coefficients $\tilde{C}_{m\nu}$, for $l=20$, $n_r=2$, and two perturbers, against $m$. Blue crosses show eigenstates at the given parameters of the EP, which are $\alpha=1.6$, $\Delta \varphi=1.199605$, $r_1/R=1.5$, and $r_2/R=1.55420$, red pluses show eigenstates with the perturber strength ratio changed to $\alpha=3$, and green dots show eigenstates with the angle between the perturbers changed to $\Delta\varphi=2$. Panels (a), (b), (c), and (d) show the expansion coefficients for states $\nu=1,2,3$ and 4, respectively.}
\label{Cm}
\end{figure*}

Figures \ref{1st WF}(a) and \ref{1st WF}(b) show the affected states in the vicinity of the EP, where they are almost identical, demonstrating the degeneracy of the eigenfunctions. The small difference between the nearly degenerate eigenfunctions arises from the fact that we cannot exactly reach an EP in the parameter space. Figures \ref{1st WF}(c) and \ref{1st WF}(d) show the states at $\Delta \varphi=2$ changed to be away from the parameters of the EP (for which $\Delta \varphi=1.199605$), and Figs.\ref{1st WF}(e) and \ref{1st WF}(f) show the states with $\alpha=3$ changed to be away from the parameters of the EP (for which $\alpha=1.6$). The eigenfunctions with altered $\Delta\varphi$ have a similar pattern to each other and to the same states at the EP but they appear rotated and are no longer degenerate. The eigenfunctions with altered $\alpha$ have a much larger change in parameter and thus are much further away from the EP; the electric fields become much stronger at one perturber than the other, although the electric field at the other perturber is still non-zero. This electric field maximum is enhanced from max$\,\lvert\mathbfcal{E}_\nu\rvert\approx0.5$ at the DP and for non-degenerate affected states to max$\,\lvert\mathbfcal{E}_\nu\rvert\approx2.25$ close to the EP. The closer the system is to an EP, the greater the enhancement of the eigenfunctions, as discussed in Sec.\ref{Purcell}, see Eq.(\ref{E-chiral}). The difference between the maxima in Figs.\ref{1st WF}(a) and \ref{1st WF}(b) is small and the observation of enhanced emission or chirality will be sufficient to observe the EP.
Away from the EP, the maximum value of $\lvert\mathbfcal{E}_1(\mathbf{r})\rvert$ for the first mode occurs at the position of one perturber; conversely, the maximum of $\lvert\mathbfcal{E}_2(\mathbf{r})\rvert$ for the second mode occurs at the position of the other perturber, see Figs.\ref{1st WF}(e) and \ref{1st WF}(f). The electric field close to an EP thus has a significantly different shape compared to the same state far away from the EP, because it has maxima at the positions of both defects instead of just one. Figures \ref{2nd WF}-\ref{last WF} in Appendix \ref{C} show the electric fields of the unaffected modes at the parameters of the EP. Because the degeneracy of the unaffected states is for them a DP not an EP, all their eigenfunctions are different. $\mathbfcal{E}_{\nu>2}(\mathbf{r}_j) = 0$ at both perturber positions for all of the unaffected states which supports the fact that their eigenvalues are not affected by the perturbation. This is because the perturbation matrix only depends on the electric fields at the perturber positions, see Sec.\ref{Orthogonal} below for a more rigorous discussion.

Another feature of WGMs made analyzable by the RSE is chirality. To investigate the chirality while retaining the symmetric formalism of the RSE, the real azimuthal function $\chi_{m}(\varphi)$ given by Eq.(\ref{chi}) is expanded into the complex azimuthal functions $ e^{im\varphi}$. This change of the basis functions transforms the expansion coefficients of the RSE in the following way:
\begin{equation}\label{CC}
\sum_{{m}=\pm |m|} \tilde{C}_{{m}\nu} e^{i{m}\varphi} = \sqrt{2} \sum_{{m}=\pm |m|} C_{{m}\nu} \chi_{{m}}(\varphi)\,.
\end{equation}
Note that as we only consider degenerate modes of the same polarization, so the general RSE index $n$ is replaced here and below with azimuthal quantum number $m$. From Eq.(\ref{CC}) we find explicitly the transformed expansion coefficients of the RSE,
\begin{equation}\label{Cmp}
\tilde{C}_{\pm \lvert m \rvert, \nu} = \frac{1}{\sqrt{2}}  \left(C_{+\lvert m \rvert, \nu} \pm i  C_{-\lvert m \rvert, \nu}\right)\,,
\end{equation}
which are not normalized by Eq.(20) but instead by 
$\sum_m \tilde{C}_{-m,\nu} \tilde{C}_{m,\nu^\prime} = \delta_{\nu\nu^\prime}$.
The sign of $m$ is changed in one of the eigenvectors compared with Eq.(20) which is equivalent to the partial conjugation required in the non-symmetric formalism of the RSE [32,49].

The modulus of the coefficients $\tilde{C}_{{m}\nu}$ are then plotted against $m$ in Fig.\ref{Cm} for the affected states ($\nu=1,2$) and two of the unaffected states ($\nu=3,4$). The coefficients are taken at the given parameters of the EP  %
and also away from the EP by separately changing the strength ratio and angle between the perturbers to $\alpha=3$ and $\Delta\varphi=2$, respectively, while keeping all other parameters the same. While the eigenvalues of the unaffected states are unchanged by variations of the perturbation parameters, the corresponding expansion coefficients $\tilde{C}_{m\nu}$ (or $C_{m\nu}$) do change. This is because a variation of the parameters will alter the affected states (both their wavenumbers and wave functions), and since all the eigenvectors of the RSE matrix are orthogonal \cite{lobanov2018resonant}, the wave functions of the unaffected states change.

Figure \ref{Cm} shows that the dominant coefficient of the degenerate affected states is much larger than the other coefficients. This implies an imbalance towards negative m which is a CW chirality. This means light has a preferred angular propagation direction in this system. Eigenstates at an EP always differ by at least a factor of $i$ and therefore have a phase difference of $\pi/2$. They have similar expansion coefficients at the parameters of the EP due to the degeneracy of eigenstates. However, they are are not precisely equal because it is impossible to exactly reach any single point (including an EP), and eigenstates are more sensitive to the parameters than eigenvalues. This sensitivity is owing to the fact that the expansion coefficients are inversely proportional to the square root of the wavenumber splitting, see Eq.(\ref{E-chiral}). The expansion coefficients near an EP in Figs.\ref{Cm}(a) and \ref{Cm}(b) are shown to be enhanced a few times compared to the unaffected states in Figs.\ref{Cm}(c) and \ref{Cm}(d), due to being in the vicinity of but not exactly at an EP. Eigenstates at an EP always differ by at least a factor of $i$ and therefore have a phase difference of $\pi/2$. The demonstrated chirality supports the idea of two WGMs exactly at an EP being both CW (or CCW). Chirality imbalance is not that strong for the unaffected states or with the affected states that are not in the proximity of an EP. EPs between WGMs thus enhance chirality and the magnitudes of their electric fields, which is relevant for light absorption applications \cite{soleymani2022chiral}. %
It may also be possible to utilize the enhanced sensitivity of an EP in chiral sensing applications \cite{wiersig2020review,sarma2015rotating,wang2019arbitrary}, although Sec.\ref{Purcell} shows that the emission spectrum has limited sensitivity. %
The chirality imbalance for a microsphere is in agreement with the phenomenological analysis by Wiersig \cite{wiersig2011structure} for a microdisk with two perturbers.

\subsection{Orthogonal transformation to two-mode basis}\label{Orthogonal}
Since the wavenumbers of only two modes ($\nu=1$ and $\nu=2$) are affected by the perturbation, while those of the remaining 18 modes are unaffected, the $20\times20$ RSE matrix problem can be effectively reduced to a $2\times2$ matrix problem, similar to that treated in Secs.\ref{l=1} and \ref{Purcell}. One can do this in a mathematically rigorous way by making an orthogonal transformation of the initial RSE matrix $H_{mm^\prime}$.

To develop this orthogonal transformation we consider the $20\times20$ RSE matrix problem Eq.(\ref{RSE}) in the form
\begin{equation}\label{Hsigma}
\sum_{m^\prime}H_{mm^\prime}(\sigma)C_{m^\prime\nu}(\sigma) = \frac{1}{\varkappa_\nu(\sigma)} C_{m\nu}(\sigma)
\end{equation}
where $\sigma$ denotes one of the free parameters of the perturbation (such as $\alpha_j$ or $r_j$), or their combination. Let us denote the unaffected states with index $\bar{\nu}$ ($\bar{\nu}=\nu$ for $3\leqslant\nu\leqslant20$). At the perturber positions ${\bf r}_j$,  their electric fields vanish, which can be written as
\begin{equation}\label{Eexp}
{\cal E}_{\bar{\nu}}({\bf r}_j)= \sum_m C_{m\bar{\nu}}(\sigma) E_m ({\bf r}_j)=0\,,
\end{equation}
using the general expansion Eq.(\ref{Ecal}) and the fact that $\varkappa_{\bar{\nu}}(\sigma)=k_0$, where $k_0$ is the unperturbed wavenumber. The latter has been demonstrated numerically in Sec.\ref{20.1} but can also be proven mathematically. In fact, it follows from Eq.(\ref{Eexp}) that $\sum_{m^\prime} V_{mm^\prime} C_{m^\prime\bar{\nu}}=0$, using the explicit form of the matrix elements, Eq.(\ref{V}). Then, omitting the argument $\sigma$ for brevity, we obtain from Eq.(\ref{Hsigma}):
\begin{equation}
\sum_{m^\prime}H_{mm^\prime}C_{m^\prime\bar{\nu}} = \frac{1}{k_0}\left(  C_{m\bar{\nu}} +
\sum_{m^\prime} V_{mm^\prime} C_{m^\prime\bar{\nu}}\right)=\frac{1}{k_0} C_{m\bar{\nu}}\,,
\end{equation}
that results in $\varkappa_{\bar{\nu}}=k_0$.

Note that Eq.(\ref{Eexp}) is a scalar equation. In fact, on the equatorial plane, the electric field has only the $\theta$ component in TE polarization for $l$ and $m$ of the same parity which is the present case treated here. Since $j$ only takes the values 1 or 2, for each state $\bar{\nu}$, Eq.(\ref{Eexp}) presents a set of two simultaneous homogeneous linear algebraic equations with constant coefficients $E_m ({\bf r}_j)$ determining 20 variables $C_{m\bar{\nu}}(\sigma)$. This pair of algebraic equations  is exactly the same for all $\bar{\nu}$ and therefore has 18 linearly independent solutions (labeled by $\bar{\nu}$) which can be chosen orthogonal to each other. By an orthogonal transformation (or matrix rotation), they can be transformed to any other orthogonal combination. Since there are two affected states, and the RSE matrix equation imposes a mutual orthogonality Eq.(\ref{norm}) of its eigenvectors, the orthogonal vectors $C_{m\bar{\nu}}(\sigma)$ are uniquely determined (up to a factor of $-1$) in such a way that all the unaffected states are not only orthogonal to each other but also to the affected modes, for the given $\sigma$. However, if $\sigma$ is changed to  $\sigma^\prime$, the affected modes change according to the change of the perturbation. This also changes the orthogonal combinations $C_{m\bar{\nu}}(\sigma^\prime)$ of the unaffected states. These eigenvectors $C_{m\bar{\nu}}(\sigma^\prime)$ satisfy the same Eq.(\ref{Eexp}),
namely,
\begin{equation}\label{Eexp2}
 \sum_m C_{m\bar{\nu}}(\sigma') E_m ({\bf r}_j)=0\,,
\end{equation}
as explained below, but at the same time are orthogonal to the affected states for the new value $\sigma^\prime$.

In the above discussion and in Eq.(\ref{Eexp2}), we have assumed that the coefficients $E_m ({\bf r}_j)$ in the algebraic equations do not change when going from $\sigma$ to $\sigma^\prime$. This is true if the varied parameters of the system include $\alpha$,  $r_1$, and $r_2$, but not $\varphi_2-\varphi_1$. In fact, $E_m ({\bf r}_j)$ do not depend on $\alpha_j$ and all have the same radial factor for a given radial position of the perturber $r_j$, according to Eq.(\ref{ETE}). This common factor, which is dependent on $r_j$, can be removed from Eq.(\ref{Eexp}), meaning the $18$ unaffected eigenfunctions vanish at $\mathbf{r}_j$ regardless of the values of $r_j$. The same is not true, however, if the angular position of the perturber $\varphi_j$ changes, as is clear from the same Eq.(\ref{ETE}). Further, if the basis system includes TM RSs, only $\alpha$ can be used for the change from $\sigma$ to $\sigma^\prime$ because of the radial dependence in Eq.(\ref{ETM}).

As derived below, the orthogonal transformation of the whole RSE matrix is
\begin{align}
\mathcal{H}_{\nu\nu^\prime}(\sigma,\sigma^\prime) &= \sum_{mm^\prime} C_{m\nu}(\sigma^\prime) H_{mm^\prime}(\sigma) C_{m^\prime\nu^\prime}(\sigma^\prime) \label{Hcal1}\\
&=\begin{pmatrix}
\mathcal{H}_{\nu\nu^\prime}^{2\times 2}(\sigma,\sigma^\prime) & \hat{0}\\
\hat{0} & \dfrac{1}{k_0}\delta_{\bar{\nu}\bar{\nu}^\prime}
\end{pmatrix},
\label{Hcal2}
\end{align}
where $\mathcal{H}_{\nu\nu^\prime}^{2\times 2}$ is the $2\times2$ matrix transformed from the $\nu=1,2$ modes, $\hat{0}$ are null matrices, and $\delta_{\bar{\nu}\bar{\nu}^\prime}$ is an $18\times18$ identity matrix. Owing to the $\hat{0}$ blocks, $\mathcal{H}_{\nu\nu^\prime}$ can be clearly truncated into independent $2\times2$  matrix $\mathcal{H}_{\nu\nu^\prime}^{2\times 2}$ and diagonal matrix $k_0^{-1} \delta_{\bar{\nu}\bar{\nu}^\prime}$.

 To prove the block-diagonal form  Eq.(\ref{Hcal2}) of the transformed RSE matrix $\mathcal{H}_{\nu\nu^\prime}(\sigma,\sigma^\prime)$, we use Eqs.(\ref{H}) and Eq.(\ref{V})  in Eq.(\ref{Hcal1}) which yields
\begin{equation}
\begin{split}
\mathcal{H}_{\nu\nu^\prime}(\sigma,&\sigma^\prime) = \frac{1}{k_0}\sum_{mm^\prime} C_{m\nu}(\sigma^\prime) \delta_{mm^\prime} C_{m^\prime\nu^\prime}(\sigma^\prime)\\
&+ \frac{1}{k_0} \sum_j \sum_{mm^\prime} \alpha_j  C_{m\nu}(\sigma^\prime) {E}_m(\mathbf{r}_j) C_{m^\prime\nu^\prime}(\sigma^\prime) {E}_{m^\prime}(\mathbf{r}_j)\, .
\end{split}
\end{equation}
 Now, using the orthonormality Eq.(\ref{norm}) in the first term and Eq.(\ref{Eexp2}) in the second term, we obtain
\begin{equation}\label{Hcal3}
\mathcal{H}_{\nu\nu^\prime}(\sigma,\sigma^\prime) = \frac{\delta_{\nu\nu^\prime}}{k_0}\,,
\end{equation}
if $\nu$ or $\nu'$, or both $\nu$ and $\nu'$ belong to the unaffected states (i.e. $\nu=\bar{\nu}$ or $\nu'=\bar{\nu}'$), which yields the block diagonalization in Eq.(\ref{Hcal2}).

One can also find a link between the RSE coefficients $C_{m\bar{\nu}}(\sigma)$ and $C_{m\bar{\nu}}(\sigma^\prime)$ for the unaffected states at different values $\sigma$ and $\sigma'$ of the generalized parameter of the system. To do so, one can write the above mentioned linear transformation as
\begin{equation}\label{trans}
C_{m\bar{\nu}}(\sigma) =\sum_{\bar{\nu}'} C_{m\bar{\nu}'}(\sigma') U_{\bar{\nu}\bar{\nu}'}\,,
\end{equation}
where
\begin{equation}\label{umatrix}
U_{\bar{\nu}\bar{\nu}'}=\sum_m C_{m\bar{\nu}}(\sigma) C_{m\bar{\nu}'}(\sigma')
\end{equation}
is an orthogonal matrix,
\begin{equation}\label{ortho}
\sum_{\bar{\nu}_1} U_{\bar{\nu}\bar{\nu}_1} U_{\bar{\nu}'\bar{\nu}_1} =
\sum_{\bar{\nu}_1} U_{\bar{\nu}_1\bar{\nu}} U_{\bar{\nu}_1\bar{\nu}'} =
 \delta_{\bar{\nu}\bar{\nu}'}\,.
\end{equation}
The above properties follow directly from the orthonormality relation Eq.(\ref{norm}) used at $\sigma$ and $\sigma'$. In particular, one can obtain the transformation matrix Eq.(\ref{umatrix}) by multiplying Eq.(\ref{trans}) with $C_{m\bar{\nu}_1}(\sigma')$, summing over $m$, and using the orthonormality  Eq.(\ref{norm}). A similar procedure results in the orthogonality condition Eq.(\ref{ortho}) and in the inverse transformation, taking the form
\begin{equation}\label{transinv}
C_{m\bar{\nu}}(\sigma') =\sum_{\bar{\nu}'} C_{m\bar{\nu}'}(\sigma) U_{\bar{\nu}'\bar{\nu}}\,.
\end{equation}

The diagonal block of the unaffected states in Eq.(\ref{Hcal2}) can then be derived as
\begin{eqnarray}
\mathcal{H}_{\bar{\nu}\bar{\nu}'}(\sigma,\sigma') &=& \sum_{mm'} C_{m\bar{\nu}}(\sigma') H_{mm'}(\sigma) C_{m'\bar{\nu}'}(\sigma')
\nonumber\\
&=& \sum_{mm'} \sum_{\bar{\nu}_1} C_{m\bar{\nu}}(\sigma') H_{mm'}(\sigma) C_{m'\bar{\nu}_1}(\sigma) U_{\bar{\nu}_1\bar{\nu}'}
\nonumber\\
&=& \frac{1}{k_0} \sum_{m} \sum_{\bar{\nu}_1} C_{m\bar{\nu}}(\sigma') C_{m\bar{\nu}_1}(\sigma) U_{\bar{\nu}_1\bar{\nu}'}
\nonumber\\
&=& \frac{1}{k_0}  \sum_{\bar{\nu}_1} U_{\bar{\nu}_1\bar{\nu}} U_{\bar{\nu}_1\bar{\nu}'} =
 \frac{1}{k_0} \delta_{\bar{\nu}\bar{\nu}'}\,,
\label{Hcal}
\end{eqnarray}
where in the second line we have used Eq.(\ref{transinv}) and in the third line Eq.(\ref{Hsigma}) for $\nu={\bar{\nu}}$, recalling that  all $\varkappa_{\bar{\nu}}(\sigma)=k_0$.

To validate this transformation numerically, the matrix $\mathcal{H}_{\nu\nu^\prime}^{2\times 2}$ is calculated using Eq.(\ref{Hcal1}) and truncated according to Eq.(\ref{Hcal2}), then numerically diagonalized. Its eigenvalues, along with that of $k_0^{-1}\delta_{\bar{\nu}\bar{\nu}^\prime}$, are compared with the eigenvalues of $H_{mm^\prime}(\sigma)$. The $K$-values for these matrices are plotted against the perturber strength ratio $\alpha$ with $\sigma^\prime=\alpha^\prime=10$ in Fig.\ref{l=20 EP}, with the same parameters used for each matrix. Numerically, we see agreement to a precision of at least $7$ decimal places between the eigenvalues of $H_{mm^\prime}$ and $\mathcal{H}_{\nu\nu^\prime}$, and a complete separation between the blocks $\mathcal{H}_{\nu\nu^\prime}^{2\times 2}$ and $k_0^{-1}\delta_{\bar{\nu}\bar{\nu}^\prime}$ within the presented range. %

Equation (\ref{Hcal2}) describes the affected system, including the EP, in a rigorous and elegant way. A two-mode approximation used to describe EPs from the perturbation of double degenerate WGMs in a microdisk was postulated by Wiersig \cite{wiersig2011structure}, whereas matrix $\mathcal{H}_{\nu\nu^\prime}^{2\times 2}$ is a rigorously derived equivalent for the spherical case, despite the modes being initially $2l+1$ degenerate. While this transformation is only demonstrated for the case of two affected modes, it is also applicable to a system with more affected states, e.g. due to more perturbers modifying the system. Conditions for EPs of order $N$ are often obtained from $N \times N$ matrices, which are usually introduced phenomenologically, see for example \cite{am2015exceptional,wu2021high,li2022high} dealing with third-order EPs. The RSE encodes the information about the system in matrix form in a rigorous way and therefore is, to our knowledge, the most suitable tool presently available for studying EPs. The RSE usually deals with large matrices, however orthogonal transformations similar to the one introduced in this section can reduce the large matrix describing the system to an $N \times N$ matrix suitable for studying $N$th-order EPs.

\section{Conclusions}

Using the resonant-state expansion (RSE), we have rigorously investigated exceptional points (EPs) in a spherical optical resonator perturbed by two point-like defects. We exploited the significant advantage of the RSE compared to other approaches in that it exactly maps Maxwell's equations onto a linear matrix eigenvalue problem, therefore facilitating the study of EPs in optical systems in terms of suitable matrices describing them. For weak perturbations, infinite matrices are efficiently truncated to minimal sizes appropriate for the study of the EP phenomenon.

We have considered dipolar whispering-gallery modes (WGMs), corresponding to the lowest angular momentum ($l=1$), as well as high-quality WGMs with large angular momentum ($l=20$) in a dielectric microsphere surrounded by vacuum and perturbed by two point-like defects breaking the symmetry, such as nanoparticles or large molecules, placed inside or outside the resonator. Reducing the RSE basis to $2l+1$ degenerate TE-polarized fundamental WGMs and using parity selection rules, the RSE equation is naturally truncated for $l=1$ and $l=20$, respectively, to $2\times 2$ and $20\times 20$ matrix problems, the latter being further reducible to an effective $2\times 2$ matrix problem by applying a rigorous orthogonal transformation. Varying the parameters of these systems, we have demonstrated existence of EPs and even exceptional arcs which are continuous lines of EPs in the parameter space. Moreover, using the RSE formalism allowed us to develop an
exact analytic solution, valid in first-order in the perturbation strength, and an explicit analytic criterion for EPs and exceptional arcs in this realistic physical system.

We used the eigenfunction expansion central to the RSE to find the perturbed electric fields of the optical modes.  We have shown in particular that for high $l$, the two coalescent modes are divergent in the vicinity of EPs and have distinct maxima at the positions of the perturbers. At the same time, all other  states from the same degenerate group have electric fields that strictly vanish at the perturber positions which results in their wavenumbers being unaffected by the perturbation.

We furthermore demonstrated explicitly that the resonant states coalescent at EPs have the maximum optical chirality that manifests itself in the form of a chiral squared-Lorentzian optical response, which we have calculated analytically for the $l=1$ states in the Purcell enhancement spectra. While this squared-Lorentzian part is only a first-order correction to the normal Lorentzian spectrum, owing to its chirality, it can be effectively measured in circular-dichroism.

We have demonstrated that the RSE is a powerful tool to study EPs in an optical system as it encodes the information about the system in matrix form in a rigorous way. The idea of the orthogonal transformation presented in this work is the separation of the optical modes affected by the perturbation from the unaffected ones in a large basis of degenerate states. This can be further developed for a study of higher-order EPs, for example, in systems perturbed by $N$ point-like perturbers. We expect that in this case a similar orthogonal  transformation would reduce the RSE equation to an $N\times N$ matrix problem suitable for studying $N$th-order EPs.

\section*{Acknowledgement}
E.A.M. thanks H. Schomerus for discussions.

\appendix
\section{Derivation of the electric fields of the resonant states of a homogeneous sphere}\label{A}
For a homogeneous sphere with a permittivity described by Eq.(\ref{basis}), Maxwell's wave equation for the electric field has the following solution \cite{arfken1999mathematical}
\begin{equation}\label{TEM}
\mathbf{E}(\mathbf{r}) =
\begin{cases}
	-\mathbf{r} \times \nabla f &\; \text{for TE everywhere},\\
	-\;\dfrac{1}{\epsilon k}\nabla \times \mathbf{r} \times \nabla f &\; \text{for TM with $r \leqslant R$}\,,\\
	-\dfrac{1}{k}\nabla \times \mathbf{r} \times \nabla f &\; \text{for TM with $r > R$}\,,
\end{cases}
\end{equation}
where the scalar function $f(\mathbf{r})$ satisfies the Helmholtz equation \cite{doost2014resonant}
\begin{align}\label{Helmholtz}
\begin{split}
\nabla^2 f + k^2 \epsilon f = 0 & \quad \text{for $r \leqslant R$\,,}\\
\nabla^2 f + k^2 f = 0 & \quad \text{for $r > R$\,.}
\end{split}
\end{align}
The proof for Eq.(\ref{TEM}) is as follows. Substituting the TE solution $\mathbf{E} = -\mathbf{r} \times \nabla f$ into the left hand side of Eq.(\ref{Max wave}) and applying the triple vector product rule, we get
\begin{align}\label{proof}
\begin{split}
-\nabla \!\times \!\nabla \!\times \!\mathbf{r} \!\times \!\nabla f &= \nabla \!\times \!\left( 3\nabla f - \mathbf{r} \nabla^2 f \right)\\
&=3 \nabla \!\times\! \nabla f - \left(\nabla^2 f \right) \nabla \!\times \!\mathbf{r} + \mathbf{r} \!\times \! \nabla \nabla^2 f\\
&= \mathbf{r} \!\times \!\nabla \nabla^2 f\,,
\end{split}
\end{align}
using the facts that $\nabla \cdot \mathbf{r}=3$, $\nabla \times \nabla=0$, and $\nabla \times \mathbf{r}=0$. Substituting Eq.(\ref{Helmholtz}) into Eq.(\ref{proof}) yields the right hand side of Eq.(\ref{Max wave}) for inside or outside the sphere. For TM polarization, we obtain the same wave equation (with a step-like permittivity) for the magnetic field $\mathbf{H}$,
\begin{equation}\label{mag wave}
\nabla \times \nabla \times \mathbf{H}(\mathbf{r}) = k^2 \varepsilon(\mathbf{r}) \mathbf{H}(\mathbf{r})
\end{equation}
(valid in the regions of constant permittivity), which is derived similarly to Eq.(\ref{Max wave}). It is satisfied for the same form of the solutions, $\mathbf{H}=- \mathbf{r} \times \nabla f$. This magnetic field can then be substituted into Ampere's law in Eq.(\ref{Max}), giving the TM part of Eq.(\ref{TEM}).

Maxwell's equations (Ampere's and Faraday's laws) also have static solutions, i.e. with $k=0$, given by $\mathbf{E}=-\nabla f$ with $\mathbf{H}=0$ for longitudinal electric polarization and by $\mathbf{H}=-\nabla f$ with $\mathbf{E}=0$ for longitudinal magnetic polarization \cite{lobanov2018resonant,muljarov2020full}.
However, owing to the truncation of the basis used in this work, they are not considered in the present calculation.

To solve Eq.(\ref{Helmholtz}), we introduce the angular part of the Laplacian,
\begin{equation}
L(\theta,\varphi) = \dfrac{1}	{\sin\theta}\dfrac{\partial }{\partial \theta}\sin\theta\dfrac{\partial}{\partial \theta} + \dfrac{1}	{\sin^2\theta}\dfrac{\partial ^2}{\partial \varphi ^2}\,,
\end{equation}
so that Eq.(\ref{Helmholtz}) becomes
\begin{equation}\label{Laplace}
\begin{split}
	\left[\dfrac{\partial}{\partial r}r^2\dfrac{\partial}{\partial r} + L(\theta,\varphi) +k^2 n_r^2 r^2\right] f(\mathbf{r}) =  0\; \; &\text{for} \;\; r \leqslant R,\\
	\left[\dfrac{\partial}{\partial r}r^2\dfrac{\partial}{\partial r} + L(\theta,\varphi) +k^2 r^2 \right] f(\mathbf{r}) = 0 \;\;  &\text{for} \;\; r > R.
\end{split}
\end{equation}
The spherical harmonics $Y_{lm}(\theta,\varphi)$ are eigenfunctions of the operator $L(\theta,\varphi)$ with the corresponding eigenvalues $l(l+1)$. They satisfy the equation \cite{stratton2007electromagnetic}
\begin{equation}
\left[ L(\theta,\varphi) -l(l+1)\right] Y_{lm}(\theta,\varphi)  = 0\,,
\end{equation}
independent of $r$. This, together with the orthogonality of the spherical harmonics, allows us to separate the variables in Eq.(\ref{Helmholtz}), representing the wave function as $f(\mathbf{r})={\cal R}_l(r)Y_{lm}(\theta,\varphi) $, where the radial part ${\cal R}_l(r)$ satisfies the radial equations,
\begin{equation}
\begin{split}
	\left[\dfrac{\partial}{\partial r}r^2\dfrac{\partial}{\partial r} +k^2 n_r^2 r^2 - l(l+1) \right] {\cal R}_l(r) = 0\;  \; &\text{for} \; r \;\leqslant R\,,\\
	\left[\dfrac{\partial}{\partial r}r^2\dfrac{\partial}{\partial r} +k^2 r^2 - l(l+1) \right] {\cal R}_l(r) = 0 \;\;  &\text{for} \;\; r > R\,,
\end{split}
\end{equation}
having a general solution
\begin{equation}\label{radial}
{\cal R}_l(r) =
	\begin{cases}
	A_1 j_l(n_r kr) + A_2 h_l^{(1)}(n_rkr) & \text{for} \; r \leqslant R\,,\\
	A_3 h_l^{(1)}(kr) + A_4 h_l^{(2)}(kr) & \text{for} \; r > R\,
	\end{cases}
\end{equation}
where $h_l^{(2)}(z)$ is the spherical Hankel function of the second kind and $A_{1,2,3,4}$ are some constants. The function $h_l^{(1)}(n_rkr)$ is diverging at $r\to0$, which makes it an unphysical solution  inside the sphere, thus $A_2=0$. Imposing outgoing boundary conditions results in $A_4 = 0$ since $h_l^{(2)}(kr)$ is the solution representing an incoming wave \cite{muljarov2020full}. The ratio $A_1/A_3$ is found by imposing Maxwell's boundary conditions and each constant is then found from the proper normalization of the RSs \cite{muljarov2018resonant}. Equation (\ref{Rl}) is Eq.(\ref{radial}) normalized this way.

Finally, we substitute $f(\mathbf{r})= {\cal R}_l(r) Y_{lm}(\theta,\varphi)$ into Eq.(\ref{TEM}) which gives the electric fields \cite{doost2014resonant} in Eqs.(\ref{ETE}) and (\ref{ETM}).\\

\section{Parity selection rules for the perturbation matrix elements}
\label{3.2}
In the case of all perturbers located in the same plane, the electric fields of unperturbed modes with equal $l$, same polarization, and a different parity of $m$ are orthogonal. The proof is as follows.

Every component of the electric field vectors in Eq.(\ref{ETE}) and Eq.(\ref{ETM}) has a factor of an associated Legendre polynomial as a function of $\cos\theta$ or its derivative with respect to $\theta$. Since all perturbers share the same plane, without loss of generality, one can choose the coordinate system in such a way that this is the equatorial plane. With $\theta=\pi/2$, the Legendre functions and their derivatives become $P_l^{\lvert m\rvert}(0)$ and $\partial P_l^{\lvert m\rvert}(0)/\partial \theta$, respectively. These polynomials follow the rule
\begin{equation}\label{P cos}
P_l^m(0) \propto \cos\left(\frac{\pi}{2}(l+m)\right)
\end{equation}
and are thus vanishing for odd $l+m$ which happens when $l$ and $m$ have opposite parity. To get a similar expression for $\partial P_l^{\lvert m\rvert}(0)/\partial \theta$, we find the $\theta$ derivative \cite{abramowitz1988handbook}
\begin{equation}\label{dP cos}
\frac{dP_l^m(\cos\theta)}{d\theta} = -\frac{l\cos\theta P_l^m(\cos\theta) - (l+m)P_{l-1}^m(\cos\theta)}{\sin\theta}
\end{equation}
and let $\theta=\pi/2$ to get
\begin{align}\label{dP cos1}
\frac{dP_l^m(0)}{d\theta} = (l+m)P_{l-1}^m(0).
\end{align}
Substituting Eq.(\ref{P cos}) into Eq.(\ref{dP cos1}), we get the equivalent expression for the derivative
\begin{equation}\label{dP cos}
\frac{dP_l^m(0)}{d\theta} \propto \cos\left(\frac{\pi}{2}(l+m-1)\right)
\end{equation}
which vanishes for odd $l+m-1$ and thus when $l$ and $m$ have the same parity.

Focusing on the Legendre function part of Eqs.(\ref{ETE}) and (\ref{ETM}), taking the scalar product of two of these fields with the same position and polarization, but different $m$, gives
\begin{equation}
\label{dot}
\begin{split}
\mathbf{E}_{m}(\mathbf{r}) \cdot \mathbf{E}_{m^\prime}(\mathbf{r}) =\, & \xi_1(r,\varphi) P_l^{\lvert m \rvert}(\cos\theta)P_{l}^{\lvert m^\prime \rvert}(\cos\theta)\\
 & +\xi_2(r,\varphi) \frac{P_l^{\lvert m \rvert}(\cos\theta)P_{l}^{\lvert m^\prime \rvert}(\cos\theta)}{\sin^2\theta}\\
&+ \xi_3(r,\varphi) \frac{\partial P_l^{\lvert m \rvert}(\cos\theta)}{\partial\theta}\frac{\partial P_{l}^{\lvert m^\prime \rvert}(\cos\theta)}{\partial\theta}
\end{split}
\end{equation}
where the functions $\xi_{1,2,3}(r,\varphi)$ are independent of $\theta$.
$P_l^{\lvert m\rvert}(0)$ vanishes when $l$ and $m$ have opposite parity, and $\partial P_l^{\lvert m\rvert}(0)/\partial \theta$ vanishes when $l$ and $m$ have the same parity. As a result, all terms in Eq.(\ref{dot}), evaluated at $\theta=\pi/2$ vanish when $m$ and $m^\prime$ have opposite parity and the states have equal $l$ and the same polarization. This rule no longer holds if the perturbers cannot be considered to be all in the same plane.

Further truncation of the even states can be achieved by neglecting the vanishing fields. On the equatorial plane, the TE electric field is vanishing when $m=0$ and $l$ is even. Looking at Eq.(\ref{ETE}), this is because $\partial \chi_0/\partial \varphi=0$ makes the $\theta$ component vanish; since $l$ and $m$ are of the same parity, $\partial P_l^0/\partial \theta=0$, making the $\varphi$ component vanish; and the radial component is always vanishing in Eq.(\ref{ETE}). This is not the case for the TM modes because of their non-zero radial component.

\section{Matrix elements and EP condition for dipolar modes} \label{explicit}
To find the explicit form of the RSE matrix elements, we find the inner products of the TE electric fields in Eq.(\ref{ETE}) for degenerate modes with $m=\pm1$ and $\theta_j=\pi/2$ with $j=1,\,2$ labeling the perturbers. Since $P_1^{\pm1}(\cos\theta) \propto \sin\theta$, the azimuthal component of the electric field, which takes the $\theta$ derivative of the Legendre function at $\theta=\theta_j=\pi/2$, vanishes. Using Eq.(\ref{H}), where the indices $n=1$ and $n=2$ denote the modes with $m=-1$ and $m=1$, respectively, the RSE matrix elements in Eq.(\ref{H condition}) take the form
\begin{equation}\label{inner}
\begin{split}
H_{11} &= \frac{1}{k_0} + \frac{1}{k_0} \sum_{j=1}^2 \alpha_j \tilde{\cal R}^2(r_j) \cos^2\varphi_j\,,\\
H_{22} &= \frac{1}{k_0} + \frac{1}{k_0} \sum_{j=1}^2 \alpha_j \tilde{\cal R}^2(r_j) \sin^2\varphi_j\,,\\
H_{12} &= -\frac{1}{k_0} \sum_{j=1}^2 \alpha_j \tilde{\cal R}^2(r_j) \cos\varphi_j\sin\varphi_j =H_{21}\,,
\end{split}
\end{equation}
where
\begin{equation}\label{tR}
\tilde{\cal R}(r)=\sqrt{\frac{3}{4\pi}} A_1^{\rm TE} {\cal R}_1(r)\,,
\end{equation}
and ${\cal R}_1(r)$ and $A_1^{\text{TE}}$ are given, respectively by Eqs.(\ref{Rl}) and (\ref{Anorm}) for $l=1$.

Substituting Eq.(\ref{inner}) into the degeneracy condition Eq.(\ref{H condition}) and using the facts that
\begin{equation}
k(H_{11} - H_{22}) = \alpha_1 \tilde{\cal R}^2(r_1) \cos(2\varphi_1) + \alpha_2 \tilde{\cal R}^2(r_2) \cos(2\varphi_2)
\end{equation}
and
\begin{equation}\label{2H12}
-2k H_{12} = \alpha_1 \tilde{\cal R}^2(r_1) \sin(2\varphi_1) + \alpha_2 \tilde{\cal R}^2(r_2) \sin(2\varphi_2)\,,
\end{equation}
we find
\begin{equation} \label{EPequ}
e^{\pm2i(\varphi_1-\varphi_2)} = - \frac{\alpha_1}{\alpha_2} \frac{\tilde{\cal R}^2(r_1)}{\tilde{\cal R}^2(r_2)}\,.
\end{equation}
With $\alpha=\alpha_2/\alpha_1$, $\Delta\varphi=\varphi_2-\varphi_1$, and Eq.(\ref{tR}) we get Eq.(\ref{degen condition}).

We also use Eq.(\ref{inner}) to derive the scaled dimensionless wavenumber $K$ at EPs. The degenerate eigenvalue $1/\varkappa$ at an EP is the mean of diagonal elements of the $2\times2$ matrix, given by the first term in Eq.(\ref{EP2 eigenvalue}). For this RSE matrix, explicitly,
\begin{equation}\label{lamEP}
\begin{split}
\frac{1}{\varkappa} = \frac{1}{k_0}\left( 1 + \frac{1}{2}\sum_{j=1}^2 \alpha_j \tilde{\cal R}^2(r_j)  \right),
\end{split}
\end{equation}
valid at the EPs and also at the DPs contained within the EA.  Using Eq.(\ref{EPequ}), we can write the perturbed wavenumber as
\begin{equation}
\varkappa = k_0 \left[  1 +  \frac{1}{2}\alpha_1 \tilde{\cal R}^2(r_1) \left( 1 - e^{\pm 2i\Delta\varphi}\right) \right]^{-1}
\end{equation}
which is substituted into Eq.(\ref{K}) to obtain Eq.(\ref{Kep}).

The EA in Fig.\ref{EA} contains a countable number of DPs where $H_{12}=0$. Combining Eqs.(\ref{2H12}) and (\ref{EPequ}), this DP condition can be written as
\begin{equation}
\tan(2\varphi_1) = \tan(2\varphi_2) \,.
\end{equation}
Since $\alpha_1 \tilde{\cal R}^2(r_1) \neq 0$, we find that $\varphi_2=\varphi_1+p\pi/2$, where $p$ is any integer, at these DPs. \\

\section{Eigenfunctions of the unaffected modes}\label{C}
\begin{figure*}
\includegraphics[scale=0.59]{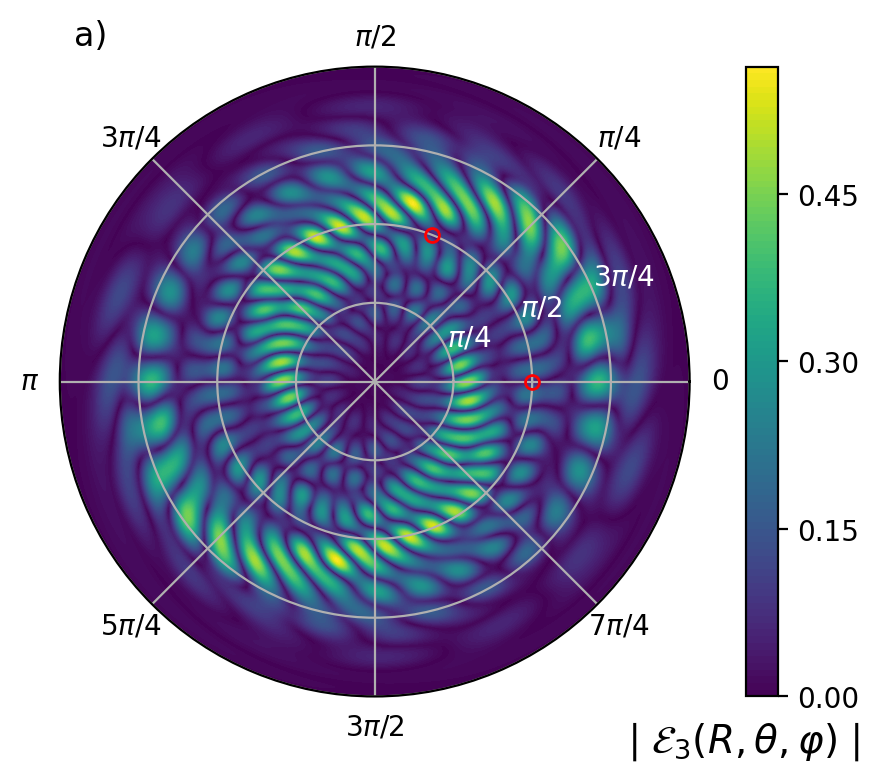}
\includegraphics[scale=0.59]{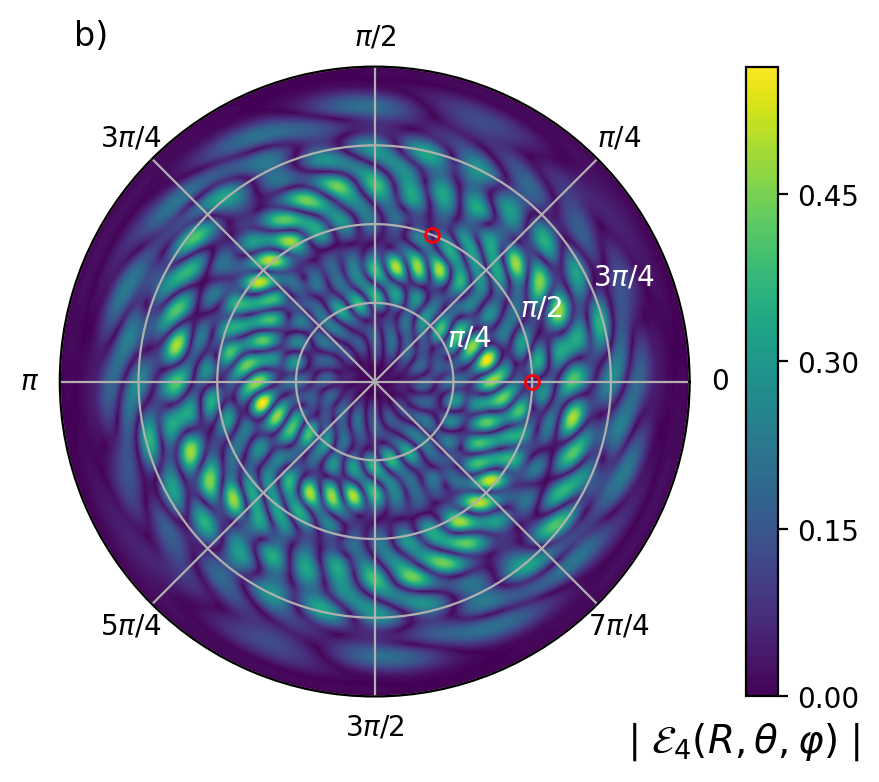}
\includegraphics[scale=0.59]{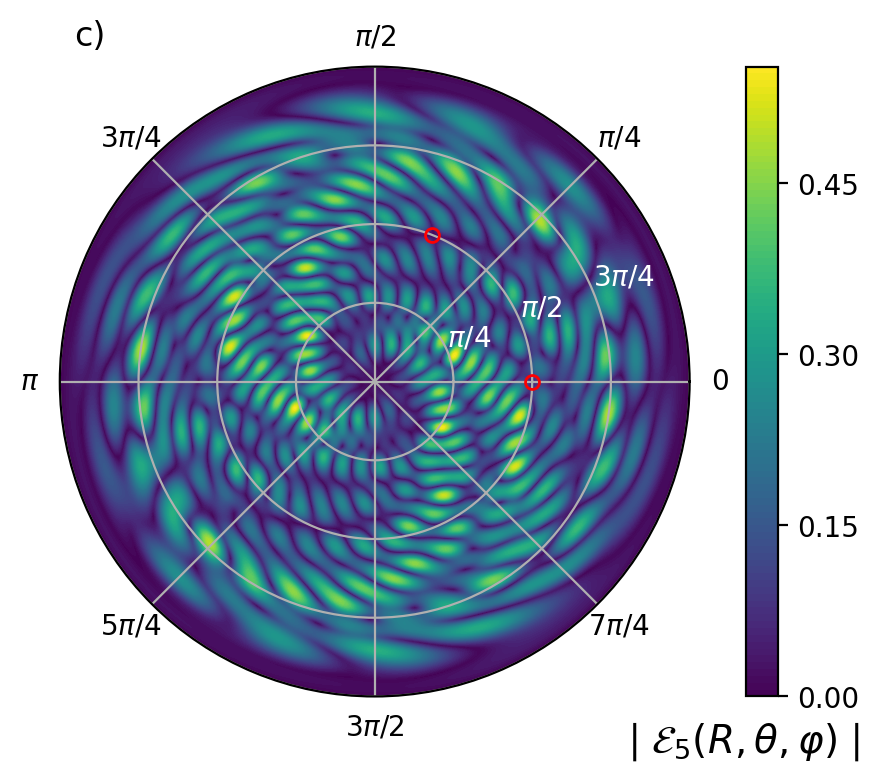}
\includegraphics[scale=0.59]{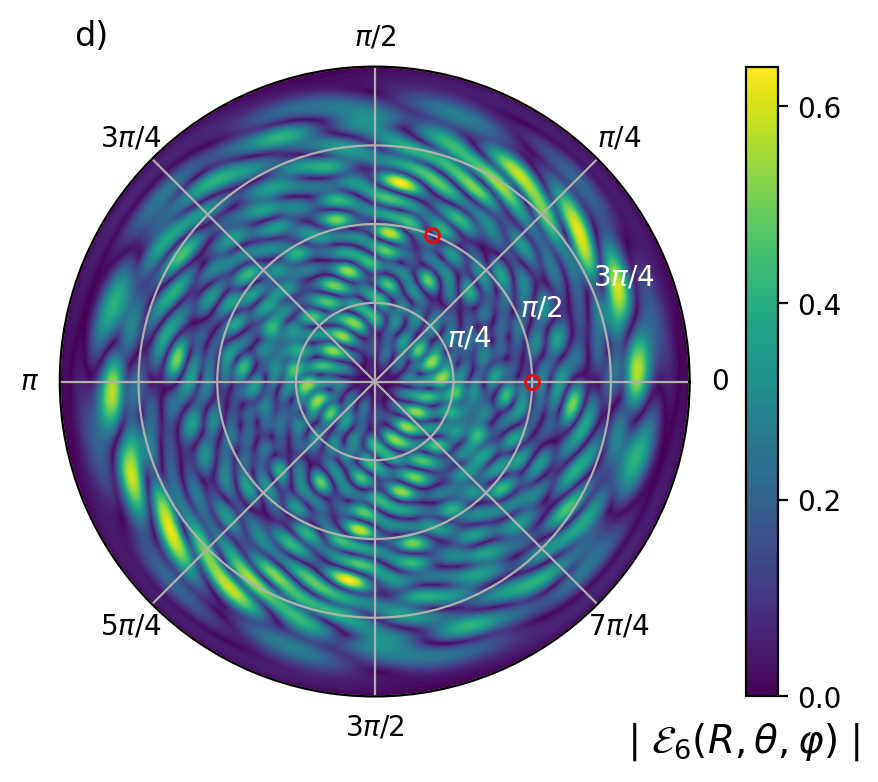}
\includegraphics[scale=0.59]{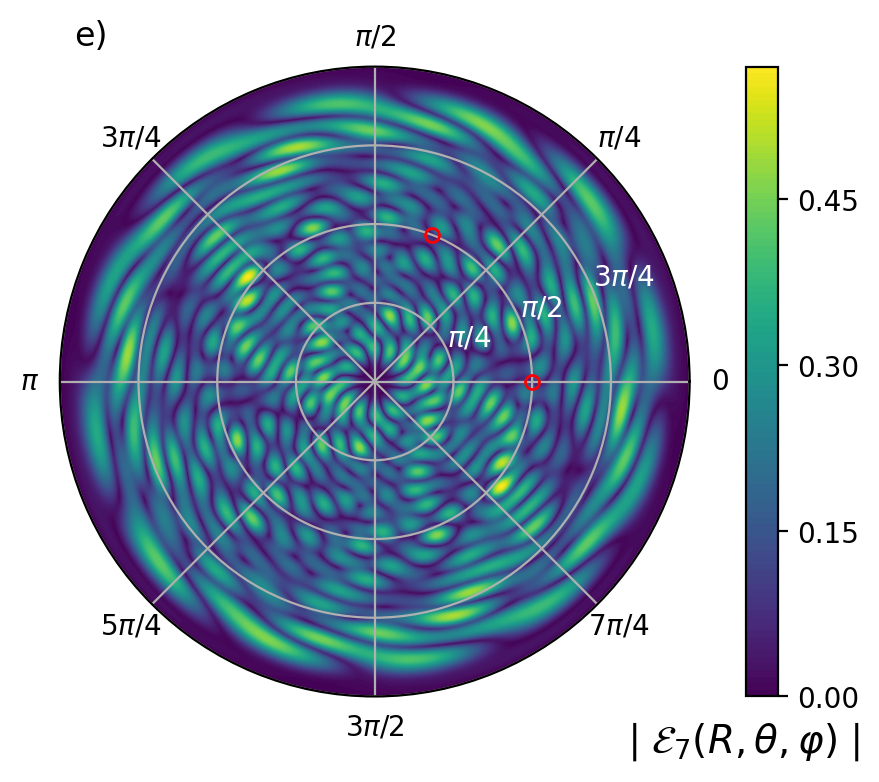}
\includegraphics[scale=0.59]{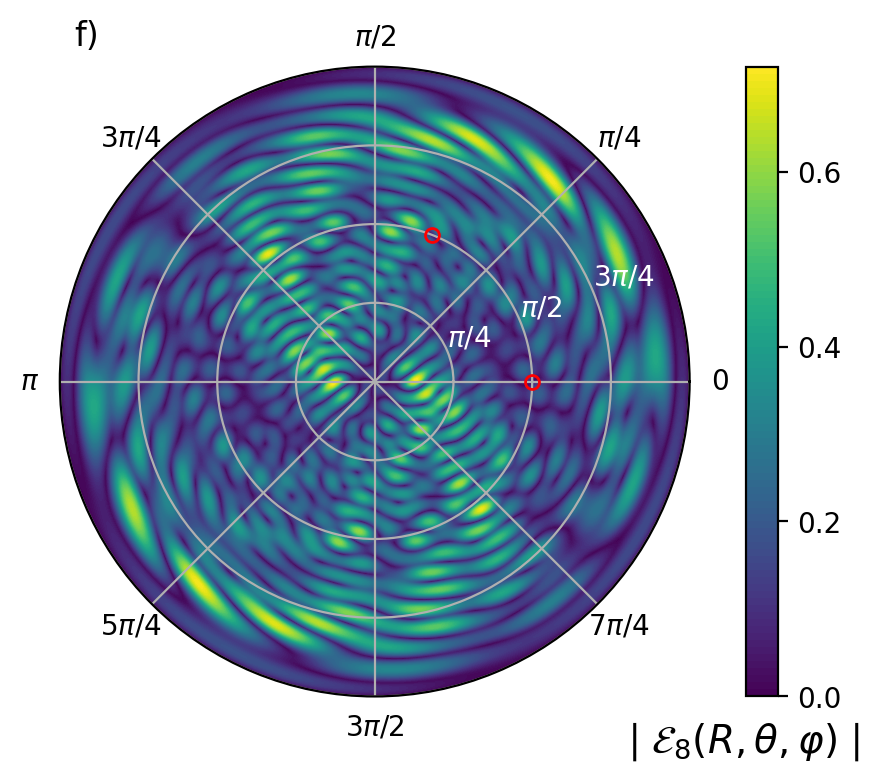}
\caption{(Color online) As Fig.\ref{1st WF}(a) but with panels (a), (b), (c), (d), (e), and (f) showing the wave functions of states $\nu=3,4,5,6,7$, and 8, respectively.
}
\label{2nd WF}
\end{figure*}

Figures \ref{2nd WF}-\ref{last WF} show the perturbed sphere-surface electric fields $\lvert\mathbfcal{E}_{\bar{\nu}}(R,\theta,\varphi)\rvert$ at the parameters of the EP in Sec.\ref{20.1} of the 18 unaffected WGMs, which are solutions of the $20\times20$ RSE matrix problem, with their eigenvalues unaffected by the perturbation.

Since the unaffected eigenfunctions are degenerate at a DP (not an EP), they make up 18 orthogonal electric fields. A DP does not enhance electric fields like an EP so these fields, which are normalized according to Eq.(\ref{norm}), have much weaker maxima than the affected states at an EP. The most significant feature of the unaffected eigenfunctions is that they vanish at the perturber positions $\mathbf{r}_j$, as it is clear from Figs.\ref{2nd WF}-\ref{last WF}.

\begin{figure*}
\includegraphics[scale=0.59]{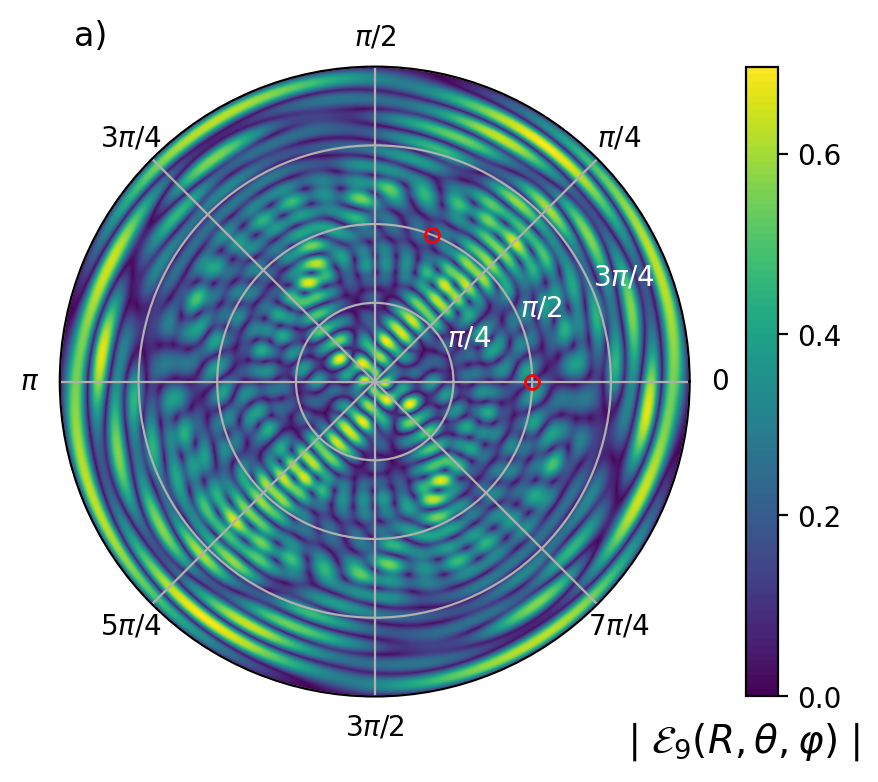}
\includegraphics[scale=0.59]{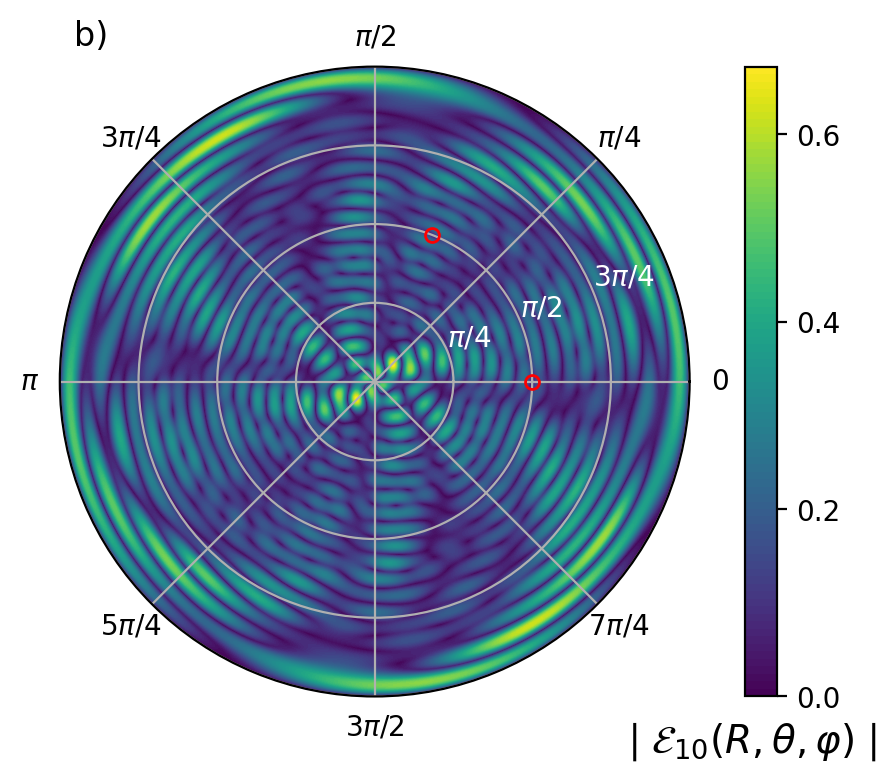}
\includegraphics[scale=0.59]{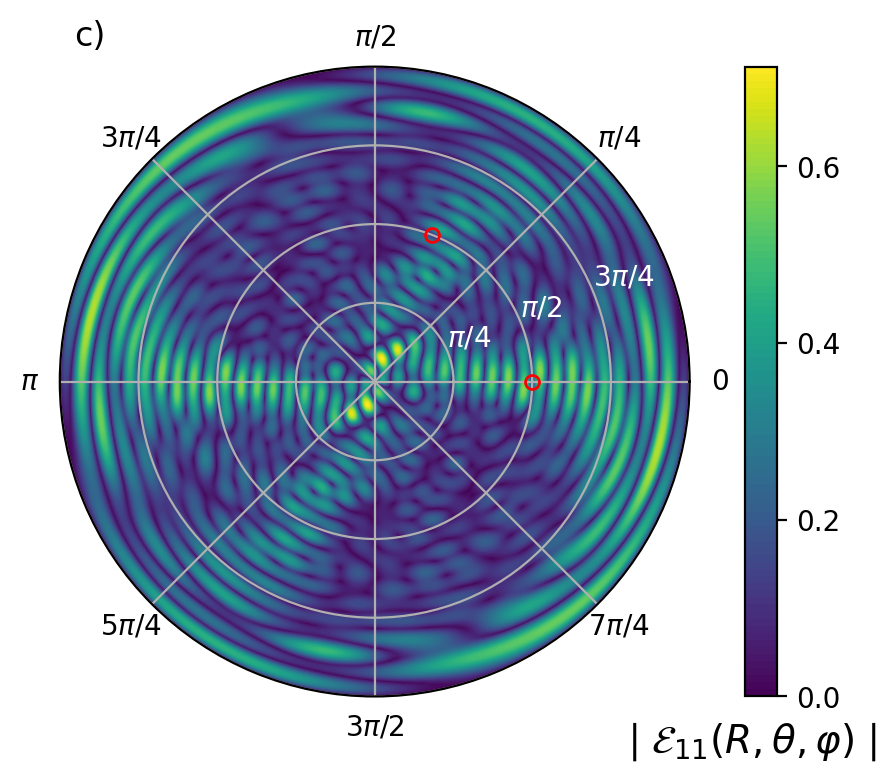}
\includegraphics[scale=0.59]{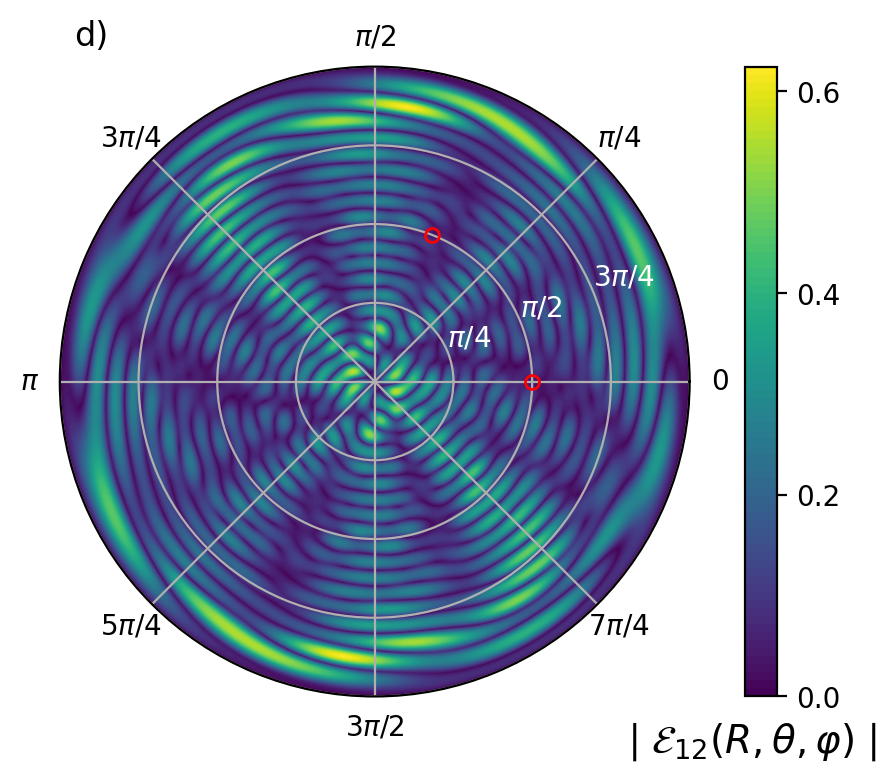}
\includegraphics[scale=0.59]{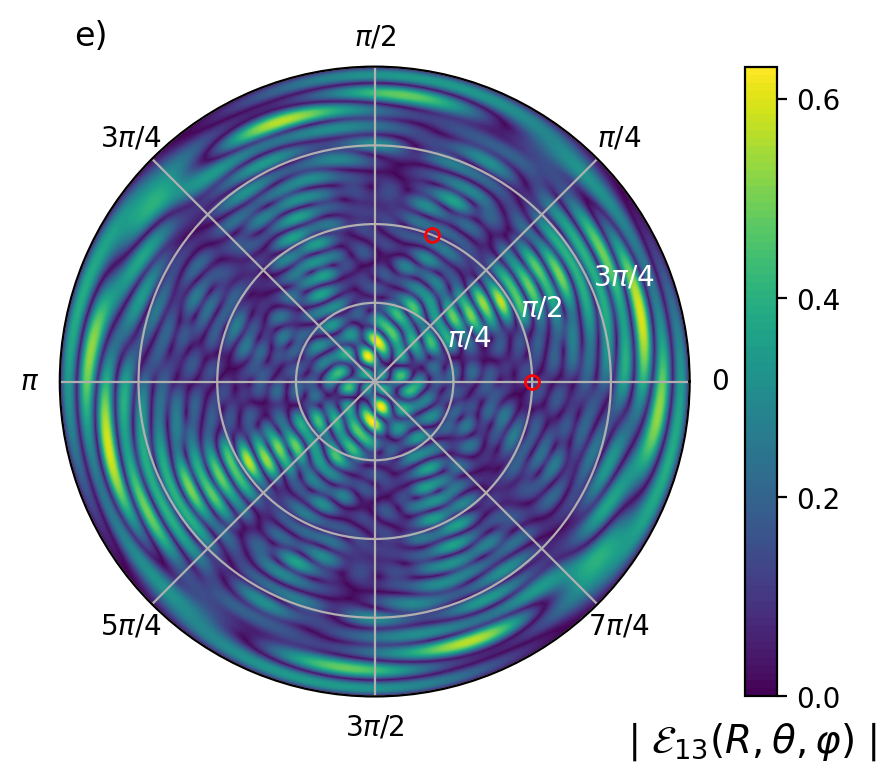}
\includegraphics[scale=0.59]{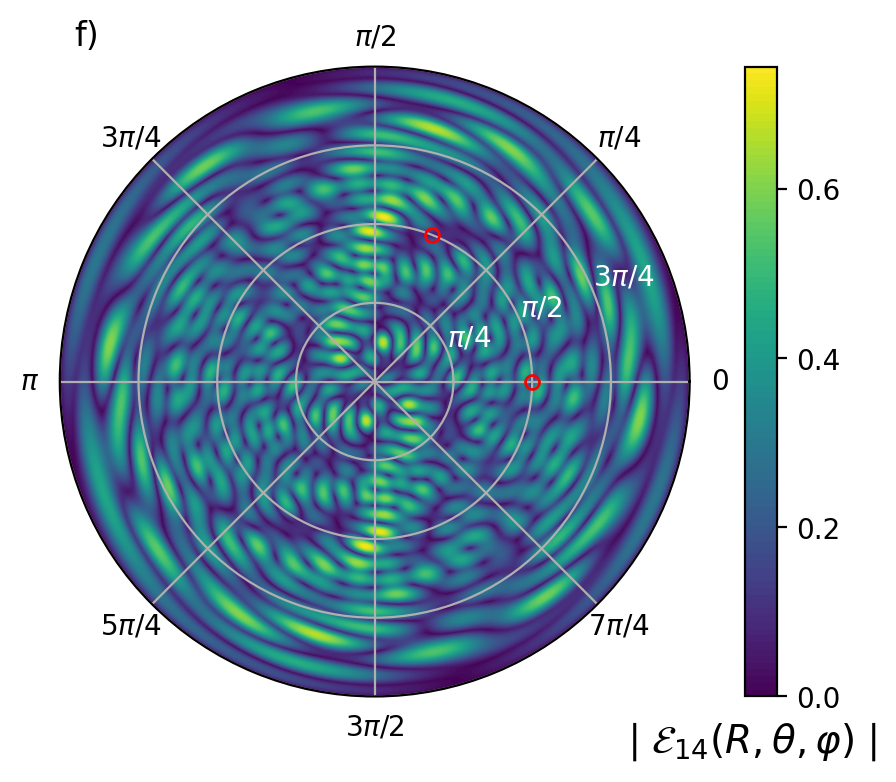}
\caption{(Color online) As Fig.\ref{1st WF}(a) but with panels (a), (b), (c), (d), (e), and (f) showing the wave functions of states $\nu=9,10,11,12,13$, and 14, respectively.
}
\end{figure*}

\begin{figure*}
\includegraphics[scale=0.59]{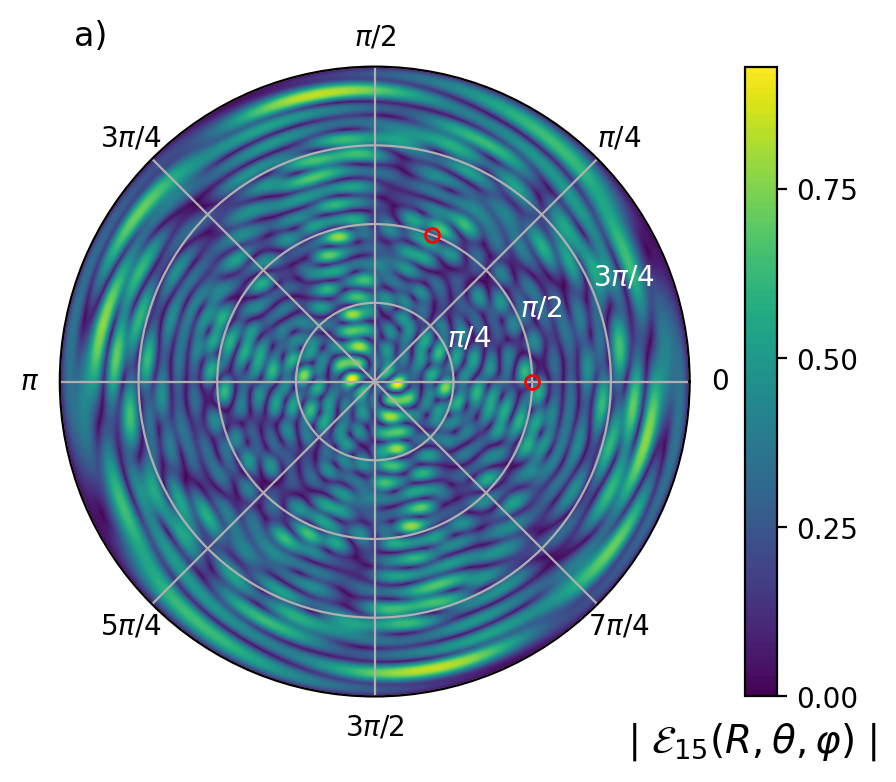}
\includegraphics[scale=0.59]{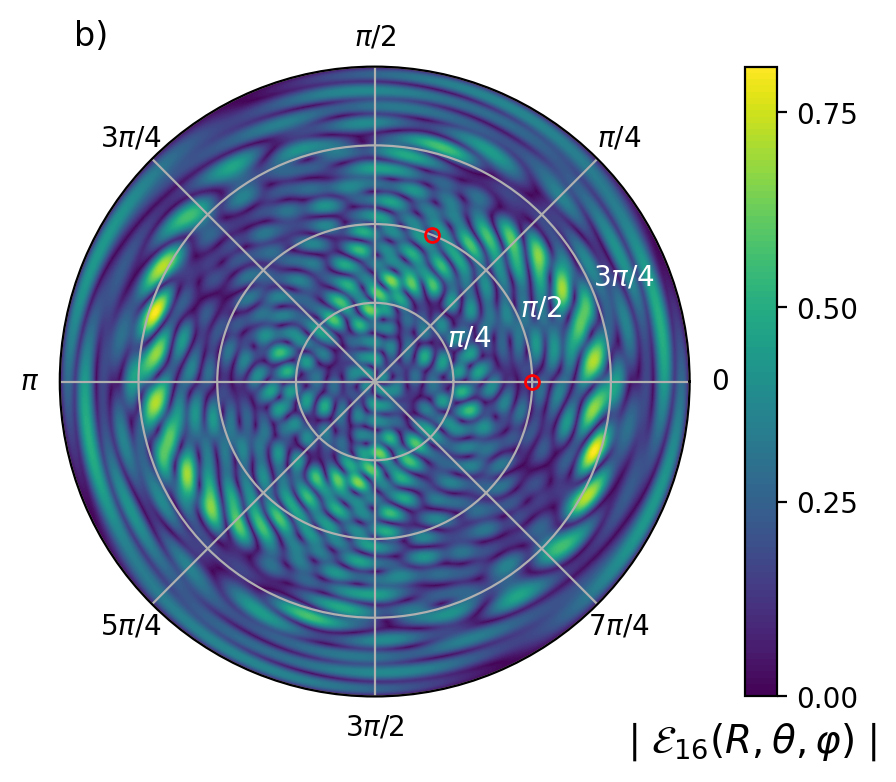}
\includegraphics[scale=0.59]{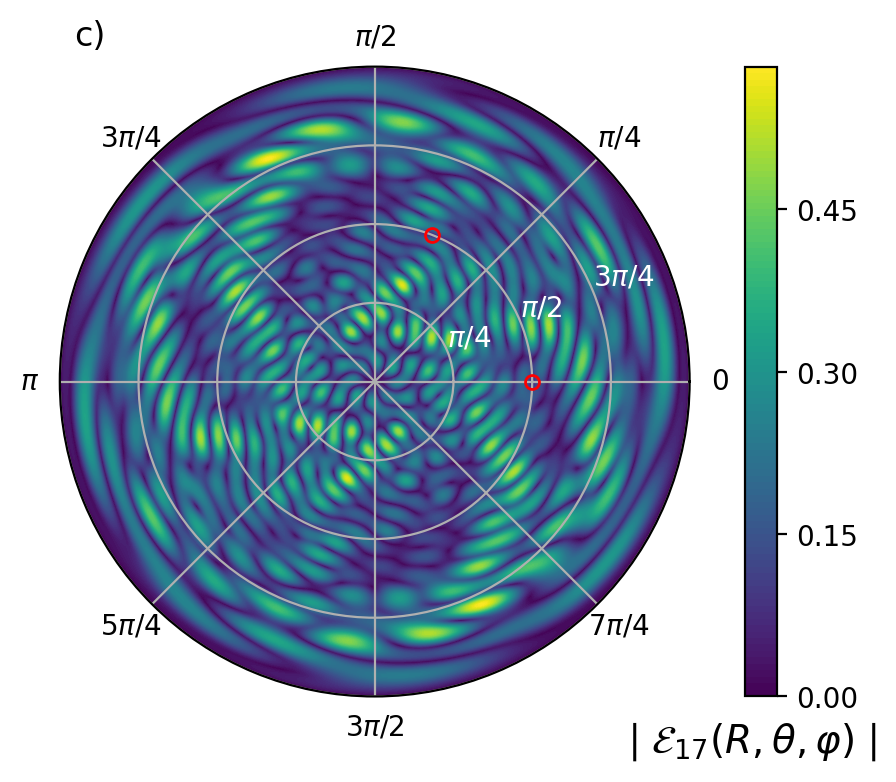}
\includegraphics[scale=0.59]{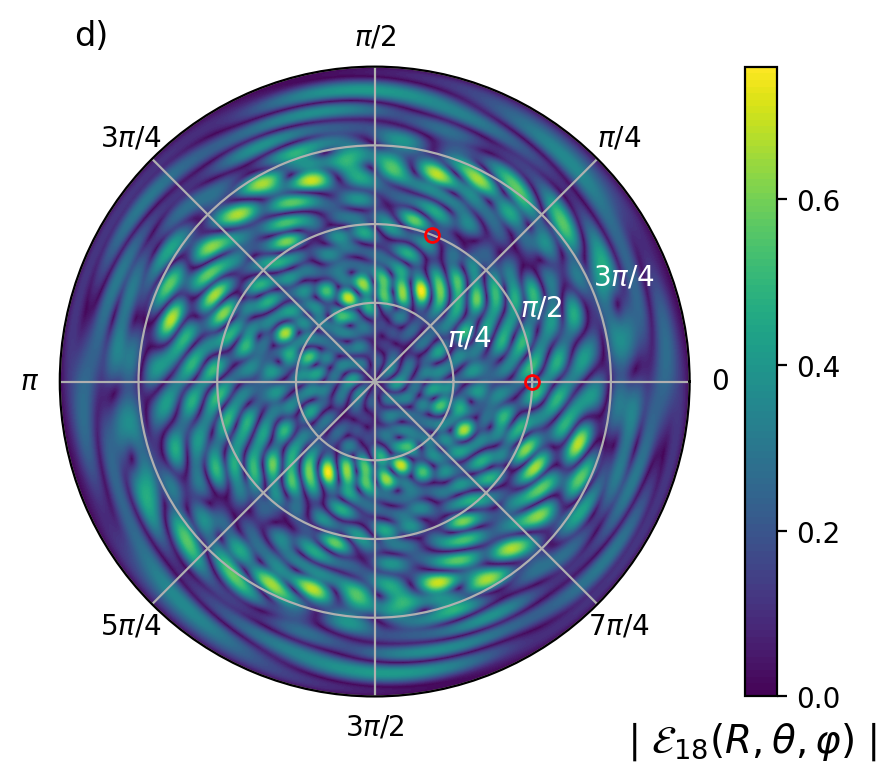}
\includegraphics[scale=0.59]{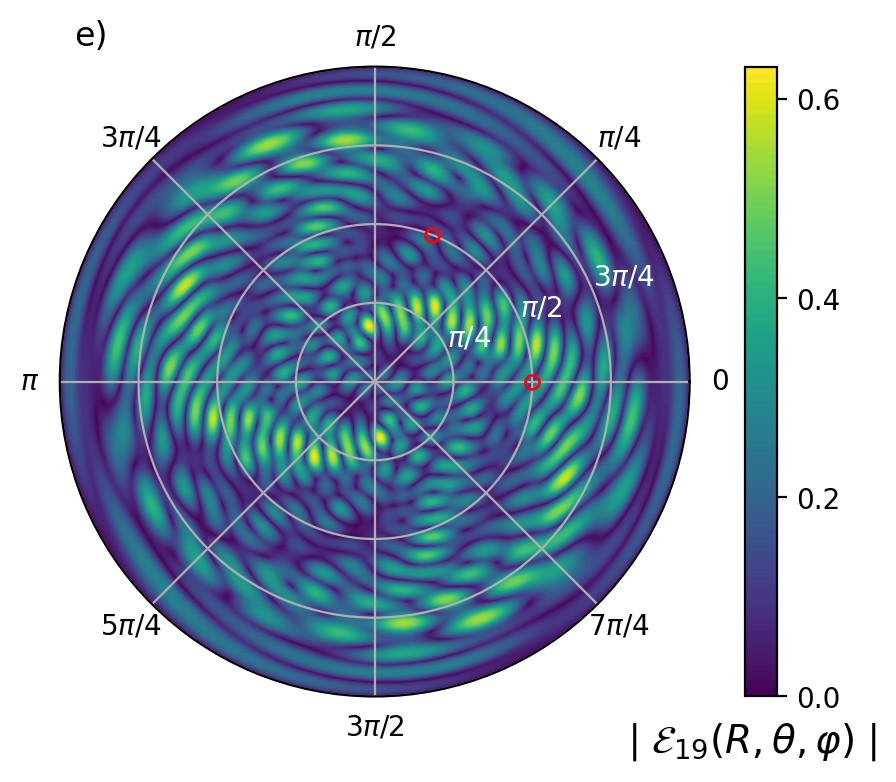}
\includegraphics[scale=0.59]{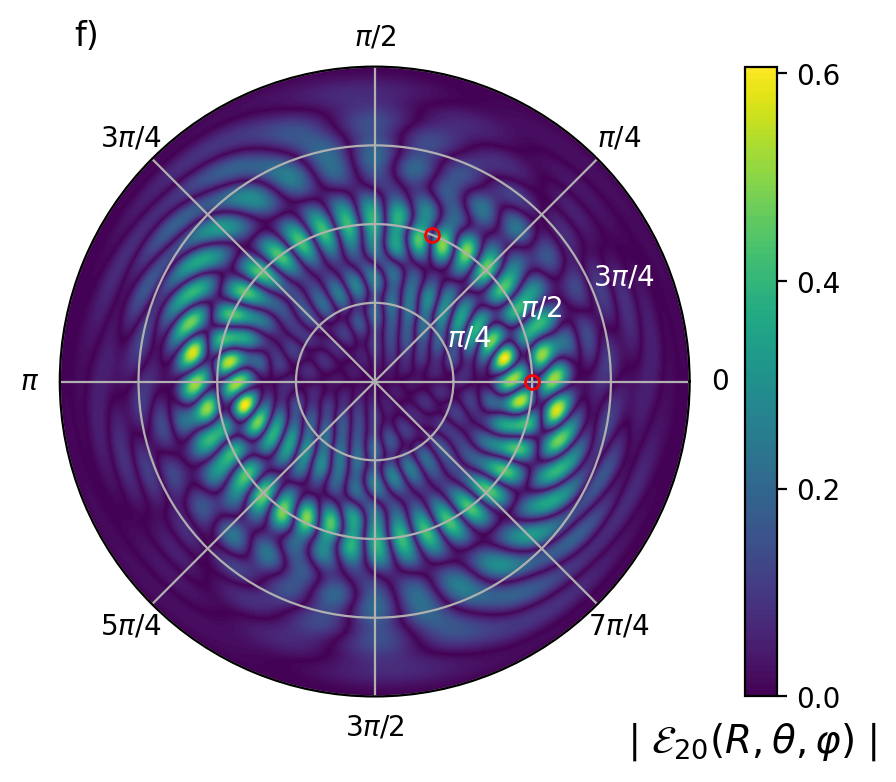}
\caption{(Color online) As Fig.\ref{1st WF}(a) but with panels (a), (b), (c), (d), (e), and (f) showing the wave functions of states $\nu=15,16,17,18,19$, and 20, respectively.
}
\label{last WF}
\end{figure*}

\bibliographystyle{apsrev4-2}

\end{document}